%% file: o3_lensing_technical_document.tex
\documentclass[usenatbib, twocolumn]{mnras}
\usepackage{calc}
\usepackage{mathtools,graphicx}
\usepackage{pifont}
\usepackage{bm}
\usepackage{microtype}
\usepackage{booktabs}
\usepackage{times}
\usepackage[varg]{txfonts}
\usepackage[utf8]{inputenc}
\usepackage[caption=false]{subfig}
\usepackage{paralist}
\usepackage[normalem]{ulem}
\usepackage{multirow}
\usepackage{etoolbox}
\usepackage{longtable}
\usepackage{xstring}
\usepackage{xparse}
\usepackage[left]{lineno}
\usepackage{blindtext}
\usepackage[nolist,nohyperlinks,printonlyused]{acronym}
\usepackage{dcolumn}
\usepackage{author_macros}
\usepackage{diagbox}
\usepackage{mathrsfs}
\usepackage{amsmath,amsfonts,amssymb,pifont,gensymb}
\usepackage{multirow,enumerate}
\usepackage{natbib}
\usepackage{slashbox}

\title[Follow-up Analyses to the O3 LVK Lensing Searches]{Follow-up Analyses to the O3 LIGO-Virgo-KAGRA Lensing Searches}
\input{authors_mnras.tex}

\date{\today}
\pubyear{2023}

\begin{document}

\maketitle

\begin{abstract}
	\input{abstract.tex}
\end{abstract}

\section{Introduction}\label{Sec:Intro}
\input{introduction.tex}

\section{Lensing Regimes and Analysis Methods}\label{Sec:TheoryIntro}
\input{lensing-theory.tex}

\section{GW190412}\label{sec:GW190412}
\input{GW190412.tex}

\section{GW191103--GW191105}\label{Sec:GW191103_GW191105}
\input{GW191103_GW191105.tex}

\section{GW191230\_180458--LGW200104\_180425 }\label{Sec:GW19130}
\input{GW191230.tex}

\section{GW200208\_130117}\label{Sec:GW200208}
\input{GW200208.tex}

\section{Conclusions and Prospects}\label{Sec:Conclusions}
\input{ConclusionsAndProspects.tex}

\section*{Acknowledgments}
\input{Acknowledgments.tex}

\section*{Data availability}
The data underlying this article will be shared in reasonable request to the corresponding authors.

\bibliographystyle{aasjournal}
\bibliography{lensing.bib}

\onecolumn
\appendix

\section{LensID GW191103--GW191105 Investigations}\label{app:lensid}
\input{appendix_lensID.tex}

\section{Discarded Targeted \textsc{PyCBC} Sub-Threshold Search Triggers}\label{app:pycbc}
\input{app_pycbcsub}

\section{Details of the Millilensing Investigation}\label{app:milli_lensing}
\input{appendix_millilensing.tex}

\end{document}

%% file: authors_mnras.tex
\author[Janquart, Wright, Goyal et al.]{\textbf{Editorial Team:}
    J. Janquart,$^{1, 2}$\thanks{E-mail: j.janquart@uu.nl} 
    M. Wright,$^{3}$\thanks{m.wright.3@research.gla.ac.uk}
    S. Goyal,$^{4}$ 
   \newauthor\textbf{Analysts:}
    J. C. L. Chan,$^{5}$
    A. Ganguly,$^{4, 6}$
    \'A. Garr\'on,$^{7}$
    D. Keitel,$^{7, 8}$
    A. K. Y. Li,$^{9}$
    A. Liu,$^{10}$
    \newauthor
    R. K. L. Lo,$^{9}$
    A. Mishra,$^{6}$
    A. More,$^{6, 11}$
    H. Phurailatpam,$^{10}$
    P. Prasia,$^{6}$
    \newauthor\textbf{Contributors:}
    P. Ajith,$^{4, 12}$
    S. Biscoveanu,$^{13, 14}$
    P. Cremonese,$^{7}$
    J.R. Cudell,$^{15}$
    J. M. Ezquiaga,$^{5}$
    \newauthor
    J. Garcia-Bellido,$^{16}$
    O. A. Hannuksela,$^{10}$
    K. Haris,$^{1, 2}$
    I. Harry,$^{8}$
    M. Hendry,$^{3}$
    S. Husa,$^{7}$
    S. Kapadia,$^{6}$
    \newauthor
    T. G. F. Li,$^{17, 18, 10}$
    I. Maga\~na Hernandez,$^{19, 20}$
    S. Mukherjee,$^{21}$
    E. Seo,$^{3}$
    C. Van Den Broeck,$^{1, 2}$
    J. Veitch$^{3}$
\\
\\
$^{1}$Institute for Gravitational and Subatomic Physics (GRASP), Department of Physics, Utrecht University,\\ Princetonplein 1, 3584 CC Utrecht, The Netherlands \\
$^{2}$Nikhef – National Institute for Subatomic Physics, Science Park, 1098 XG Amsterdam, The Netherlands\\
$^{3}$SUPA, School of Physics and Astronomy, University of Glasgow, Scotland\\
$^{4}$International Centre for Theoretical Science, Tata Institute of Fundamental Research, Bangalore---560089, India\\
$^{5}$Niels Bohr International Academy, Niels Bohr Institute, Blegdamsvej 17, DK-2100 Copenhagen, Denmark\\
$^{6}$The Inter-University Centre for Astronomy and Astrophysics, Post Bag 4, Ganeshkhind, Pune 411007, India\\
$^{7}$Departament de F{\'i}sica, Universitat de les Illes Balears, IAC3--IEEC, E-07122 Palma, Spain\\
$^{8}$University of Portsmouth, Institute of Cosmology and Gravitation, Portsmouth PO1 3FX, United Kingdom\\
$^{9}$LIGO, California Institute of Technology, Pasadena, CA 91125, USA\\
$^{10}$Department of Physics, The Chinese University of Hong Kong, Shatin, New Territories, Hong Kong\\
$^{11}$Kavli Institute for the Physics and Mathematics of the Universe, 5-1-5 Kashiwanoha, Kashiwa-shi, Chiba 277-8583, Japan\\
$^{12}$Canadian Institute for Advanced Research, CIFAR Azrieli Global Scholar, MaRS Centre, West Tower, 661 University Ave, Toronto, ON M5G 1M1, Canada\\
$^{13}$LIGO Laboratory, Massachusetts Institute of Technology, 185 Albany St, Cambridge, MA 02139, USA\\
$^{14}$Department of Physics and Kavli Institute for Astrophysics and Space Research, Massachusetts Institute of Technology, \\ 77 Massachusetts Ave, Cambridge, MA 02139, USA\\
$^{15}$STAR Institute, B\^atiment B5, Universit\'e de Li\`ege, Sart Tilman B4000 Li\`ege, Belgium\\
$^{16}$Instituto de F{\'\i}sica Te{\'o}rica UAM/CSIC, Universidad Autonoma de Madrid, 28049 Madrid, Spain\\
$^{17}$Department of Physics and Astronomy, KU Leuven, Celestijnenlaan 200D, B-3001 Leuven, Belgium\\
$^{18}$Department of Electrical Engineering (ESAT), KU Leuven, Kasteelpark Arenberg 10, B-3001 Leuven, Belgium\\
$^{19}$University of Wisconsin-Milwaukee, Milwaukee, WI 53201, USA\\
$^{20}$McWilliams Center for Cosmology, Department of Physics, Carnegie Mellon University, Pittsburgh, PA 15213, USA\\
$^{21}$Department of Astronomy \& Astrophysics, Tata Institute of Fundamental Research, Homi Bhabha Road, Mumbai, 400005, India\\
}

%% file: abstract.tex
\noindent 
Along their path from source to observer, gravitational waves may be gravitationally lensed by massive objects leading to distortion in the signals. Searches for these distortions amongst the observed signals from the current detector network have already been carried out, though there have as yet been no confident detections. However, predictions of the observation rate of lensing suggest detection in the future is a realistic possibility. Therefore, preparations need to be made to thoroughly investigate the candidate lensed signals. In this work, we present some follow-up analyses that could be applied to assess the significance of such events and ascertain what information may be extracted about the lens-source system by applying these analyses to a number of O3 candidate events, even if these signals did not yield a high significance for any of the lensing hypotheses. These analyses cover the strong lensing, millilensing, and microlensing regimes.  Applying these additional analyses does not lead to any additional evidence for lensing in the candidates that have been examined. However, it does provide important insight into potential avenues to deal with high-significance candidates in future observations.

%% file: introduction.tex
Gravitational lensing of gravitational waves (GWs) happens when they pass nearby a massive object and the deformation of space-time caused by that object modifies their propagation. The observed modifications depend on the exact properties of the lens and include repeated events, phase shifts, changes in amplitude, beating patterns and distortions~\citep{Ohanian:1974ys, Thorne:1982cv, Deguchi:1986zz, Wang:1996as, Nakamura:1997sw, Takahashi:2003ix}.

When the lens is an extended high-mass object (e.g.\ a galaxy or galaxy cluster), the GW frequency evolution is unaffected as the geometric optics limit applies and results in multiple signals---called images---with different magnification, phase shift, and time delay~\citep{Wang:1996as, Dai:2017huk, Ezquiaga:2020gdt}. The relative time delays between the images range from minutes to years depending on the scale of the lens with shorter time delays for galaxies~\citep{Ng:2017yiu, Li:2018prc, Oguri:2018muv}, and longer ones for galaxy clusters~\citep{Smith:2017mqu, Smith:2018gle, Smith:2018kbc, Smith:2019dis, Robertson:2020mfh, Ryczanowski:2020mlt}. This lensing regime is referred to as strong lensing and the predicted rates imply non-negligible chances of detection for current ground-based detectors, with around one event in a thousand being strongly lensed~\citep{Ng:2017yiu, Li:2018prc, Oguri:2018muv, Wierda:2021upe, Xu:2022jif, LIGOScientific:2023bwz}. 

For lighter compact lensing objects (e.g.\ individual stars) significant frequency-dependent wave optics effects occur in the waveform~\citep{Deguchi:1986zz, Nakamura:1997sw, Takahashi:2003ix, Cao:2014oaa, Jung:2017flg, Lai:2018rto, Christian:2018vsi, Dai:2018enj, Diego:2019lcd, Caliskan:2022hbu}. Fields of objects such as stars can also lead to these effects, usually with more complex characteristics than for a single object~\citep{Diego:2019rzc, Pagano:2020rwj, Cheung:2020okf, mishra2021, Meena:2022unp}. This regime of lensing is known as microlensing to refer to the small size of the lens compared to the GW wavelength.

In between these two scales of lenses, is a region in which the multiple images produced by the lens have a time separation of the order of milliseconds, making the different images overlap which produces similar beating patterns to microlensing. Objects in this scale include, for instance, galactic sub-halos. With appropriate analysis tools~\citep{Liu:2023ikc} and the ability to consider this scale in the geometric optics regime, the different images can be resolved and analyzed, leading to information about the lensing object at the root of the phenomenon.

The search for GW lensing is motivated by the diverse scientific opportunities its observation would offer. Examples include, but are not limited to, precise localisation of the source~\citep{Hannuksela:2020xor, Yu:2020agu}, characterisation of the lens~\citep{Lai:2018rto, Diego:2019lcd, Oguri:2020ldf}, precision cosmology~\citep{Sereno:2011ty, Liao:2017ioi, Cao:2019kgn, Li:2019rns, Hannuksela:2020xor}, statistical cosmology~\citep{Xu:2022jif}, and tests of general relativity~\citep{Baker:2016reh, Collett:2016dey, Fan:2016swi, Goyal:2020bkm, Ezquiaga:2020gdt, goyal2023probing}.

The LVK collaboration has searched for strong lensing and for microlensing signatures in the following LVK observing runs: O1--O2~\citep{Hannuksela:2019kle}, O3a~\citep{LIGOScientific:2021izm}, and the full O3 run~\citep{LIGOScientific:2023bwz}, yielding no confident signatures. In parallel, other searches have been performed, confirming that no lensing features have been confidently detected so far~\citep{McIsaac:2019use, Li:2019osa, Dai:2020tpj, Liu:2020par}.

Nevertheless, in these searches, interesting candidates have been found. This is the case, for example, for GW190412 that shows some support for being a type II image, the GW191103--GW191105 pair for strong lensing---discarded only after the inclusion of both the population priors and selection effects---, and GW200208\_130117 which displays some features which are compatible, at low significance, with microlensing~\citep{LIGOScientific:2023bwz}. Although, ultimately, not confirmed as lensed, such events contain features representative of signatures one could find in genuinely lensed events. It is therefore important to see what sort of follow-up analyses one could do on such events to have a better grasp on their significance, and to extract a maximum of information about the systems. 

Possible avenues to achieve this in the future are presented in this work by applying them to these interesting O3 candidates. We follow up on strongly-lensed candidates by making a background distribution of simulated unlensed events in order to compute each candidate's false-alarm probability (FAP). We also compare the candidates to the most recent simulations as to see if we can identify the lens that could be at the root of such a lensed event. Additionally, we look for lensed electromagnetic (EM) counterparts by cross-matching with galaxy-lens catalogues.  Moreover, since some fainter counterparts are likely to be present in a strongly-lensed multiplet, we also follow up on an additional strongly-lensed candidate containing a supra-threshold event GW191230\_180458 and a weaker ``sub-threshold'' event LGW200104\_184028 identified for investigation by a new method~\citep{Goyal_in_prep}. We analyze this pair in more details in this work, showing that it is an intriguing pair but is unlikely to be lensed. We also analyse the most significant candidate microlensing events using different lens models, inferring the parameters of the lens models at the same time. We compare these models to investigate which is most likely. Moreover, we analyze the most significant of these microlensing candidates with a millilensing framework to see if the signatures could come from a source in this lensing regime. We do not report any additional evidence for lensing but outline some important next steps to further deal with a possibly lensed event.

We stress that whilst the events discussed in this paper may be treated as though they were lensed, they do not display significant evidence for lensing~\citep{LIGOScientific:2023bwz}. The goal of this work is to demonstrate the methodologies that can be used to dig deeper in the case of genuinely lensed events and to better assess the importance of candidates. To represent this, we refer to the events as ``lensed candidates'' in what follows. Additionally, since the events and event pairs analyzed in this work have been selected because they present interesting features, it is often the case that they lead to higher Bayes factors. However, this is generally not enough to claim lensing, and we would also require to have posteriors converging to a given value of the lensing parameters or a high significance compared to a background before considering an event as lensed. 

The rest of the paper is structured as follow. In Sec.~\ref{Sec:TheoryIntro}, we introduce GW lensing and its different regimes. Then, in Sec.~\ref{sec:GW190412}, we show the results for different new analyses performed on GW190412, an event flagged with some support for a type II image. Next, in Sec.~\ref{Sec:GW191103_GW191105}, we analyze the GW191103--GW191105 event pair, found to have some characteristics resembling the ones expected for strongly-lensed event pairs. We continue in Sec.~\ref{Sec:GW19130} by analyzing another event pair, GW191230--LGW200104, a pair made of a supra and a sub-threshold event. In Sec.~\ref{Sec:GW200208}, we analyze GW200208, an event found to have a mild support for microlensing in past searches. We then give our conclusions and prospects in Sec.~\ref{Sec:Conclusions}.

%% file: lensing-theory.tex
\begin{figure*}
\centering
\includegraphics[keepaspectratio, width=0.33\textwidth]{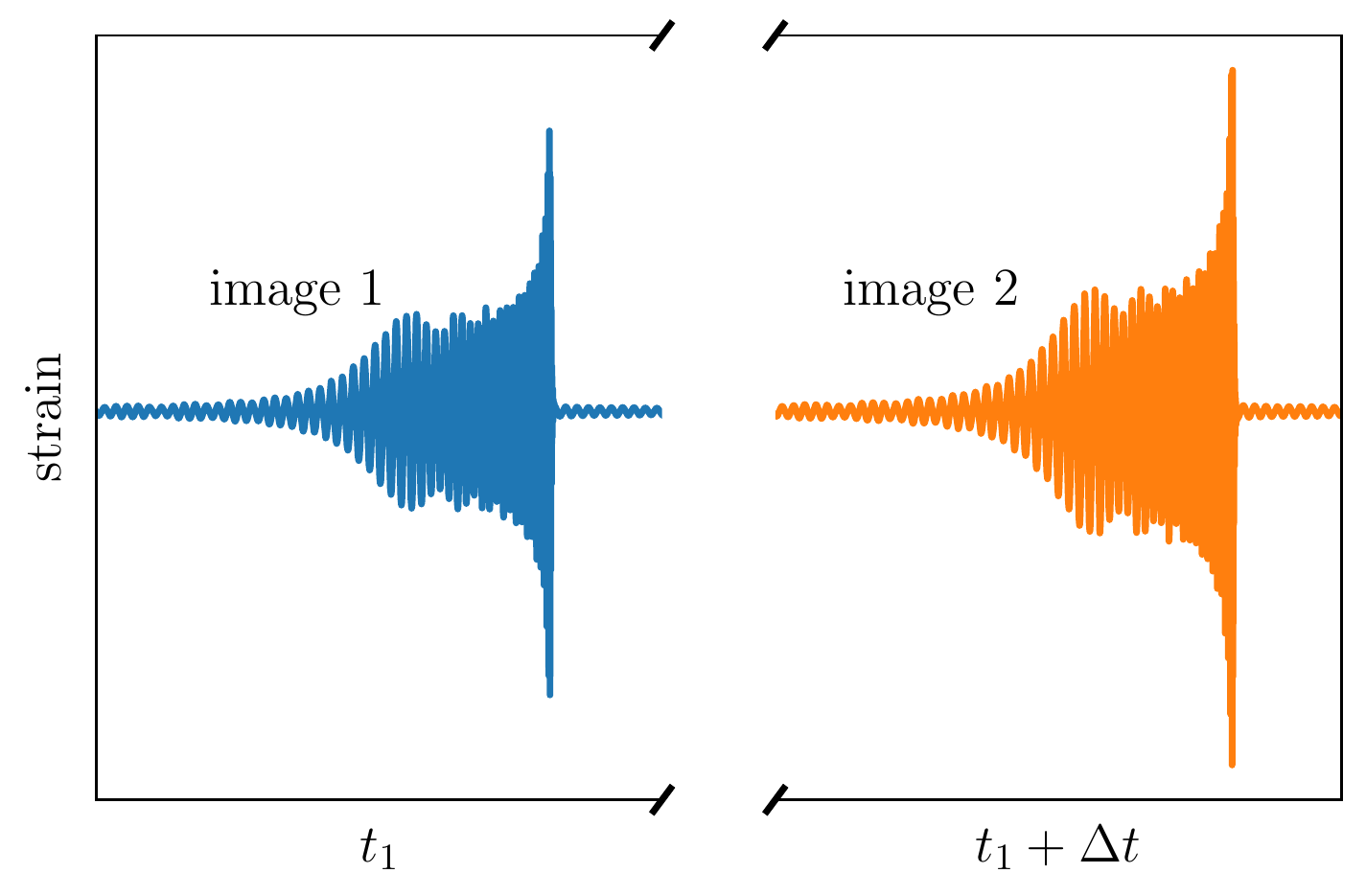}
\includegraphics[keepaspectratio, width=0.33\textwidth]{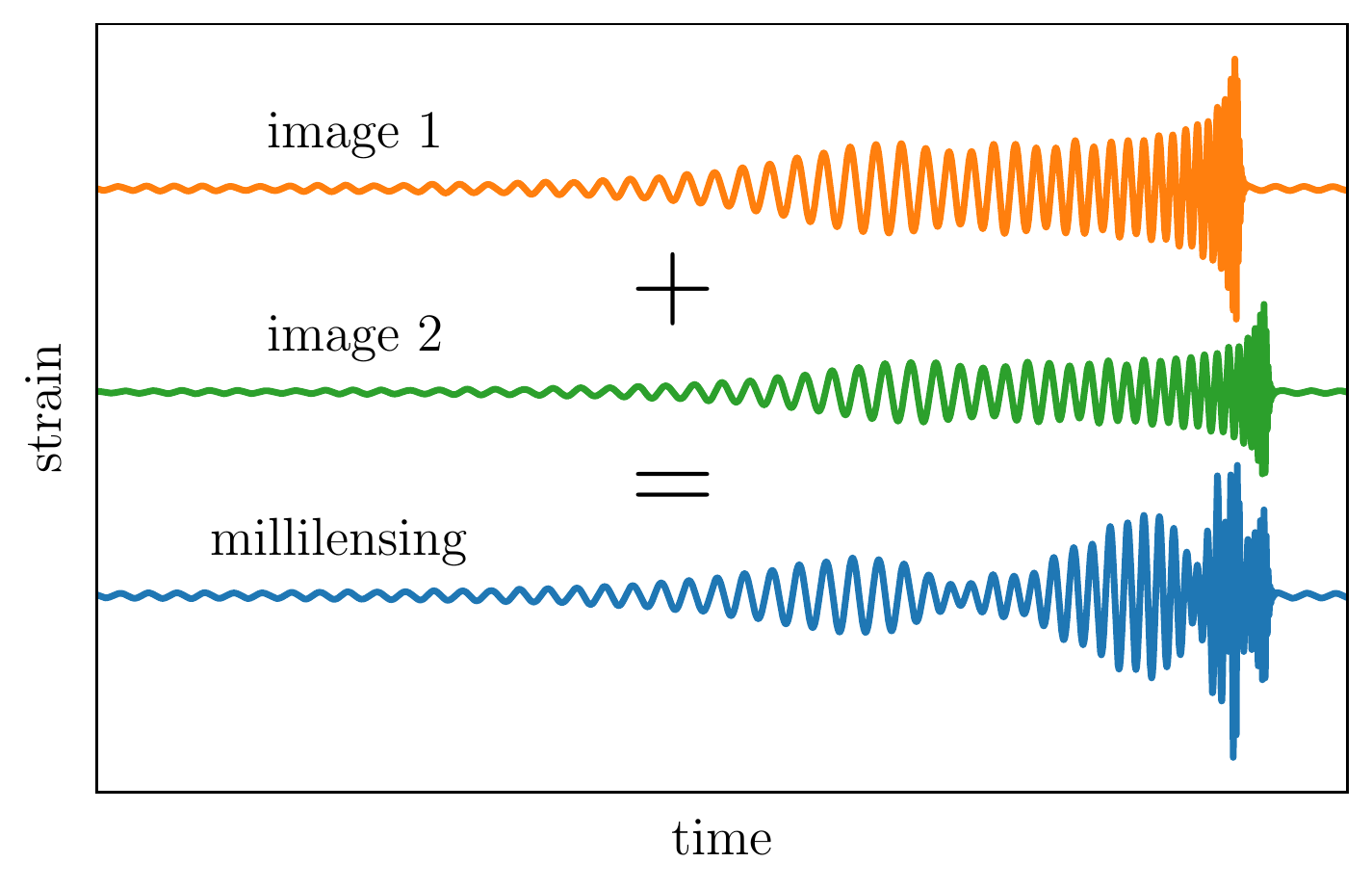}
\includegraphics[keepaspectratio, width=0.33\textwidth]{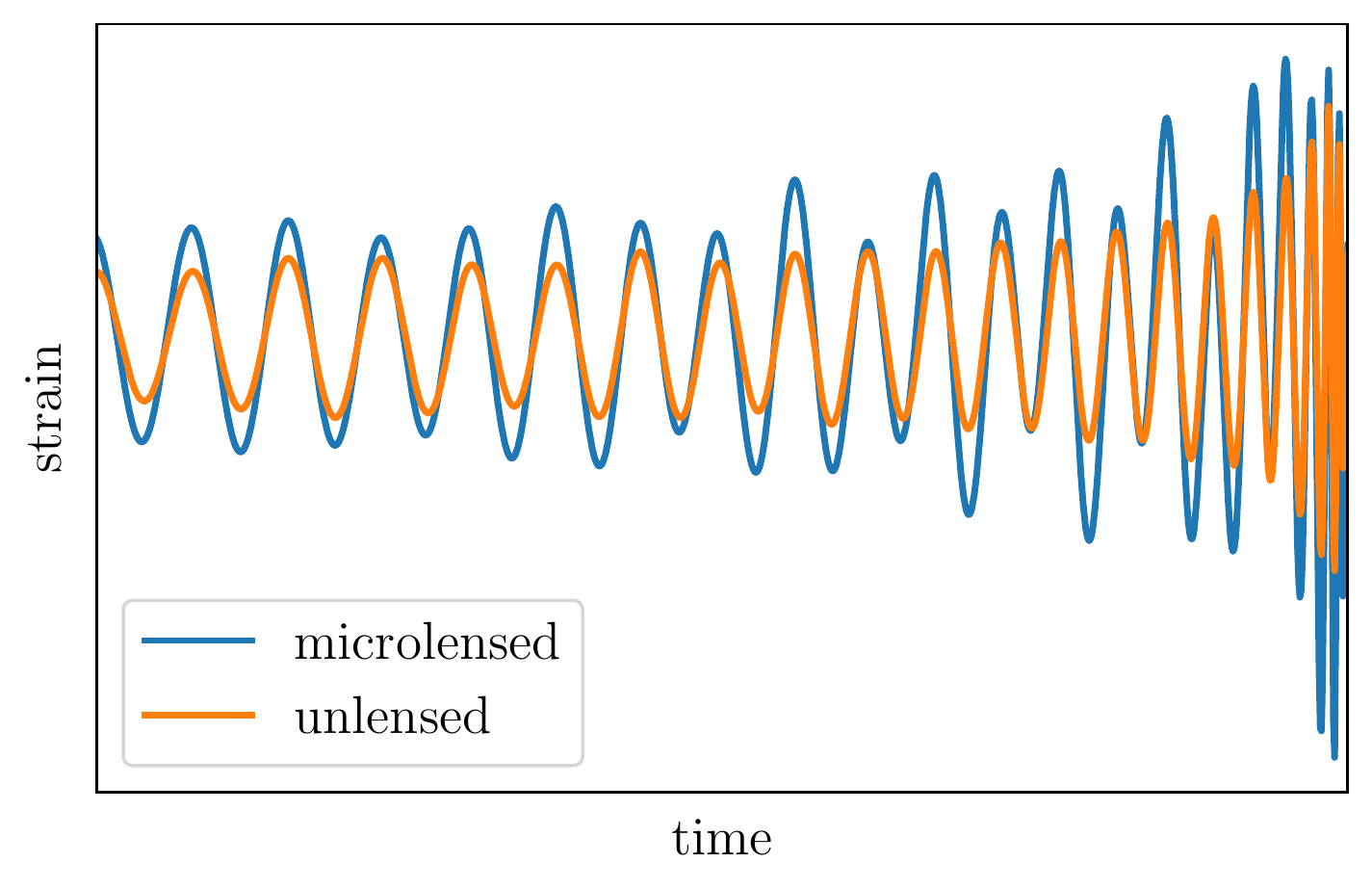}
\caption{A representation of the effects of the different types of lensing considered in this work on gravitational wave strain. From left to right, we have strong lensing---where one has multiple distinct images---, millilensing---where one has multiple images with a time separation such that they overlap, giving a modulated signal in the detector---, and microlensing---where one has frequency-dependent beating patterns.}\label{fig:representation_lensing}
\end{figure*}

Depending upon the characteristics of the lens and the configuration of the lens-source system, the effects of the space-time distortions on the GW will be different. Categories of lensing are usually grouped depending upon whether \emph{geometric optics} is valid, or \emph{wave optics} must be taken into account. The former grouping contains \emph{strong lensing} and \emph{millilensing}, whereas the latter grouping contains \emph{microlensing}. A representation of the effect of each of these types of lensing on GWs is given in Fig.~\ref{fig:representation_lensing}. Typically, in the wave optics regime, we see one single distorted waveform, while in the geometric optics regime, we get multiple images, possibly sufficiently separated in time to be distinguished from one another. 

\subsection{Strong Lensing}
When the GW travel path is close enough to a massive lens, gravitational lensing leads to several---possibly detectable---images having the same frequency evolution. This is called the strong lensing regime. The number of images and their specific characteristics depend on both the parameters of the lens and the lens-source configuration. Each of the images experiences a magnification, time delay, and phase shift compared to the unlensed waveform~\citep{1992grle.book.....S} and their strain, $h_{\mathrm{L}}^{j}(f)$, may be described by modifying the unlensed strain, $h_{\mathrm{U}}(f)$, such that

\begin{equation}\label{eq:StronglyLensedGW}
    h_{\mathrm{L}}^{j}(f; \theta, \mu_j, \Delta t_j, n_j) = \sqrt{|\mu_j|} \, e^{2i\pi f \Delta t_j - i \pi \mathrm{sign}(f) n_j} h_{\mathrm{U}}(f; \theta) \, ,
\end{equation}
where $\mu_j$ is the magnification factor related to the inverse of the determinant of the lensing Jacobian matrix, $\Delta t_j$ is the time delay related to the different geometrical path taken by the wave and the Shapiro delay~\citep{shapiro:1964gre}, and $n_j$ is the Morse factor which may take one of three distinct values, $0, 1/2$, and $1$ corresponding to the so-called type of the image which is, respectively, I, II, and III.\@ $\theta$ represents the usual compact binary coalescence (CBC) source parameters. In this paper, we only consider the lensing of binary black holes (BBHs). \@ Whilst the magnification and the time delay do not cause any changes to the waveform morphology, the three values of Morse phase may contribute distinct features. Most notable are type II images, where the overall de-phasing between the modes can lead to distortions in the waveform when there is a significant contribution from sub-dominant modes~\citep{Dai:2017huk, Ezquiaga:2020gdt, Wang:2021kzt, Janquart:2021nus, Vijaykumar:2022dlp}. By contrast, type I images have no extra shift at all, and type III images lead to a sign flip in the GW phase---completely degenerate with a $\pi/2$ shift in polarisation angle. 

This characterisation of strong lensing led to two distinct ways of searching for such events in the O3 data~\citep{LIGOScientific:2023bwz}: looking only for the type II image distortion in single images and looking for pairs of events that are compatible with the strong lensing hypothesis. For the multiple event search case, two main scenarios can be identified; when the images are sufficiently magnified so as to be detected by the usual CBC search pipelines, termed ``supra-threshold'', and when one or more of the images is not sufficiently magnified or is demagnified so as the resultant signal is termed ``sub-threshold''. In the latter case, one can look for the possible lensed counterpart of a supra-threshold event by building an event-specific template bank and searching the data~\citep{McIsaac:2019use, Li:2019osa}. So far, these searches have yielded no compelling evidence for strong lensing image pairs.  A new method to identify lensed sub-threshold candidates is proposed in~\citet{Goyal_in_prep}, and used in this work. To avoid the problem of performing computationally-expensive parameter estimation (PE) for many merely potential candidates, it relies on the matched filter-estimated chirp masses and the rapidly produced \textsc{Bayestar} skymaps~\citep{Singer_2016}. The matched filter estimations also provide millisecond-level precision on the event's time delay relative to the supra-threshold counterpart allowing the use of the lens model to further assess the probability of the pair being lensed~\citep{Haris:2018vmn}.  Here, we follow up on the most promising candidate from this analysis using the PE-based pipelines.

In this work, several methods are used to analyze the interesting lensing candidates and compute Bayes factors for lensed versus unlensed hypotheses: $\mathcal{B}^{L}_{U} = \mathcal{Z}_{L}/\mathcal{Z}_{U}$, where $\mathcal{Z}_{H} = P(d_1, d_2 |H)$ is the evidence under the hypothesis $H$ ($H = L$ for lensed or $H = U$ for unlensed), and $d_i = n_i + h_i^{H}$ is the data stream for the $i^{\mathrm{th}}$ image, made of a noise ($n_i$) and a GW ($h_i^{H}$) component. In this work, we adopt the same conventions as those used in~\citet{Lo:2021nae, Janquart:2021qov}, referring to the evidence ratio as the Bayes factor when it includes both the population and selection effects, and the coherence ratio when these are not included.

The first analysis method is called posterior overlap~\citep[PO,][]{Haris:2018vmn}. Since the frequency evolution of lensed images is unchanged, the detector frame parameters should match (except for those modified by lensing), there should be a significant overlap between the posteriors obtained for these images under the unlensed hypothesis. Therefore, one can compute a detection statistic comparing the evidence for the lensed and unlensed hypotheses
\begin{equation}\label{eq:POblu}
	\mathcal{B}^{\textrm{overlap}} = \int d\theta \frac{P(\theta | d_1)P(\theta | d_2)}{P(\theta)}  ,
\end{equation}
where $P(\theta | d_{i})$ is the posterior obtained from data $i$, and $P(\theta)$ is the prior. 

Typically, kernel-density estimators (KDEs) are performed on the posteriors in Eq.~\eqref{eq:POblu} for a subset of parameters (component masses, spins, inclination angle and sky location), and $\mathcal{B}^{\textrm{overlap}}$ is computed using those KDEs. The posteriors used often come from the usual unlensed PE~\citep{Veitch:2014wba, Ashton:2018jfp}.

Another method used is joint parameter estimation (JPE), where one performs the joint inference of the lensed images, linking them through the lensing parameters~\citep{McIsaac:2019use, Dai:2020tpj, Liu:2020par, Lo:2021nae, Janquart:2021qov, Janquart:2023osz}. These pipelines have two different ways to tackle the problem. Some compute the full joint evidence $p (d_1, d_2 | H_L)$ at once, such as those outlined in~\cite{Liu:2020par} and~\cite{Lo:2021nae}. Here, we use the \textsc{Hanabi} pipeline from~\cite{Lo:2021nae}. The alternative approach is to instead consider the evidence for the second image as conditional on that of the first~\citep{Janquart:2021qov, Janquart:2023osz}. This makes the computation faster and is equivalent to JPE under the lensed hypothesis. However, some level of approximation is added by doing sub-sampling to improve the speed. The pipeline undertaking this method used within this work is called \textsc{GOLUM}~\citep{Janquart:2021qov, Janquart:2023osz}. The analysis done using the pipeline for the first image also directly offers the possibility to search for type II images when higher-order modes are present~\citep{Ezquiaga:2020gdt, Wang:2021kzt, Lo:2021nae, Janquart:2021nus, Vijaykumar:2022dlp, LIGOScientific:2023bwz}.

In addition to matching purely on the observational parameters of the system, one can also use models of the lens to inform the strong-lensing detection process~\citep{Haris:2018vmn, Lo:2021nae, Wierda:2021upe, More:2021kpb, Janquart:2022zdd, hanabi_astro}. Lensed events do not only have matching frequency-domain evolution but they are also linked via the lensing parameters. Their measured values can be used to assess how likely it is for the observed events to be lensed for a given model. To do this, one compares the probability of having the apparent lensing parameters under the lensed and unlensed hypotheses. This may be done for all of the lensing parameters, or a subset of them. To obtain the probabilities, an unlensed population of BBH mergers is constructed using given population models (mass, redshift, spin, \dots distributions) and the phase differences, relative magnifications, and time delays are computed between these events. In parallel, the same process is performed on a lensed population, produced from a BBH population and a lens population following a specific model such as a Singular Isothermal Sphere (SIS~\citep{Witt:1990})~\citep{Haris:2018vmn} or a Singular Isothermal Ellipsoid (SIE~\citep{Koopmans2009})~\citep{Wierda:2021upe, More:2021kpb} for the galaxy lenses. From the two computed distributions, a probability density function can be obtained via, for example, KDE reconstruction. It is then possible to evaluate the ratio

\begin{equation}\label{eq:lens_params_prob_ratio}
	S^{\mathrm{gal}} = \frac{P\left(\Phi | \mathrm{H}_{\mathrm{L}}\right)}
				{P\left(\Phi | \mathrm{H}_{\mathrm{U}}\right)},
\end{equation}
where $\mathrm{H}_{\mathrm{L}}$ and $\mathrm{H}_{\mathrm{U}}$ designate the lensed and unlensed hypotheses respectively, and $\Phi$ is the set of lensing parameters under consideration.

Specific examples of the statistics computed with this method for this work are $\mathcal{R}^{\mathrm{gal}}$~\citep{Haris:2018vmn} using specifically the time delay, and $\mathcal{M}^{\mathrm{gal}}$~\citep{More:2021kpb} which may use all or a subset of the lensing parameters. Both models can be used either with an SIS or an SIE lens model. These statistics are used to select candidates to be followed up by the more extensive analyses or are multiplied with the detection statistics evaluating the match between the parameters. Though model dependent, this in general decreases the risk of false alarm detections~\citep{Haris:2018vmn, Wierda:2021upe, Janquart:2022zdd}.

One can also formally combine the effect of these lensing statistics with the coherence ratio. To do this, one reweights the evidence under the lensed and unlensed hypothesis with a weight computed based on the lens model~\citep{Janquart:2022zdd}. This model-reweighted coherence ratio thus evaluates the ratio of lensed and unlensed evidence for a given lens model directly relating the observed binary parameters and the lensing parameters for a given model. Additionally, the above lensing statistics can also be used when computing the selection effects to obtain the final Bayes factor.

Alternatively, one can consistently incorporate the information from a lens and a source population model~\citep{Lo:2021nae}, where the lens and the source population model affect both the probability of observing a given set of data, in this case $\left( d_{1}, d_{2} \right)$, under the lensed and the unlensed hypothesis. Specifically, the lens population model informs the joint probability distribution on the magnification, the image type, and the time delay between images, as well as the optical depth for strong lensing, while the source population model informs the distribution of the (true) redshift and the source parameters of a lensed source. This was already done in~\cite{LIGOScientific:2021izm} and~\cite{LIGOScientific:2023bwz} using the simple SIS lens model.

In practice, it is difficult to write down an analytical form for the above-mentioned joint probability distribution from a lens model except for some simple lens models (e.g.\ the SIS model), and instead one usually resorts to constructing a surrogate that approximates the probability density function, such as the aforementioned KDE technique. However, it can be computationally expensive to use KDE-based schemes to construct an estimate for the probability density from a catalog of simulated lensed images that contains many (e.g.\ millions of) samples, which in turn is evaluated over a set of (roughly tens of thousands of) posterior samples.

In this work, we use the probability density surrogate described in~\cite{hanabi_astro} that fits the joint probability density on the magnification and the image type conditioned on the time delay between images from a catalog of mock lens images used in~\cite{Oguri:2018muv} using a normalization-flow-based method~\citep{denmarf}. The underlying strong lensing model adopted in the simulation is a population of galaxy-scale SIE lenses with external shear. The lens-redshift-dependent velocity dispersion function is constructed from hybridizing the velocity dispersion measurement for the local Universe derived from the Sloan Digital Sky Survey Data Release 6~\citep{10.1111/j.1365-2966.2010.16425.x} with the Illustris simulation result for the velocity dispersion function at higher lens-redshifts~\citep{10.1093/mnras/stv1986}. The ellipticity and the external shear follow a Gaussian distribution and a log-normal distribution respectively with additional detail found in~\cite{Oguri:2018muv}.

\subsection{Millilensing}

When the mass of the lens and its extent is reduced, the different images produced by lensing can start overlapping. In such a case, they interfere, and one gets only one image exhibiting beating patterns. This is called millilensing~\citep{Liu:2023ikc}, which is expected to take place for lens masses in the range $M^{z}_{\textrm{Lens}}\in [10^2, 10^6] M_\odot$ for which the geometric optics approximation continues to hold. Therefore, the observed GW signal ($h_{\mathrm{Milli}}$) that results from the sum of the different images produced can be expressed as 
\begin{align}\label{eq:MillilensingWF}
h_{\mathrm{Milli}}(f; \theta, &\left\{d^\mathrm{eff}_j, t_j, n_j \right\}_{j = \{1, \ldots , K\}}) = \nonumber \\
& \bigg( \sum_{j = 1}^{K} \frac{d_L}{d^\mathrm{eff}_j} e^{2i\pi f t_j - i \mathrm{sign}(f) \pi n_j}\bigg) \, h_{\mathrm{U}}(f; \theta) \, ,
\end{align}
where $\theta$ represents the set of usual BBH parameters, $K$ is the total number of signals produced by lensing, and $\{d^\mathrm{eff}_j, t_j, n_j\}$ is the set of lensing parameters for image $j$: the effective luminosity distance\footnote{Note here that the magnification $\mu_j$ and the effective luminosity distance $d_j^{\mathrm{eff}}$ are linked through the source unbiased luminosity distance $d_L$ as $\mu_j = (d_L/d_j^{\mathrm{eff}})^2$.}, relative time delay and Morse factor, respectively. Note that the signal for each image has the same frequency evolution as the unlensed signal. However, the interaction between them makes for a non-trivial total lensed signal (see middle panel of Fig.~\ref{fig:representation_lensing} for an illustration).

To search for such signals, one needs to analyze the GW signal assuming that several lensed images are interfering with each other. Usually, the number of signals is not known beforehand. Therefore, it can either be fixed in the search or it can be a variable one tries to infer~\citep{Liu:2023ikc}.

\subsection{Microlensing}

For lens-source systems such that the wavelength of the GW is comparable to the Schwarzschild radius of the lens, frequency-dependent modulation of the amplitude occurs.  Observing such a phenomenon can offer insights into the nature of the lens itself~\citep{Takahashi:2003ix, Cao:2014oaa, Jung:2017flg, Lai:2018rto, Christian:2018vsi, Dai:2018enj, Diego:2019lcd, Diego:2019rzc, Pagano:2020rwj, Cheung:2020okf, mishra2021, Meena:2022unp}. Expected lenses in which this regime is applicable and could be detected by current ground-based detectors---objects with masses up to $\sim10^5 M_{\odot}$---include stellar-mass objects and intermediate-mass black holes. However, it is unlikely that a GW travelling to Earth would cross paths with only one such object as they are most often found in larger structures. As a consequence, a more realistic microlensing scenario would be the case of one or more microlenses embedded within a larger macrolens, such as a galaxy or galaxy cluster. In this case, the effect on the unlensed waveform is much more complicated~\citep{Diego:2019rzc, Cheung:2020okf, 10.1093/mnras/staa278, mishra2021, Yeung:2021roe}. This scenario is often unaccounted for because its modeling requires very computationally expensive modifications to the standard waveform models. In addition, this can also lead to a joint effect with strong lensing leading to multiple microlensed images~\citep{Seo:2022xxc}.

To maintain computational tractability, in~\citet{Hannuksela:2019kle, LIGOScientific:2021izm, LIGOScientific:2023bwz} the microlensing search has been performed using an isolated point mass model. Regardless of the exact model of the lens, the lensed ($h_{\mathrm{Micro}}$) and unlensed ($h_{\mathrm{U}}$) waveforms are linked as

\begin{equation}\label{eq:microlensed-isolated-point-source}
	h_{\mathrm{Micro}}(f; \theta, M^z_{\mathrm{Lens}}, y) = h_{\mathrm{U}}(f; \theta) \times F(f; M^{z}_{\textrm{Lens}}, y),
\end{equation}

where, as before, $\theta$ are the parameters defining the unlensed GW signal, $y$ is the dimensionless source position, $M^{z}_{\mathrm{Lens}}$ is the redshifted lens mass, and $F(f; M^{z}_{\mathrm{Lens}}, y)$ is the amplification factor which is dependent upon both the frequency and the mass-density profile of the lensing object (more details can be found in~\cite{Takahashi:2003ix} for example), hence on the lens model. 

Whilst the isolated point mass model has been used for its simplicity and consequent speed, there are other density profiles that may describe microlenses. These include, but are not limited to, the SIS~\citep{Witt:1990}, the SIE~\citep{Koopmans2009}, or the Navarro-Frenk-White (NFW)~\citep{NFW:1997} profiles. Efforts have been made to incorporate some of these models into microlensing GW analyses~\citep{Wright:2021cbn} to determine more information about the lensing object in the event of microlensing detection. This work will use these models to analyse the data and recover information about potential lenses that could be at the origin of a lensed event system. 

To explore these multiple models, microlensing candidates are analysed using \textsc{Gravelamps}~\citep{Wright:2021cbn}. This algorithm performs PE analyses of the GW data by jointly inferring the source and lens parameters, assuming a particular lens model. Therefore, it can not only extract information on the lens, but also perform lens-model selection by comparing the evidence obtained for different models and see which one is favoured based on the data. In addition, to cross-check the results obtained with this pipeline, the data is also analysed with the \textsc{GWMAT} pipeline~\citep{GWMAT2023}, a similar but independent \textsc{python}/\textsc{cython} analysis package implementing the point lens model.

%% file: GW190412.tex
GW190412~\citep{LIGOScientific:2020stg} is a well-known event as it is, alongside GW190814~\citep{LIGOScientific:2020zkf}, one of the events with significant higher order modes (HOMs) detected during O3. It is identified as the coalescence of a $\sim 30 \, M_{\odot}$ black hole, with a $\sim 8 \, M_{\odot}$ one. During the O3 lensing search~\citep{LIGOScientific:2023bwz}, this event was found to be the most likely candidate for being a type II image. However, the evidence ($\log_{10}(\mathcal{B}^{\mathrm{II}}_{\mathrm{I}}) = 0.61$ and $\log_{10}(\mathcal{B}^{\mathrm{II}}_{\mathrm{III}}) = 0.30$) is inconclusive and could be related to other effects as well, such as noise or waveform effects. In this section, we investigate possible origins of this feature. In particular, we re-analyze the data searching for type II images using other waveform models, and see if the observed feature could be related to microlensing effects.

\subsection{Check for Waveform Systematics}

For any astrophysical inference about CBCs from GW data,
it is crucial to understand the possible systematic errors due to approximations in the waveform models used or whether observed features could not originate from the model used. Since full numerical relativity simulations are only available as a reference for a few points in parameter space, the most convenient way to study the impact of waveform systematics is to compare results from different models. PE runs for GWTC data releases have always used at least two waveforms from independent modelling approaches and additional studies on events of special interest regularly compare larger numbers of models~\citep[see, e.g.,][]{LIGOScientific:2016ebw,LIGOScientific:2020ufj,Colleoni:2020tgc,Mateu-Lucena:2021siq,Hannam:2021pit,Estelles:2021jnz}.

In~\citet{LIGOScientific:2023bwz}, the type II image (Morse factor of 0.5) search was performed with the \texttt{IMRPhenomXPHM} waveform~\citep{Pratten:2020ceb}, including HOM and precession effects. A feature similar to the observed one ---meaning a peak around a value of 0.5 in the Morse factor posterior--- was recovered in various scenarios. For example, by injecting a signal with the maximum likelihood parameters in simulated noise with a given waveform family and recovering with \texttt{IMRPhenomXPHM}. This seemed to indicate that the feature was consistent with a real signal, at least given the used waveform.

Here, we re-analyse the data using two other commonly-used waveforms: \texttt{IMRPhenomPv3HM}~\citep{Khan:2020kde} and \texttt{SEOBNRv4PHM}~\citep{Ossokine:2020kjp}. These two waveforms encapsulate the same type of physics as \texttt{IMRPhenomXPHM} with HOMs and precession included. The analyses are performed using the same setup as the one used for the Type II image search performed in~\citet{LIGOScientific:2023bwz}.

\begin{figure}
\includegraphics[keepaspectratio, width=\linewidth]{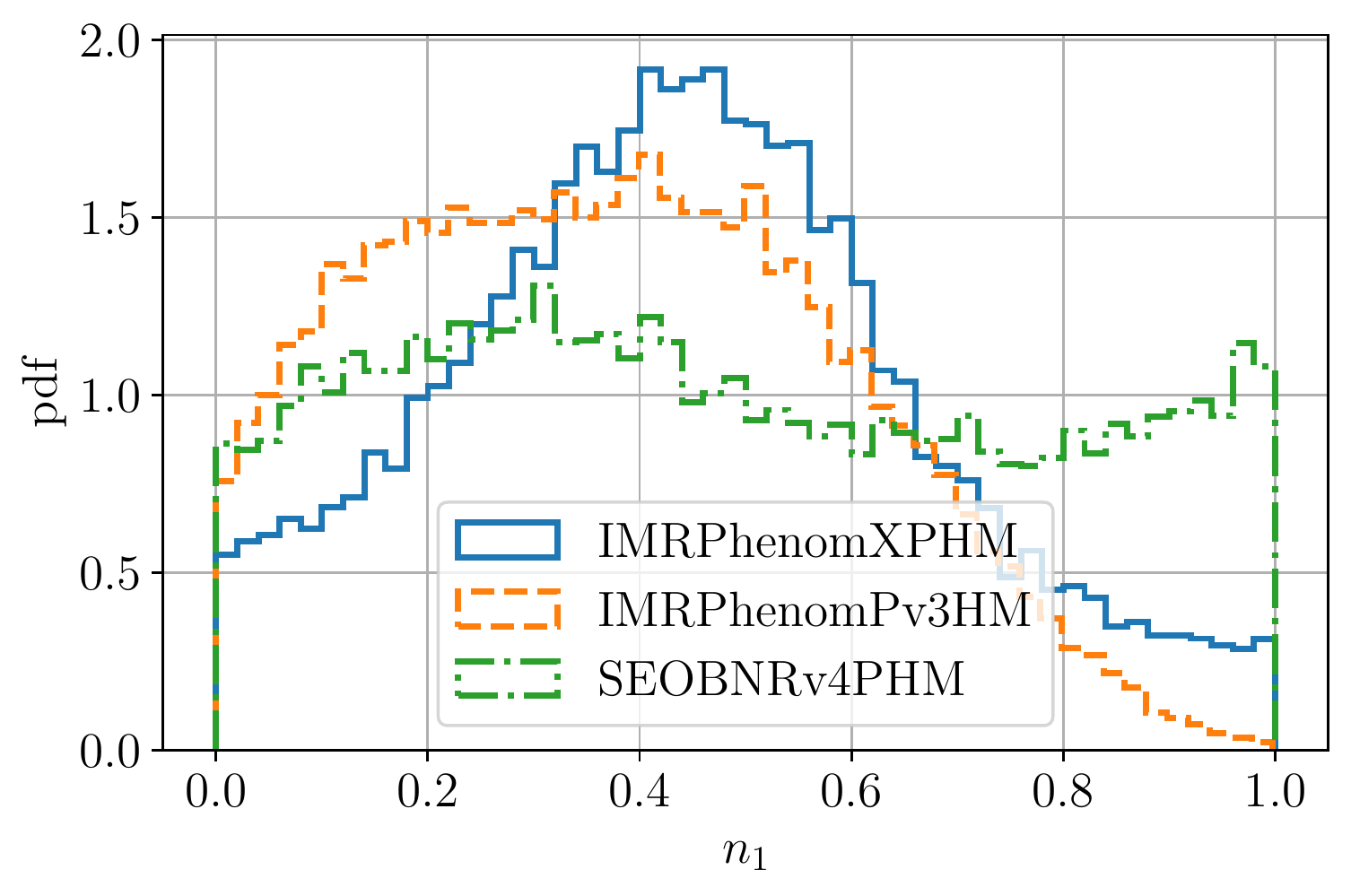}
\caption{Recovery of the Morse factor for the GW190412 event with different waveforms: \texttt{IMRPhenomXPHM}, \texttt{IMRPhenomPv3HM}, and \texttt{SEOBNRv4PHM}. The support for a type II image is present for the two waveforms from the \texttt{Phenom} family. However, the feature is less prominent for the \texttt{IMRPhenomPv3HM} waveform, and only marginally present for the \texttt{SEOBNRv4PHM} waveform. Therefore, the observed feature is probably spurious and the event is not a type II image.}\label{fig:n_phase_post_different_wfms}
\end{figure}

The posterior recovery of the Morse factor ($n_1$), allowing it to take any value from 0 to 1 instead of discrete,  for the different waveforms is shown in Fig.~\ref{fig:n_phase_post_different_wfms}. The posterior peaks at 0.5 using \texttt{IMRPhenomXPHM}.  When analyzing the data with the \texttt{IMRPhenomPv3HM} waveform  we still observe a peak but it is less prominent. On the other hand, with the \texttt{SEOBNRv4PHM} waveform the posterior distribution has an earlier maximal value, is much broader, and is lacking a pronounced peak. This is different from the results seen with the \texttt{Phenom}-family waveforms. Therefore, the observed feature seems to come from a combination of noise and waveform effects, and the event is probably not a type II image.

\subsection{Microlensing Analysis}

The possible signs of de-phasing that generated initial interest in GW190412 as a type II lensing candidate may also be a mistaken interpretation of frequency-dependent microlensing effects. This motivates an analysis of the event using the \textsc{Gravelamps} pipeline to determine if there is any potential favouring of this interpretation of the signal's features.

\begin{figure}
	\includegraphics[width=\linewidth]{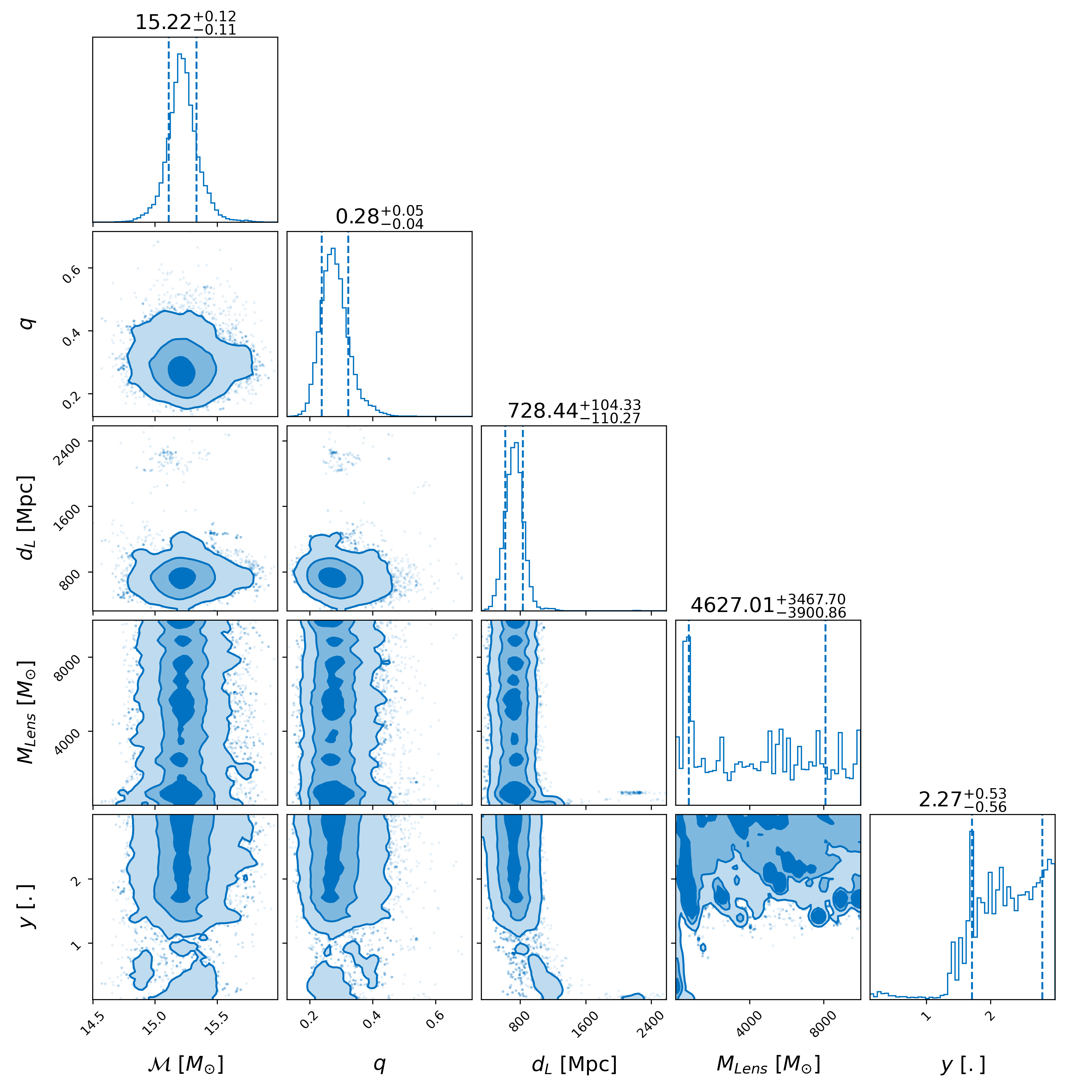}
	\caption{Posteriors for a subset of the parameters including the detector frame chirp mass and mass ratio, the luminosity distance, the lens frame lens mass, and the dimensionless displacement of the source from the optical axis (i.e.\ the source position). These posteriors were produced during the point mass microlensing analysis done for GW190412 using \textsc{Gravelamps}. As can be seen, similarly to GW191103 (shown in Fig.~\ref{fig:191103-microlensing-sis}), the lensing parameter posteriors are extremely broad and uninformative. This is consistent with the expectations for a non-microlensing event.}\label{fig:190412-microlensing-posteriors}
\end{figure}

Looking at the results of the analysis for GW190412 shown in Fig.~\ref{fig:190412-microlensing-posteriors}\footnote{The redshifted lens mass $M_{\textrm{Lens}}^{z}$ $(1 + z_{\textrm{Lens}})M_{\textrm{Lens}}$ where $M_{\textrm{Lens}}$ is the lens frame lens mass. In the \textsc{Gravelamps} analyses the redshift of the lens is calculated based on the lens being half way between the source and the observer.} marks the event as unique amongst those analysed for this paper in that it favours the point mass lensing model over the SIS model with $\log_{10}(\mathcal{B}^{\rm{L}}_{\rm{U}})$ values of $0.6$ and $0.4$ respectively. This remains quite marginal preference for the microlensing hypothesis in either case, although it is worth noting that this support is higher than for GW191103 or GW191105, the events analysed in a strong-lensing context in Sec.~\ref{Sec:GW191103_GW191105}. This is consistent with the apparent signs of de-phasing being present only in GW190412. Whether the correlation holds would need to be confirmed with additional examinations of simulations or additional type II lensing candidates.

However, whilst support for this event is higher in terms of the raw Bayes factors, the posteriors for the lensing parameters need to be examined. Fig.~\ref{fig:190412-microlensing-posteriors} presents these posteriors for the marginally more preferred point mass lensing model. The source posteriors are tightly constrained but the lensing parameters are extremely broad, leading to the conclusion that this event does not indicate any particular signs of microlensing either, and again the apparent features could be related to noise or waveform issues.

%% file: GW191103_GW191105.tex
GW191103 and GW191105 were BBHs detected during the third observing run~\citep{LIGOScientific:2021djp}. In the main LVK analyses, the standard treatment of the signals revealed nothing out of the ordinary for these events. However, when treating the events as potential lensing candidates, the pair display some intriguing characteristics. There is a notable amount of overlap between some of the reported source parameters, such as the sky location and masses; see Fig.~\ref{fig:PO_11031105} for a representation of the posteriors. Moreover, the two events have about two days delay between their merger times which is consistent with galaxy-scale lenses~\citep{Wierda:2021upe, More:2021kpb}. However, in the LVK lensing search, these events were ultimately discarded once the JPE Bayes factor had been computed~\citep{LIGOScientific:2023bwz}, meaning that the observed overlap is unlikely to be coming from a lensed BBH and is more likely to be coincidental. Nevertheless, as was stated in the introduction, in the following analyses we have disregarded this and treated the event as though it were a lensed pair. 

\subsection{Posterior Overlap Candidate Identification}
\label{subsec:PO_191103_191105}

\begin{figure*}
\centering
\includegraphics[width=0.49\linewidth]{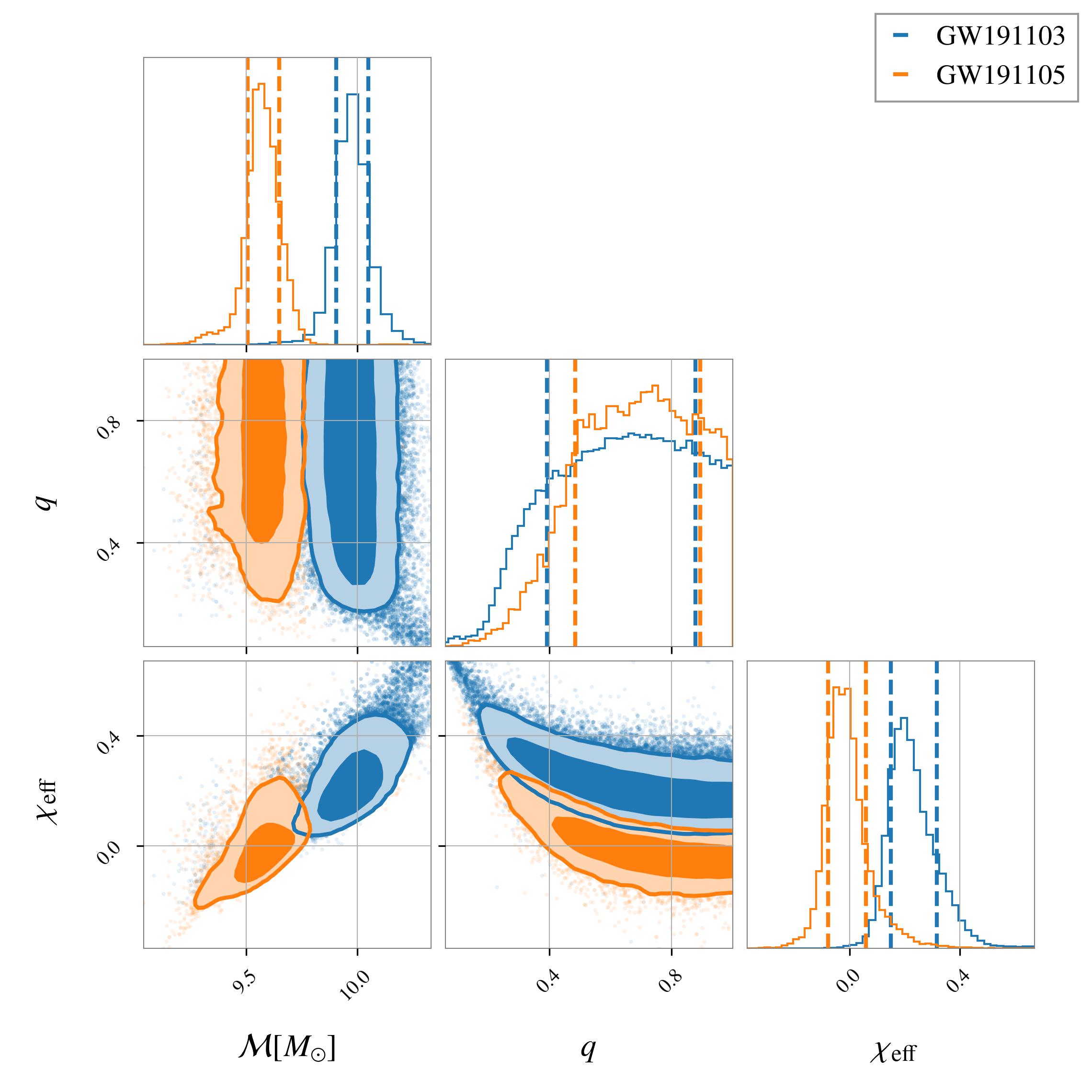}
\includegraphics[width=0.49\linewidth]{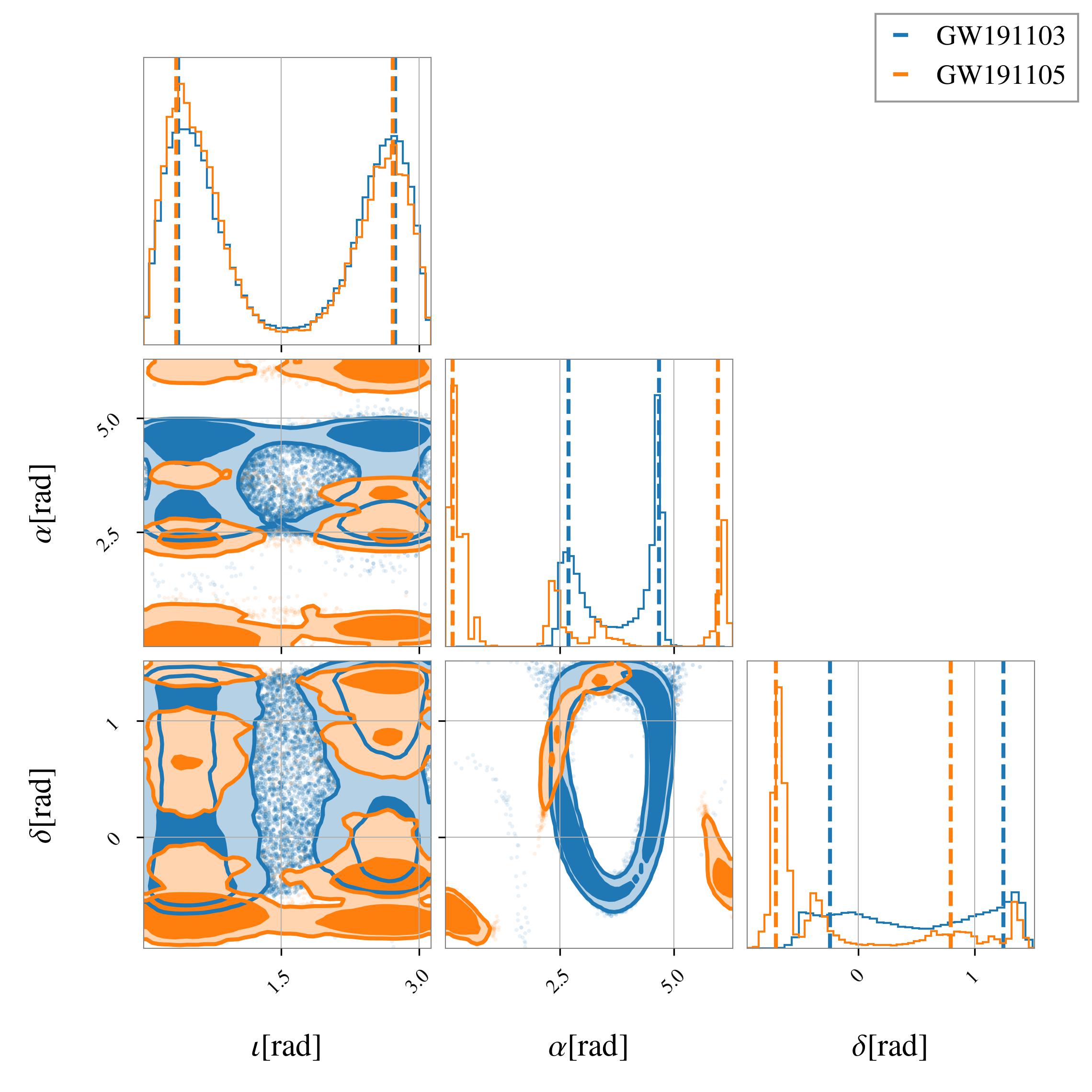}
\caption{Posteriors for some of the parameters obtained using the \texttt{IMRPhenomXPHM} waveform for GW191103 (blue) and GW191105 (orange). The overlap in the extrinsic parameters (e.g.\ sky location) is larger than that for the intrinsic parameters (e.g.\ detector-frame chirp mass and spins).}\label{fig:PO_11031105}
\end{figure*}

During the O3 strong lensing search, the PO analysis and a machine learning pipeline, \textsc{LensID}~\citep{PhysRevD.104.124057}, were used to identify potential lensing candidate pairs from the detected events. The top $1\%$ of these---approximately a hundred pairs---were then passed on to \textsc{GOLUM}~\citep{Janquart:2021qov, Janquart:2023osz} for filtering, and then to \textsc{hanabi}~\citep{Lo:2021nae} for final analysis. The GW191103--GW191105 pair was identified as one of the most likely candidates by the PO analysis using the posteriors of some of the parameters obtained with the \texttt{IMRPhenomXPHM} waveform~\citep{Pratten:2020ceb} released publicly on the Gravitational Wave Open Science Centre (GWOSC)~\citep{RICHABBOTT2021100658, GWTC-3-release}, whereas \textsc{LensID}---which uses Q-transform images and \textsc{Bayestar}~\citep{Singer_2016} skymaps---had not classified it as a candidate. Appendix~\ref{app:lensid} discusses possible reasons for the non-identification of the pair by the machine-learning based pipeline. For PO, the pair ended up having the highest overall statistic: $\log_{10} \mathcal{B}^{\mathrm{overlap}} = 3.03$ and $\log_{10} \mathcal{R}^{\mathrm{gal}} = 1.14$ for the SIS model giving a total of $4.17$. Fig.~\ref{fig:PO_11031105} shows the posteriors of both events where one may see the varying degrees of overlap between the events.

\begin{figure}
\centering
\includegraphics[width=\linewidth]{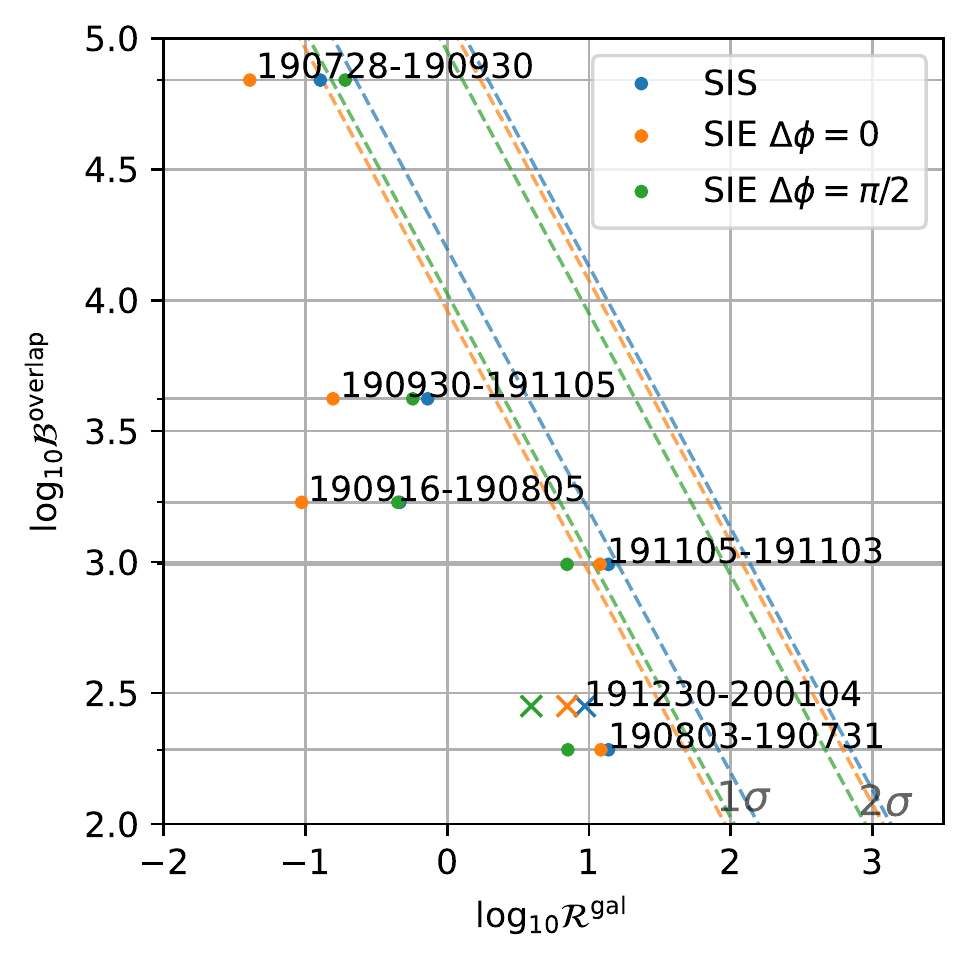}
\caption{The highest ranked candidate strong lensing pairs from the PO analysis considering all the event pairs found based on the O3 data (dots)~\citep{LIGOScientific:2023bwz} and the supra-sub pair analyzed in this work. The dashed lines correspond to the $1\sigma$ and $2\sigma$ confidence levels for the combined PO statistic ($\mathcal{B}^{\mathrm{overlap}} \times \mathcal{R}^{\mathrm{gal}}$) with different lensing models computed from the background simulations. We note that beside GW191103--GW191105, the pair analyzed in this work, GW190728--GW190930 is also close to $1\sigma$ for PO. However, the pair has been discarded in previous searches with a lower overall significance than GW191103--GW191105~\citep{LIGOScientific:2021izm}. Therefore, it is not considered for further analyzes in this work.}\label{fig:PO_candidates}
\end{figure}

Fig.~\ref{fig:PO_candidates} shows the candidate event pairs identified by PO analysis on the $\log \mathcal{B}^{\mathrm{overlap}}$--$\log \mathcal{R}^{\mathrm{gal}}$ plane considering both the SIS and SIE galaxy models. The choice of model affects only the $\log \mathcal{R}^{\mathrm{gal}}$ value. The PO analysis is marginalised over phase and is, therefore, insensitive to the relative Morse phase ($\Delta \phi$) between the two events. As a result of this insensitivity, the SIE cases $\Delta \phi=0$ and $\Delta \phi=\pi/2$ are considered separate models, hence we compute the $\mathcal{R}^{\mathrm{gal}}$ expected distributions for each case.

Posteriors of events detected by the LVK detectors can overlap by random coincidence meaning that unlensed pairs could also give high Bayes factors. For this reason, a background injection study with $\sim 1000$ unlensed events (the combinations of which yield about half a million pairs) is done to calculate the FAP~\citep{Caliskan:2022wbh} of the observed log Bayes factor for the candidate pair. The FAP per-pair ($\mathrm{FAP}_{\mathrm{PP}}$) for the candidate, hence the number of unlensed events with a Bayes factor higher than the one observed for the pair of interest, is found to be 1 in 10,000. Taking into consideration that a total of $\sim 68$ BBH events were detected in O3 the total FAP (given by $\mathrm{FAP} = 1 - {\left( 1 - \mathrm{FAP}_{\mathrm{PP}} \right)}^{N_{\mathrm{pairs}}}$) is found to be 0.3 i.e.\ a significance of slightly above $1\sigma$. As seen in the figure the time delay for this event pair is more compatible with an SIE with $\Delta \phi=0$ as compared to an SIE with $\Delta \phi=\pi/2$ and SIS.\@ After this step, to extract more information about the event pair, it is passed to more extensive pipelines for further investigation.

\subsection{Waveform Systematics Study on Posterior Overlap Candidate Identification}\label{sec:1103-1105_blusys}

At the time of writing, no dedicated studies of waveform systematics have
been conducted for gravitational lensing analyses.
As an initial check, we report on a comparison of PO calculations
on single-event PE performed with different waveforms.
This is a practical first step as the single-event PE is significantly cheaper computationally
than JPE,
and detailed studies of waveform systematics on the latter are left for future work. The results presented here for the GW191103--GW191105 pair
are an excerpt from a larger pioneer study on waveform model systematics in GW lensing that will be published separately~\citep{Garron:2023gvd}.

The PO statistics reported in~\citet{LIGOScientific:2023bwz}
and used to initially qualify the GW191103--GW191105 pair as sufficiently interesting for further follow-up
were based on the \texttt{IMRPhenomXPHM} waveform~\citep{Pratten:2020ceb}.
Besides the posterior samples for this waveform, the data release~\citep{GWTC-3-release} for GWTC-3~\citep{LIGOScientific:2021djp} contains samples obtained with the \texttt{SEOBNRv4PHM} waveform~\citep{Ossokine:2020kjp}.
For this study, we performed additional PE runs on GW191103 and GW191105
for several other models,
using \textsc{parallel\_Bilby}~\citep{Ashton:2018jfp,Smith:2019ucc}
with the~\textsc{dynesty} sampler~\citep{10.1093/mnras/staa278}, using settings and priors consistent with the GWTC-3 \texttt{IMRPhenomXPHM} runs~\citep{LIGOScientific:2021djp}.

The additional waveforms covered are
three variants from the same family of frequency-domain phenomenological waveforms as \texttt{IMRPhenomXPHM},
as well as one time-domain phenomenological waveform:
\begin{itemize}
 \item \texttt{IMRPhenomXAS} is an aligned-spin frequency-domain waveform for dominant-mode-only GW emission~\citep{Pratten:2020fqn};
 \item \texttt{IMRPhenomXHM} is an aligned-spin frequency-domain waveform including HOMs~\citep{Garcia-Quiros:2020qpx};
 \item \texttt{IMRPhenomXP} is a frequency-domain waveform allowing for spin precession but for dominant-mode-only GW emission~\citep{Pratten:2020ceb};
 \item \texttt{IMRPhenomTPHM} is a time-domain waveform allowing for spin precession and including HOMs~\citep{Estelles:2021gvs}.
\end{itemize}
The three reduced-physics \texttt{IMRPhenomX} waveforms allow us,
in comparison with the most complete family member \texttt{IMRPhenomXPHM},
to check if neglecting any of these features has a significant impact
on the PEs for each event,
and hence on their overlap.
In addition, the \texttt{IMRPhenomTPHM} waveform
shares its time-domain nature with \texttt{SEOBNRv4PHM}
but much of its modelling approach with \texttt{IMRPhenomXPHM},
making it an ideal tool to further cross-check for consistency between different modelling strategies.

We have followed the original KDE-based calculation from~\citet{Haris:2018vmn} to compute the PO statistic $\mathcal{B}^{\mathrm{overlap}}$,
with the modification of computing sky overlaps and intrinsic-parameter overlaps separately and then multiplying them,
as done in~\citet{LIGOScientific:2023bwz}.

\begin{table}
\centering
\begin{tabular}{lr}
\hline
\hline
Waveform & $\log_{10}(\mathcal{B}^{\mathrm{overlap}})$ \\
\hline
\hline
\texttt{IMRPhenomXAS} & 3.37 \\
\texttt{IMRPhenomXHM} & 3.48 \\
\texttt{IMRPhenomXP} & 3.08 \\
\texttt{IMRPhenomXPHM} & 3.03 \\
\texttt{IMRPhenomTPHM} & 2.70 \\
\texttt{SEOBNRv4PHM} & 2.65 \\
\hline
\end{tabular}
\caption{\label{table:blusys}
PO statistic values for the GW191103--GW191105 pair using different waveform models in the single-event PE.}
\end{table}

Table~\ref{table:blusys} lists the $\mathcal{B}^{\mathrm{overlap}}$ resulting from comparing the posteriors from both events with each waveform. 
There are changes up to factors of $\sim 6$ in overlap statistics,
with \texttt{IMRPhenomXHM} producing the highest $\mathcal{B}^{\mathrm{overlap}}$ and \texttt{SEOBNRv4PHM} resulting in the lowest.
We first notice that the aligned-spin waveforms produce the highest $\mathcal{B}^{\mathrm{overlap}}$. 
These have fewer free parameters than the precessing models and hence also different prior volumes. 
By inspecting the posteriors we find that the aligned-spin waveforms prefer $a_2$ closely peaked to zero for both events which 
gives a high contribution to the overlap, while the precessing waveforms have broader distributions in this parameter, compensating with 
the additional freedom in the tilt angles.
In addition, the two time-domain waveforms produce lower $\mathcal{B}^{\mathrm{overlap}}$ than the frequency-domain waveforms.
However, these changes are not significantly larger than  
expected from other sources such as prior choice and KDE implementation details,
and all results are consistently in favour
of the lensing hypothesis.
Hence, we conclude that waveform choice does influence the PO method to some degree, 
but that for this specific event pair it
is sufficiently robust under waveform choice,
in the sense that all results agree qualitatively on identifying the pair as possibly lensed
and interesting for follow-up.

One caveat on this type of study is that a full interpretation of $\mathcal{B}^{\mathrm{overlap}}$
requires extensive simulation studies on both lensed and unlensed pairs,
and the respective distributions could easily be different for different waveforms.
However, in~\citet{LIGOScientific:2023bwz}, this factor was used purely as a ranking statistic. So, as long as the number does not drop strongly for any of the considered waveforms, we can conclude that the identification of the candidate pair is robust under waveform choice.

\subsection{Compatibility with Lensing Models}

\begin{figure}
    \centering
    \includegraphics[keepaspectratio, width=\linewidth]{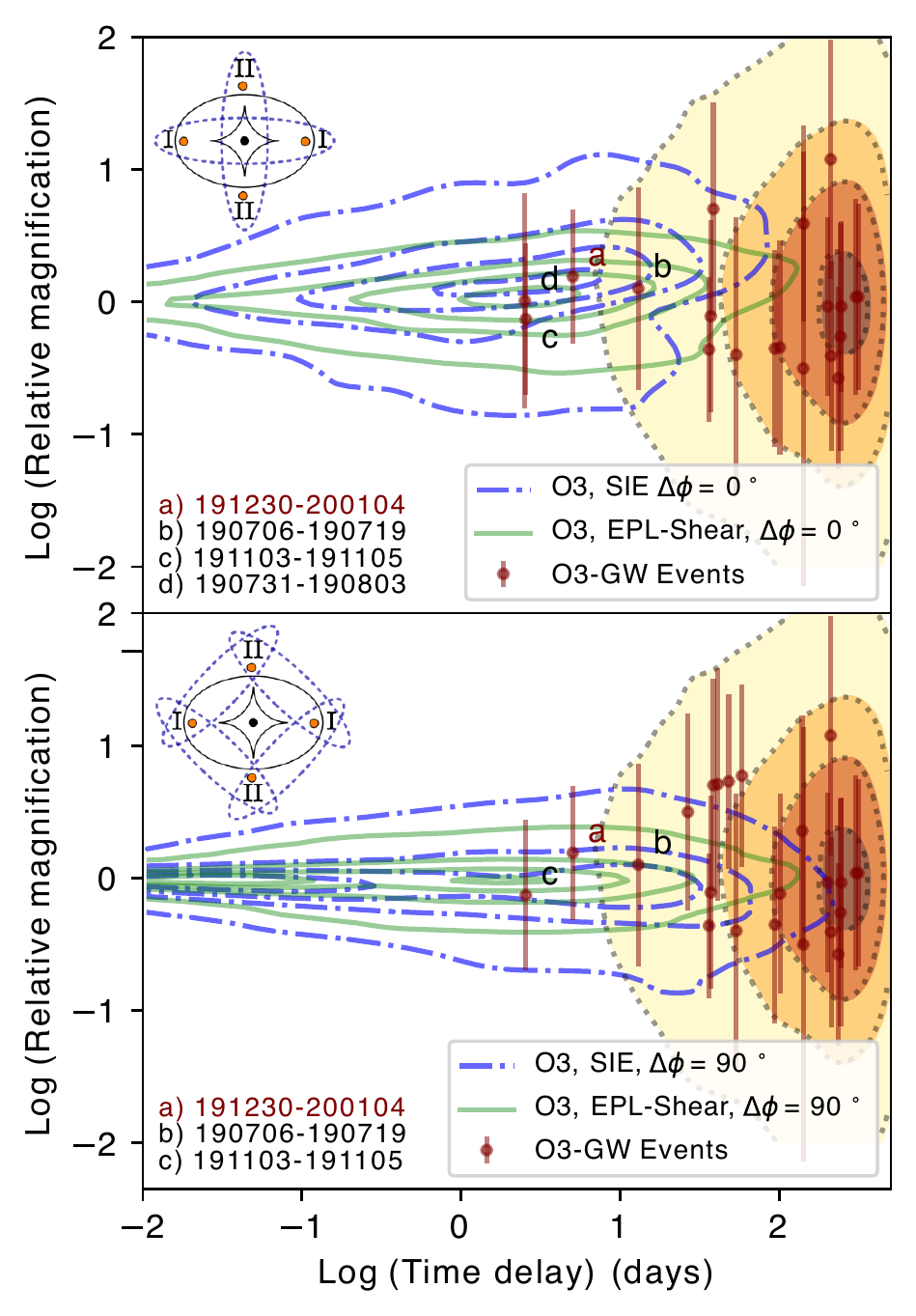}
    \caption{Comparison of the posterior for the observed relative magnification and time delays for the O3 event pairs with the expected distributions for the lensed population of mergers from~\citet{More:2021kpb} (dashed blue, using an SIE model) and from~\citet{Wierda:2021upe} (solid green, using an elliptical model with a power low density profile and external shear (EPL-Shear)) and unlensed population (yellow-orange-red), both assuming galaxy lenses. Overlayed in brown are the observed values for selected O3 event pairs, and the letters mark the event pairs more compatible with the lensing hypothesis. Written in brown, and denoted with letter a, is a pair made of a supra-threshold and a sub-threshold event, and further analysed in this work. The top panel corresponds to the expected distribution when the two images are of the same type, i.e., there is no phase difference between the two images (see top-left illustration), while the bottom panel corresponds to a configuration where the two image types differ, i.e., there is a $\pi/2$ shift between the images. Most of the observed event pairs are well outside the lensed distribution. The GW191103--GW191105, GW190706--GW190719, and GW101230--LGW200104 pairs are more compatible with the lensed hypothesis than with the unlensed one. In particular, the GW191103--GW191105 pair lies in a higher probability density region than the other pairs. One also sees that the GW191230--LGW200104 pair ---made of a supra and a sub-threshold event--- lies in a higher density region, even if it is less important than the GW191103--GW191105 pair. This pair is discussed further in Sec.~\ref{Sec:GW19130}.
      \label{fig:CompaMgal}
    }
\end{figure}

Once an event has been identified as a potential candidate through the aforementioned PO or machine learning searches, it can be followed up by other pipelines. However, an additional check can be made by comparing the observed lensing parameters with the ones predicted by specific lensing models~\citep{Haris:2018vmn, Lo:2021nae, Wierda:2021upe, More:2021kpb, Janquart:2022zdd}. In this work, we focus on galaxy lenses. A comparison of the lensing parameters observed in the O3 search with the catalogue presented in~\citet{More:2021kpb} and~\citet{Wierda:2021upe} is given in Fig.~\ref{fig:CompaMgal}. 
Most of the events are compatible with the values expected for unlensed events. Noticeably, two supra-threshold event pairs have lensing parameters that are more consistent with a lensed hypothesis: GW191103--GW191105, which is the one the most on the left and therefore the most favoured, and GW190706--GW190719. 
One can also compute the $\mathcal{M}^{\rm{\mathrm{gal}}}$~\citep{More:2021kpb} statistics for the pairs. This number is the ratio of the probabilities for observing the lensing parameters under the lensed and unlensed hypotheses given by the lensing catalog though it does not account for the rate of lensing\footnote{Such a catalogue is obtained through extensive lensed and unlensed populations simulations~\citep{Haris:2018vmn, Wierda:2021upe, More:2021kpb}.}. For GW191103--GW191105, we find $\log_{10}(\mathcal{M}^{\rm{\mathrm{gal}}}) = 1.3$, while for GW190706--GW190719, $\log_{10}(\mathcal{M}^{\rm{\mathrm{gal}}}) = 0.8$. This shows that in the two cases, based on the catalogue, the observed lensing parameters agree more with the expected values for the lensed hypothesis than for the unlensed one.

Whilst such a comparison is valuable to gain information on the chances of lensing from specific models, it does not account for the compatibility of the binary parameters. Since the lensed hypothesis is favored over the unlensed one both based on PO and lensing statistics, we need to further ascertain the lensed nature by turning to JPE methods.

\subsection{Joint Parameter Estimation Based Investigations}

In the case of events being genuine lensed images, in addition to the lensing parameters being compatible with at least one lensing model, parameters whose estimation are unaffected by the lensing---e.g.\ sky location, component masses---should be the same between the events. For the GW190706--GW190719 pair, the \textsc{GOLUM} analysis gives $\log_{10}(\mathcal{C}^\mathrm{L}_\mathrm{U}) = 0.6$, which is a value slightly favouring lensing, but still well within the values one can expect for unlensed events~\citep{Janquart:2022zdd}. On the other hand, for GW191103--GW191105, we find $\log_{10}(\mathcal{C}^\mathrm{L}_\mathrm{U}) = 2.6$. However, despite this higher value, this does not guarantee that the signals are not merely coincidentally similar~\citep{Janquart:2022zdd}. Here, we follow up only on the GW191103--GW191105 pair, which has the highest coherence ratio observed in the O3 pairs~\citep{LIGOScientific:2023bwz}, with the methodologies described in \cite{Janquart:2022zdd} and \cite{Lo:2021nae} to include information from a lens model into the analyses.

We use the method from~\citet{Janquart:2022zdd} to include a lens model in the detection statistic coming from the \textsc{GOLUM} pipeline~\citep{Janquart:2021qov, Janquart:2023osz}, modifying the coherence ratio so that it also accounts for the observed lensing parameters and a given lens model. This reduces the risks of false alarms in lensing searches and makes the detection of genuinely lensed pairs more robust. This also enables one to compare the compatibility of the observation with different lens models, constraining the nature of the lens. 

To explore the event's significance, we make an injection study by generating an unlensed background. We simulate 250 unlensed BBHs, sampling their masses from the \textsc{powerlaw + peak} distributions; the spins and redshift are also sampled from the latest LVK observations, using the maximum likelihood parameters to generate the distributions~\citep{LIGOScientific:2021psn}. The sky location is sampled from a uniform distribution over the sphere of the sky. The inclination is uniform in cosine, and the phase and polarizations are sampled uniformly on their domain. The merger times are set uniformly throughout the third observing run. For each set of parameters, we randomly associate a real event from the GWTC-3 catalogue and assume the same observation conditions (the same detectors are online, and the noise is generated from the event's PSD). The 250 events selected are such that their network SNR is higher than 8\footnote{Whether an event can be considered to be of astrophysical origin is not dependent only on its SNR, and recent GWTC papers also put a threshold on the probability $p_{\textrm{astro}}$~\citep{pastro_technical, LIGOScientific:2021djp}. Here, we consider the SNR threshold sufficient to assess detectability.}.

Based on this population, we can compute the $\textrm{FAP}_{\textrm{PP}}$ for the coherence ratio for a given lens model. Table~\ref{tab:ClusGW191103GW191105} summarizes the detection statistic and the $\textrm{FAP}_{\textrm{PP}}$ for the analysis with and without model.

\begin{table}
    \centering
    \begin{tabular}{l c r}
    \hline
    \hline
    Statistic & $\log_{10}$ value & $\textrm{FAP}_{\textrm{PP}}$ \\
    \hline 
    \hline
      $\mathcal{C}_U^L$ & 2.5 &  $2.0\times10^{-3}$\\
      $\mathcal{C}_{\mathcal{M}_{\mu, t}}$ & 2.4 & $1.6\times10^{-3}$ \\ 
      $\mathcal{C}_{\mathcal{M}_{t}}$ & 2.9 & $9.8\times10^{-4}$\\
    \hline
    \end{tabular}
    \caption{Values of the detection statistic obtained using the \textsc{GOLUM} framework for the GW191103--GW191105 lensed candidate pair without lens model ($\mathcal{C}_U^L$), and with an SIE lens, with ($\mathcal{C}_{\mathcal{M}_{\mu, t}}$) and without ($\mathcal{C}_{\mathcal{M}_{t}}$) relative magnification accounted for. The $\textrm{FAP}_{\textrm{PP}}$ is decreased when using an SIE model.}\label{tab:ClusGW191103GW191105} 
\end{table}

As a first step, we can verify the  $\textrm{FAP}_{\textrm{PP}}$  for the event pair when considering the coherence ratio. In this case, $\textrm{FAP}_{\textrm{PP}} = 2.007\times10^{-3}$, meaning that for our 250 unlensed BBHs, the chance is 1 in 500 that these events are not a genuine lensed pair. Thus, based only on the match between the parameters, the probability for these events to originate from two unrelated unlensed events is relatively high. Statistically, this means that the combination of only 33 randomly-selected unlensed BBH mergers is capable to make at least one pair with at least the same coherence ratio as the GW191103--GW191105 pair. 

Including a lens model helps study which astrophysical object could be at the origin of the lensed event. Here, we consider an SIE model~\citep{Koopmans2009, More:2021kpb}, which is a good model for a galaxy lens. We consider the case where we account for the time delay and the relative magnification in the model, and the case where we only consider the time delay. The combinations of the coherence  ratio and the lensing statistics are written $\mathcal{C}_{\mathcal{M}_{\mu, t}}$ and $\mathcal{C}_{\mathcal{M}_{t}}$ for the case with and without relative magnification respectively.  The values for detection statistics and the FAP for the two cases are given in Table~\ref{tab:ClusGW191103GW191105}. The inclusion of the SIE model, with and without the relative magnification, decreases the FAP, meaning that the candidate pair becomes more significant. It is the case even if the SIE model with the time delay and the relative magnification slightly decreases the coherence ratio. The reduction in FAP is slightly larger when only the time delay is considered. This is because the observed relative magnification is slightly outside the highest-density regions for the two models under the lensed hypothesis. Moreover, the overlap between lensed and unlensed populations for this parameter is high, making it less helpful to discriminate between the two situations. The results have also been cross checked using an SIE-based catalog from~\citet{Wierda:2021upe}, giving the same conclusions.

We note here that whilst the $\textrm{FAP}_{\textrm{PP}}$ seem relatively small for the SIE model, it is insufficient to claim the pair to be lensed. The smallest value found is $9.8\times10^{-4}$. Whilst this is an improvement over the original $\textrm{FAP}_{\textrm{PP}}$, it repreresently only an increase from the combinations of 33 randomly-selected unlensed BBH mergers to the combinations of 47 such mergers to, statistically, have a pair display a detection statistic higher or equal to the one reported for the observed pair. Consequently, whilst  GW191103--GW191105 displays some interesting behaviours, these are insufficiently significant to claim a first strong lensing detection.

Additionally, we repeat the Bayes factor calculation comparing the probability ratio of the lensed versus the unlensed hypothesis as described in~\cite{LIGOScientific:2023bwz} using the more realistic lens population model described in~\cite{Oguri:2018muv} (see~\cite{hanabi_astro} and also Sec.~\ref{Sec:TheoryIntro}) using \textsc{hanabi} \citep{Lo:2021nae}. We use the same set of source population models as in~\cite{LIGOScientific:2023bwz}, e.g.~the \textsc{powerlaw + peak} model for the source masses from the GWTC-3 observations~\citep{LIGOScientific:2021psn} and three models for the merger rate density: Madau-Dickinson~\citep{Madau:2014mmd}, $\mathcal{R}_{\mathrm{min}}(z)$, and $\mathcal{R}_{\mathrm{max}}(z)$. Table~\ref{tab:191103_191105_hanabi} shows the log-10 Bayes factors computed using the three merger rate density models with the simple SIS lens model reported in~\cite{LIGOScientific:2023bwz} and the SIE + external shear model reported in~\cite{hanabi_astro}. We see that the values calculated using the SIE + external shear model are consistently higher than those using the SIS model, indicating that the pair is more consistent with a more realistic strong lensing model. Still, the $\log_{10}\mathcal{B}^L_U$ values are negative, and therefore the event pair is most likely unlensed.

\begin{table*}
\centering
\begin{tabular}{lccc}
\hline 
\hline
\backslashbox{Lens model}{Merger rate density} & Madau-Dickinson & $\mathcal{R}_{\mathrm{min}}(z)$ & $\mathcal{R}_{\mathrm{max}}(z)$ \\
\hline 
\hline
  SIS \citep{LIGOScientific:2023bwz} 	& $-3.27$ & $-3.21$ & $-2.33$ \\
  SIE + external shear \citep{hanabi_astro} & $-2.60$ & $-2.46$ & $-1.28$ \\
\hline
\end{tabular}
\caption{ $\log_{10} \mathcal{B}^{\mathrm{L}}_{\mathrm{U}}$ for the GW191103--GW191105 pair from \textsc{hanabi} assuming three different merger rate density models and two different lens models. The values computed using the SIS model are reproduced from~\protect\cite{LIGOScientific:2023bwz} for the sake of comparison. We see that the values with the SIE + external shear model are consistently higher than that with the SIS model, indicating a higher compatibility of the pair with a more realistic strong lensing model. However, since the values remain negative, the event is still most likely to be unlensed considering a more realistic lensing population with the most recent population models.}
\label{tab:191103_191105_hanabi}
\end{table*}

\subsection{Electromagnetic Follow-Up}
\label{subsec:GW191103_GW191105_EM_follow}

In the case of a genuinely lensed GW signal, the light emitted by the host galaxy should also be lensed~\citep{Hannuksela:2020xor, Wempe:2022zlk}. As a result of this, in the case of a high-significance lensing candidate, one practical avenue would be to initiate an EM follow-up search. Identifying the host galaxy would be a way to verify the lensed nature of the signal independently.

The EM counterpart of a signal may be searched for in two ways: cross-matching of the joint GW sky localisation with EM catalogues containing known lens-source systems or a dedicated EM search on a per-event basis. Both of these make use of the improved GW sky localistation from the observation of multiple images which provide a virtually extended detector network~\citep{Dai:2020tpj, Liu:2020par, Lo:2021nae, Janquart:2021qov, Janquart:2023osz}. A dedicated, per-event, EM search could be peformed by looking for lenses within the sky localisation area and performing lens reconstruction to try to identify the specific lens at the origin of the observation~\citep{Hannuksela:2020xor, Wempe:2022zlk}.

\begin{figure}
\centering
\includegraphics[keepaspectratio, width=\linewidth]{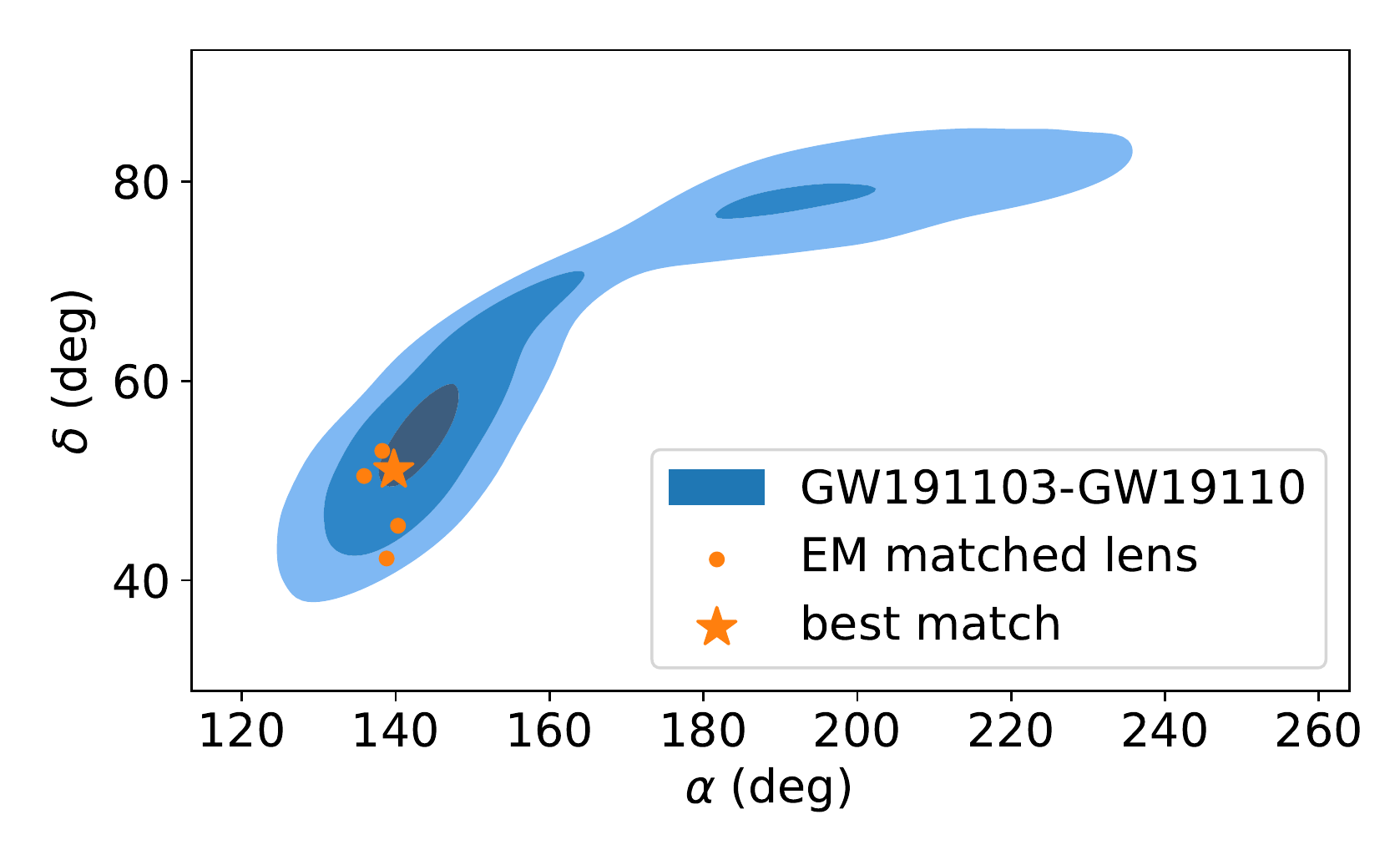}
\caption{From dark to lighter, the 10\%, 50\%, and 90\% confidence sky localisation for the GW191103--GW191105 pair. Overlaid are the cross-matched 5 candidates from the Master Lens Database.}\label{fig:1103_1105_skyloc}
\end{figure}

We cross-matched the GW191103--GW191105 pair with a few lens catalogues from optical surveys such as SuGOHI~\citep{sugohi_i, sugohi_ii, sugohi_iii, sugohi_iv, sugohi_v, sugohi_vi, sugohi_vii, sugohi_viii}, AGEL~\citep{tran2022}  and the Master Lens Database~\citep[MLD,][]{moustakas2012}. Whilst no matches were found from the SuGOHI and AGEL catalogues, the grade A and grade B lenses\footnote{In this context, grade A lenses have a higher observational quality and accuracy than grade B ones.} selected from MLD at galaxy scales showed 5 matches---4 doubly lensed systems and 1 quadruply lensed system (see Fig.~\ref{fig:1103_1105_skyloc}). Two of the doubles are predicted to have time delays $>50$~days based on the best-fit lens mass models~\citep{johnston2003, paraficz2009}. For the remaining double, we infer a time delay of $\sim$20 days given the redshifts of the lens galaxy and the source as well as the velocity dispersion~\citep{brewer2012} under the assumption of an SIS lens mass distribution. All of these time delays are too long to be consistent with the 2-day time gap of the GW191103--GW191105 pair. 

Lastly, we determine the time delays expected for the quadruplet SDSS~J0918+5104 using the best-fit mass model and results from~\citet{ritondale2019}. The expected time delay for the closest pair in SDSS~J0918+5104 is around $\sim 0.5-1$~day. Given the uncertainties in the lens model, this ballpark estimate of time delay could be consistent with that of the GW191103--GW191105 pair. However, a more detailed mass modelling analysis and exploring different physical scenarios for the offset between the host galaxy and the GW source can lead to slightly different degrees of compatibility. Still, the predicted relative magnification is both qualitatively and quantitatively consistent since the latter GW event is found to be weaker than the former GW event according to the \textsc{Hanabi} analysis. Furthermore, the closest pair of SDSS~J0918+5104 images corresponds to a minimum (Type I) and a saddle point (Type II) suggesting a phase shift of $\Delta \phi=\pi/2$. This is somewhat less favored but yet plausible for the GW191103--GW191105 pair based on the PO analysis (see Fig.~\ref{fig:PO_candidates}). 

We note that the cross-matching analysis is limited by both the incompleteness of the databases of known EM lenses and the algorithms used to find matching lenses. A more detailed investigation (see~\citet{Wempe:2022zlk} for an example of how to investigate the link between EM and GW lensed systems) is warranted to assess the probability of a lens like SDSS~J0918+5104 to be a genuine EM counterpart of the GW191103--GW191105 pair, if the latter is considered to be lensed. In the future, having dedicated EM follow-up of the  lensed event candidates using optical telescopes could also help to gather more information about the lens and localise the source to the host galaxy.

\subsection{Microlensing Analysis}

Whilst interest in the GW191103--GW191105 pair was triggered from the strong lensing perspective, it is worth paying additional attention to whether either of the single events in the pair displays any signs of frequency-dependent effects associated with microlensing. As has been noted above, the most likely microlensing scenario is a microlens embedded within a macrolens. In such a scenario, one or both of the individual signals could display the signatures of the microlens~\citep{mishra2021, Seo:2022xxc}. To determine if that scenario is plausible both GW191103 and GW191105 have been analysed using the \textsc{Gravelamps} pipeline to determine the evidence for an isolated point mass or SIS microlens. 

\begin{figure}
    \includegraphics[width=\linewidth]{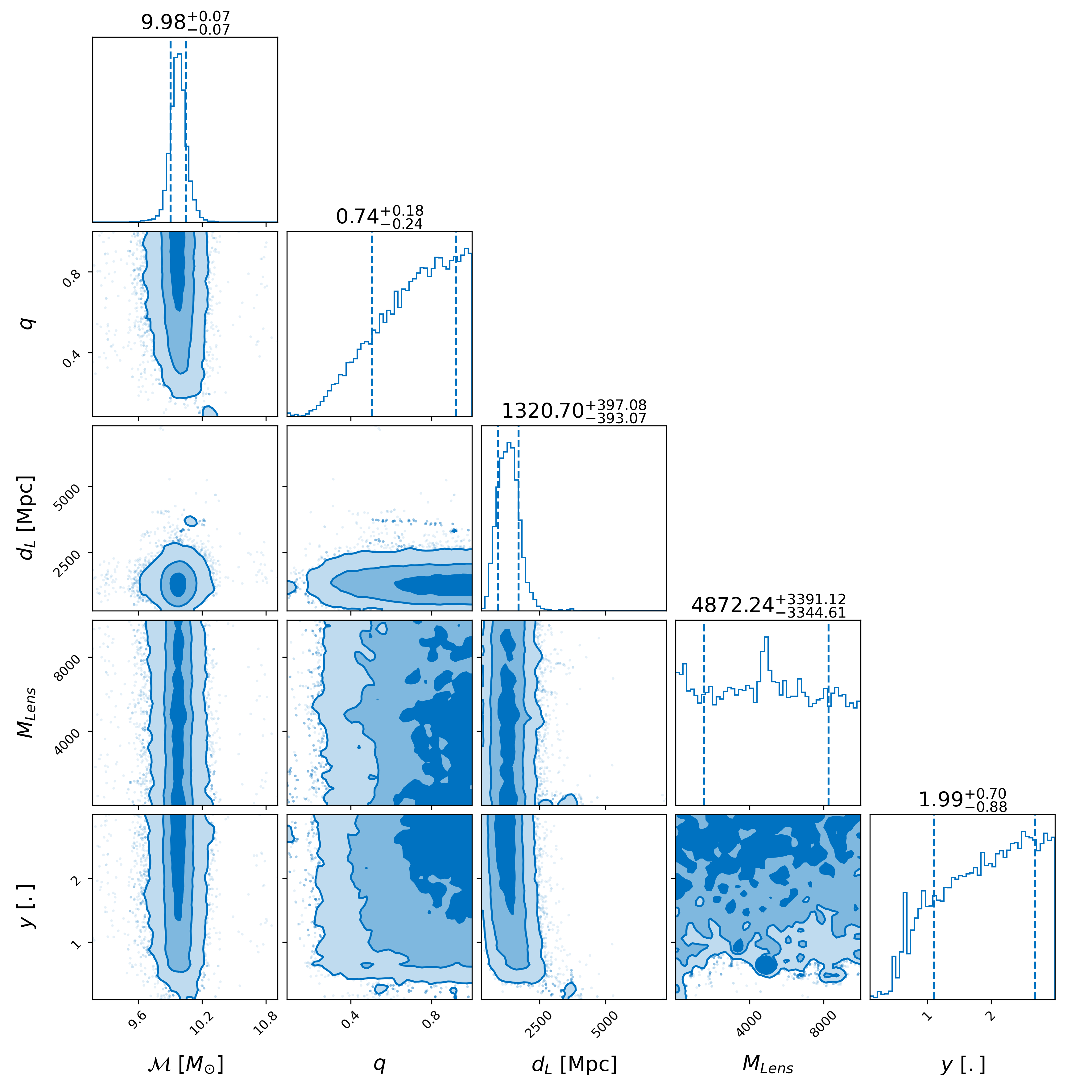}
    \caption{Posteriors for a subset of detector-frame source parameters and the lensing parameters produced during the \textsc{Gravelamps} microlensing analysis of GW191103. As can be seen, whilst the source posteriors are well constrained, the lensing parameter posteriors are extremely broad and uninformative. This shows there are no observable microlensing features in this signal.}\label{fig:191103-microlensing-sis}
\end{figure}

The main result of this analysis is that neither event shows particular favouring for either the point mass or the SIS microlensing models over the unlensed model. The preferred case is that of an SIS microlens in GW191103 which produces a $\log_{10}(\mathcal{B}^{\rm{L}}_{\rm{U}}) = 0.38$. In the point lens case, however, support drops to a $\log_{10}(\mathcal{B}^{\rm{L}}_{\rm{U}})$ of $0.09$ meaning that neither case posits strong evidence. To further compound this, the posteriors for the SIS case are given in Fig.~\ref{fig:191103-microlensing-sis}. The source parameters' posteriors are well constrained, but those of the lensing parameters are extremely broad and uninformative. This is consistent with the expectation of an unlensed event and, combined with the marginal Bayes factor, leads to the conclusion that there are no observable microlensing signatures within this event.

The case of GW191105 is similar, albeit with even less favouring for the lensing models. Here, $\log_{10}(\mathcal{B}^{\rm{L}}_{\rm{U}}) = 0.21$ for the SIS case, and it is $-0.35$ for the point mass lens model. With even the more optimistic of the two models having a lower favouring, as well as a repetition of the posterior behaviour for the lensing parameters, one again concludes that there are no observable microlensing signatures within this event either.
 
\subsection{Targeted Sub-threshold Search}

Whilst lensing may produce multiple images, it is not guaranteed that all of the images will be detected. However, if it can be ascertained that a detected signal (or signal pair) is lensed, this allows deeper investigation for events below the detection threshold used for standard searches. Reciprocally, finding a sub-threshold counterpart to images with a low probability of being lensed could increase the support for the lensing hypothesis. As such, we conducted searches for sub-threshold lensed counterparts to GW191105 and GW191103 individually in the O3 data with the event-specific template banks constructed out of the intrinsic parameter posterior samples \citep{Li:2019osa}. These searches yielded 7 triggers for GW191105 and 15 triggers for GW191103 above the false-alarm rate (FAR) threshold of 1 in 69 years as defined in~\citet{LIGOScientific:2023bwz}. None of these were reported as a potential lensed counterpart to any of the GW191103 and GW191105 events. One of the interesting triggers found was LGW191106\_200820 which arrived just about a day after GW191105, agreeing with galactic scale lens models. However, this trigger was ruled out as a lensed counterpart to a GW191103--GW191105 pair as the overlap in the sky location is poor and the evidence for the event being a real event is very small. It was thus concluded that no promising candidates for an additional sub-threshold counterpart image for GW191103--GW191105 was found within the O3 data.

%% file: GW191230.tex
During the O3 sub-threshold lensing counterpart search, the \textsc{TESLA} pipeline~\citep{Li:2019osa} based on the \textsc{GstLAL} software~\citep{2017PhRvD..95d2001M, 2021SoftX..1400680C} found roughly 470 triggers which could be potential strong lensing counterparts to the supra-threshold events. Of these, two had a FAR lower than 1 in 69 years~\citep{LIGOScientific:2023bwz} though none were found to have support for the lensing hypothesis and all were ultimately discarded.  An alternative method for identifying the sub-threshold triggers as possible lensed counterparts to supra-threshold events,  developed in~\citet{Goyal_in_prep}, uses the \textsc{Bayestar} localisation skymaps, matched-filter chirp mass estimates and the time delay priors to rank all the supra-sub pairs. It identifies the sub-threshold event termed LGW200104\_180425\footnote{Here, we follow the usual naming convention, adding an L at the start of the event name to specify it is a sub-threshold candidate.  Therefore the name of the sub-threshold trigger is LGWYYMMDD\_hhmmss, where YY is the year, MM the month, DD the day, hh the hour, mm the minutes and ss the second in UTC time.} as a possible lensed counterpart to the supra-threshold GW191230\_180458 event. It is the most promising supra-sub pair according to this method as it has significant sky and mass overlap, coupled with apparent lensing parameters matching the expected values for a galaxy lensed models (see~\citet{Goyal_in_prep} for more detail). In the rest of this section, we denote the supra- and the sub-threshold events GW191230 and LGW200104, respectively.

LGW200104 was detected with both LIGO detectors with an SNR of 6.31 in Hanford and 4.94 in Livingston. The \textsc{GstLAL} matched-filter estimates on its chirp mass place it at $67.39M_{\odot}$ with the individual component masses being $82.48M_{\odot}$ and $72.71M_{\odot}$. These high component masses combined with the faintness of the signal contribute to a very low $p_{\textrm{astro}}$ of 0.01 during usual unlensed super-threshold searches, where the event was found with the \textsc{SPIIR}~\citep{Luan:2011qx, Chu:2020pjv} and \textsc{cWB}~\citep{Klimenko:2015ypf} pipelines, signifying a significant lack of probability of the event being a genuine detection. Likewise, the FAR found for this event during the super-threshold searches is $4824/\mathrm{yr}$, also favouring a terrestrial origin for the signal~\citep{Kapadia:2019uut}. Since the sub-threshold searches have a more focused template bank, they also reduce the FAR for the events when they are in the correct region of the parameter space~\citep{McIsaac:2019use, Li:2019osa}. Therefore, the FAR for the event decreases to $6.59/\mathrm{yr}$ when it is found with the \textsc{TESLA} pipeline~\citep{Li:2019osa}, still higher than the threshold used for following-up on sub-threshold events in O3~\citep{LIGOScientific:2023bwz}. In keeping with the analyses done within this work, whilst we do not claim that the event is both genuine and genuinely lensed, we treat it as though it were. Consequently, we investigate the pair using the lensing identification tools used for supra-threshold pairs.

\subsection{\textsc{PyCBC} Sub-Threshold Search}

To further verify this candidate and look for sub-threshold counterparts, an independent search pipeline, based on \textsc{PyCBC}~\citep{Usman:2015kfa,Davies:2020tsx}, 
has been applied~\citep{McIsaac:2019use}. In contrast to the \textsc{GstLAL}-based \textsc{TESLA} pipeline~\citep{Li:2019osa}, the \textsc{PyCBC}-based approach uses a single template based on the posterior distribution of each target event. This search method was previously applied to O1--O2 data~\citep{McIsaac:2019use} and O3a data~\citep{LIGOScientific:2021izm}. Whilst this search was not deployed across the totality of the O3b data, we have applied it as a cross-check on the chunk of data containing 
LGW200104 starting on 2019/12/03 at 15:47:10, and ending on 2020/01/13 at 10:28:01, looking for counterparts to GW191230.

For the template, we selected the maximum-posterior point
from the \texttt{IMRPhenomXPHM} samples for GW191230 released with GWTC-3~\citep{GWTC-3-release},
from a KDE after removing transverse spins,
as to obtain the parameters for a single \texttt{IMRPhenomXAS} template with the following values: $m_1 = 82.48 M_{\odot}$, $m_2 = 72.71 M_{\odot}$, $a_1 = - 0.0037$, and $a_2 = 0.026$.
After running \textsc{PyCBC} over the chunk
using the same clustering steps as in~\citet{McIsaac:2019use}, 
the results are collected and events are ranked by the inverse of their false alarm rate (IFAR).

For the examined chunk, we found 5 candidates above an IFAR threshold of 1 month,
with 2 previously known GW events topping the list, one being GW191230 itself.
To check the correlation of the remaining 3 events with the target supra-threshold event,
we performed a sky overlap estimation of each pair, following the idea described in~\cite{Wong:2021lxf}.
The results are shown in Table~\ref{table:skyoverlap_above_IFAR_1month}. The sky overlap is computed as the fractional overlap between the sky map obtained using parameter estimation for GW191230 and the sky map produced using \textsc{Bayestar}~\citep{Singer_2016} for each sub-threshold candidate. Since these two methods do not match exactly, it leads to a 75\% overlap for the supra-threshold event with itself.

\begin{table*}
\centering
\begin{tabular}{lllllllll}
\hline
\hline
Rank & Name& Event & $\Delta T$ [days] & IFAR [yr] & SNR & 90\%  CR Overlap\\
\hline
\hline
0 & LGW191222\_033537  & GW191230 & 8.60 & 125822.11 & 10.99 &  0.00\\
1 & LGW191230  & GW191230 & 0.00 & 312.15 & 10.11 &  0.75\\
2 & LGW191212\_220841  & GW191230 & 17.83 & 0.57 & 16.38 &  0.00\\
3 & LGW191214\_055524  & GW191230 & 16.51 & 0.10 & 7.16 &  0.02\\
4 & LGW200104  & GW191230 & 5.02 & 0.09 & 8.02 &  0.62 \\
\hline
\end{tabular}
\caption{\textsc{PyCBC} targeted sub-threshold results for counterpart candidates to GW191230 ranked by IFAR. From left to right, the columns represent the event, the time delay compared to the supra-threshold event used to make the template, the inverse false-alarm rate (IFAR), the signal-to-noise ratio (SNR), and the 90\% confidence region (CR) overlap for the sky posteriors.}\label{table:skyoverlap_above_IFAR_1month}
\end{table*}

\begin{figure}
    \includegraphics[keepaspectratio, width=0.49\textwidth]{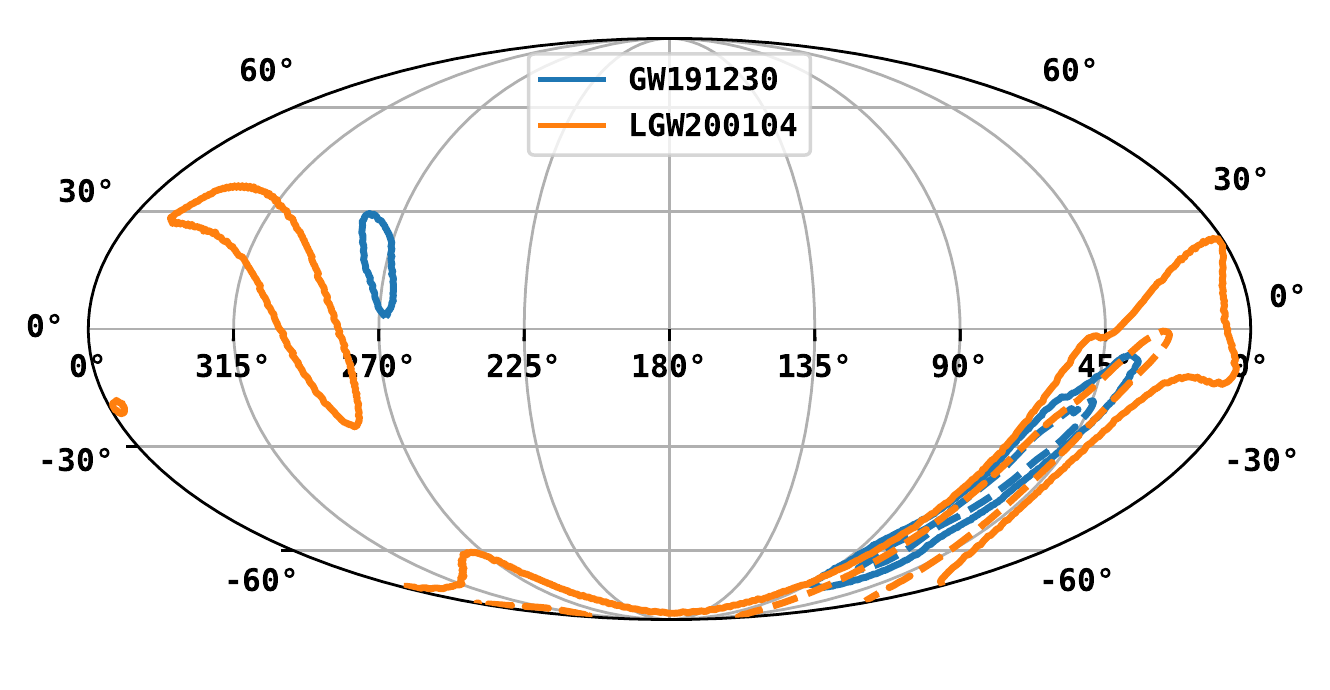}
    \caption{Overlaid LGW200104 and GW191230 skymaps with 90\% and 50\% confidence regions. 
    \label{fig:skymap_overlap_200104_184028}
    }
\end{figure}

In interpreting sub-threshold search results,
one has to take into account that there is a good chance that, 
in addition to the potential counterpart images,
there will be candidates originating from instrumental glitches
or also from different, weaker, GW events that were not identified in previous searches. 
Here, the candidate corresponding to LGW200104 is ranked
fifth (including GWTC events)
with an IFAR of 0.09\,years.
Its network SNR is recovered as 8.02
(with an SNR of 6.31 and 4.94 in H1 and L1, respectively)
and its sky localization overlap with GW191230 is 62\%.
The sky overlap map is given in Fig.~\ref{fig:skymap_overlap_200104_184028}.

The third-ranked event has a higher SNR, but no sky overlap with GW191230
and can be clearly identified as a glitch since there is simply an excess in power for all frequencies at a given time and no time-frequency evolution similar to that expected for a CBC signal.
The fourth-ranked event is clearly a case of a scattered-light glitch~\citep{Soni:2021lde, Soni:2021def, tolley2023archenemy}. Appendix~\ref{app:pycbc} further details these two events. In the end, the \textsc{PyCBC} sub-threshold search also finds LGW200104 as the most plausible lensed GW sub-threshold counterpart to GW191230 consistent with the \textsc{GstLAL} pipeline~\citep{Li:2019osa} and the ranking method proposed in~\citet{Goyal_in_prep}.

\subsection{Posterior Overlap Analyses}

\begin{figure*}
\centering
\includegraphics[width = 0.49\linewidth]{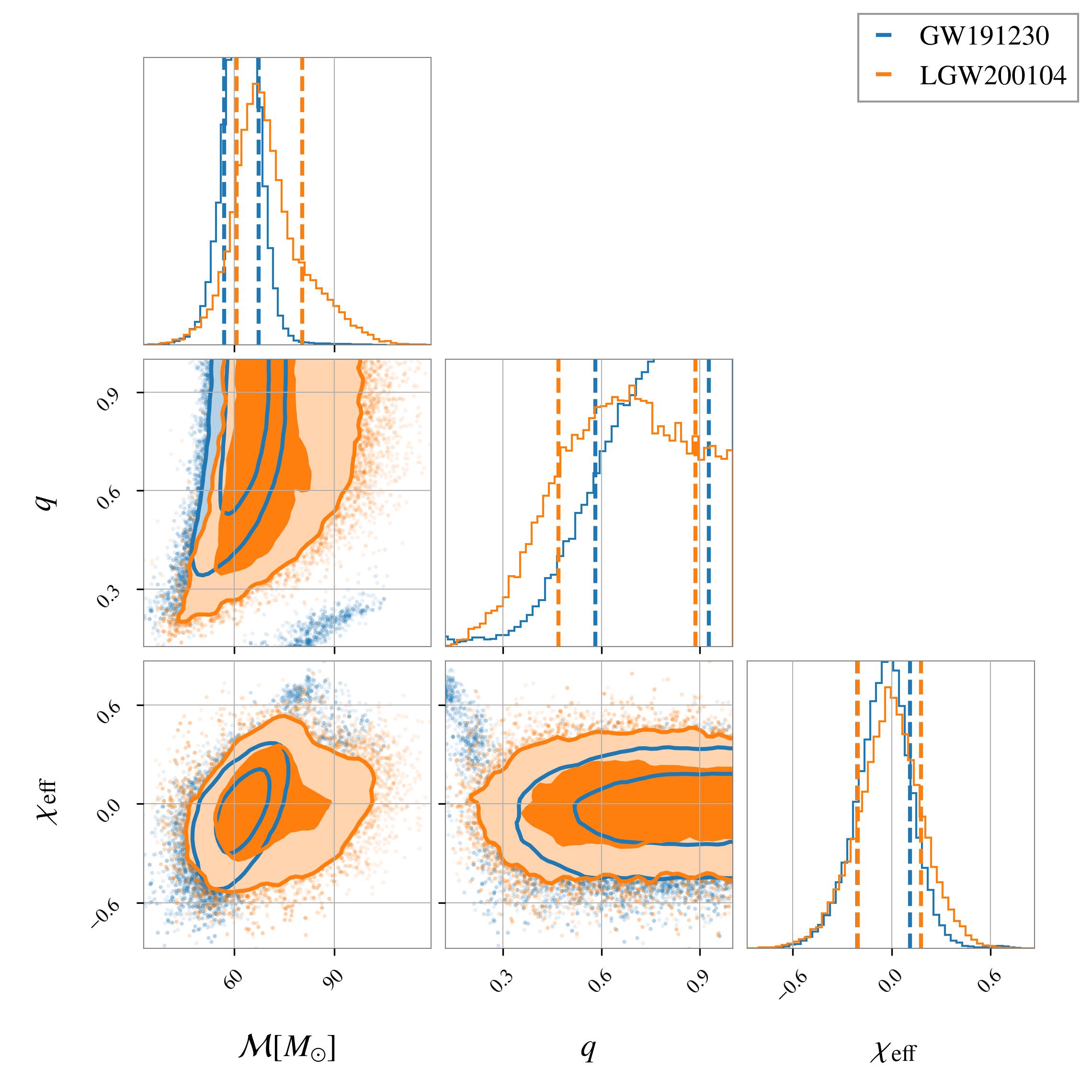}
\includegraphics[width = 0.49\linewidth]{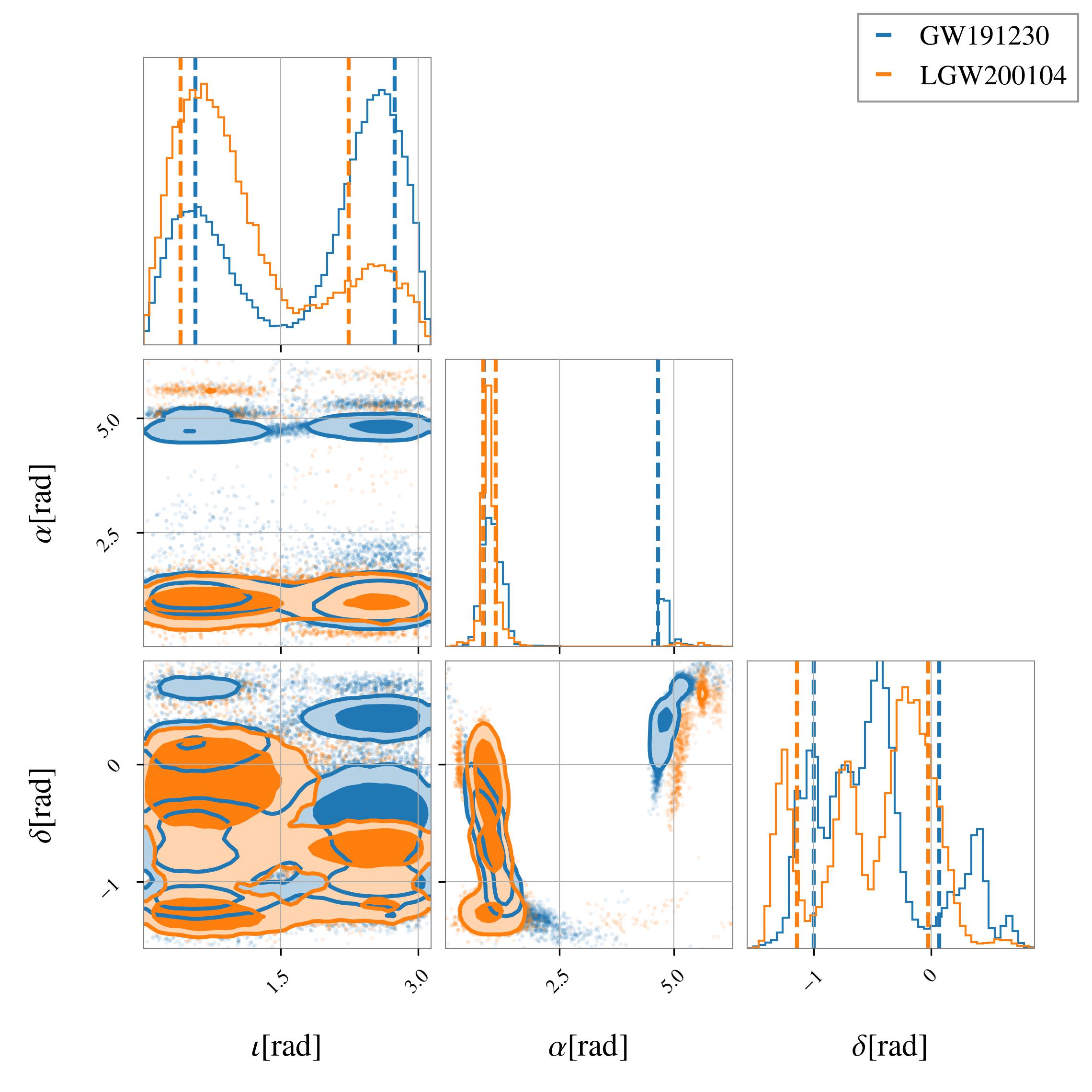}
\caption{Posteriors for GW191230 (blue) and LGW200104 (orange). The posteriors, though broad, have significant overlap for both the intrinsic (left) and extrinsic (right) parameters.}\label{fig:PO_supersub}
\end{figure*}

From the PO analysis this pair has $\log_{10} \mathcal{B}^{\textrm{overlap}} = 2.45$. Since the combined SNR of the sub-threshold trigger is close to 8, it is reasonable to treat the event pair the same way we did for other candidates. Using the same time delay priors as for the supra-threshold events we find $\log_{10} \mathcal{R}^{\textrm{gal}} = 0.97$ which makes the log of the overall PO statistic $3.43$. Fig.~\ref{fig:PO_supersub} shows the posteriors for LGW200104 and GW191230. Visually, the degree of overlap in both extrinsic and intrinsic parameters is high. However the intrinsic parameters posteriors are broader as compared to GW191103--GW191105. For events having high masses in the detector frame, such as these, the number of cycles in the waveform within the LIGO--Virgo frequency band is smaller. This leads to broader posteriors which in turn reduce the overlap statistics, while increasing the rate of coincidental overlaps~\citep{Cutler:1994ys, PhysRevD.47.2198}. In addition, lensed events are more likely to have higher detector frame masses than unlensed events due to the their magnification. Hence, it is a challenge to identify high-mass lensed candidates. Including the population priors and selection effects might help~\citep{Haris:2018vmn,Lo:2021nae, Janquart:2022zdd}.

We also compute the significance of the pair using the supra-threshold background introduced in Sec.~\ref{subsec:PO_191103_191105} and find it to be  $\lessapprox 1\sigma$, as shown in Fig.~\ref{fig:PO_candidates}.  This implies that this pair, though not conclusively lensed,  is one of the most significant candidates amongst all the O3 event pairs. 

To look for potential waveforms systematics, we perform the same analysis as in Sec.~\ref{sec:1103-1105_blusys} using results from 
\textsc{parallel\_Bilby} runs with the same five waveform models in the posterior overlap calculation. 
The results are shown in Table~\ref{table:blusysGW191230-LGW20104} and
we find relatively consistent results. 
However, we again notice that the aligned-spin waveforms produce higher 
$\mathcal{B}^{\textrm{overlap}}$, by a factor of $\sim$6.
In this case, both $a_1$ and $a_2$ peak towards zero for 
the aligned-spin models for the two events, leading to a better overlap. 
However, all waveforms agree on identifying this pair as 
consistent with lensing.

\begin{table}
\centering
\begin{tabular}{lr}
\hline
\hline
Waveform & $\log_{10}(\mathcal{B}^{\textrm{overlap}})$ \\
\hline
\hline
\texttt{IMRPhenomXAS} & 3.20 \\
\texttt{IMRPhenomXHM} & 3.13 \\
\texttt{IMRPhenomXP} & 2.52 \\
\texttt{IMRPhenomXPHM} & 2.45 \\
\texttt{IMRPhenomTPHM} & 2.55 \\
\hline
\end{tabular}\caption{\label{table:blusysGW191230-LGW20104}
Posterior-overlap factors for the GW191230--LGW200104 pair using different waveform models in the single-event PE.}
\end{table}

The PO analysis can quickly identify the lensed candidates but it does not take into account the full correlation between the data streams, the selection effects, and the lensing parameters. Hence, the candidates are passed on to JPE pipelines for further investigations.

\subsection{Joint Parameter Estimation Based Investigation}
\label{sub:JPE_GW191230}

After discovering the candidate with the sub-threshold searches and confirming interest with PO, it was analysed in more detail using \textsc{GOLUM}~\citep{Janquart:2021qov, Janquart:2023osz} with the samples of the supra-threshold event as the prior for the sub-threshold one. The evidence of this run can be compared to the results of a standard unlensed \textsc{Bilby}~\citep{Ashton:2018jfp} investigation to yield the coherence ratio. In this case, the run yielded $\log_{10}(\mathcal{C}^{L}_{U}) = 1.1$. This is lower than that calculated for GW191103--GW191105. However, in this case, one of the two images is very close to the limit of a detectable event and this may impact the coherence ratio. By itself, the coherence ratio also is still high enough to favour the lensing hypothesis. To initially investigate the pair's significance, it was compared with the same background as outlined in Sec.~\ref{Sec:GW191103_GW191105}. This results in a $\mathrm{FAP}_{\mathrm{pp}}$ of 1.4\% and thus a FAP of 0.6 which indicates the event is consistent with a coincidental unlensed background event. However, the background resulting in this FAP consisted entirely of supra-threshold events and the exact effects of sub-threshold events in such studies have not been deeply explored. 

\begin{figure}
\centering
\includegraphics[keepaspectratio, width=\linewidth]{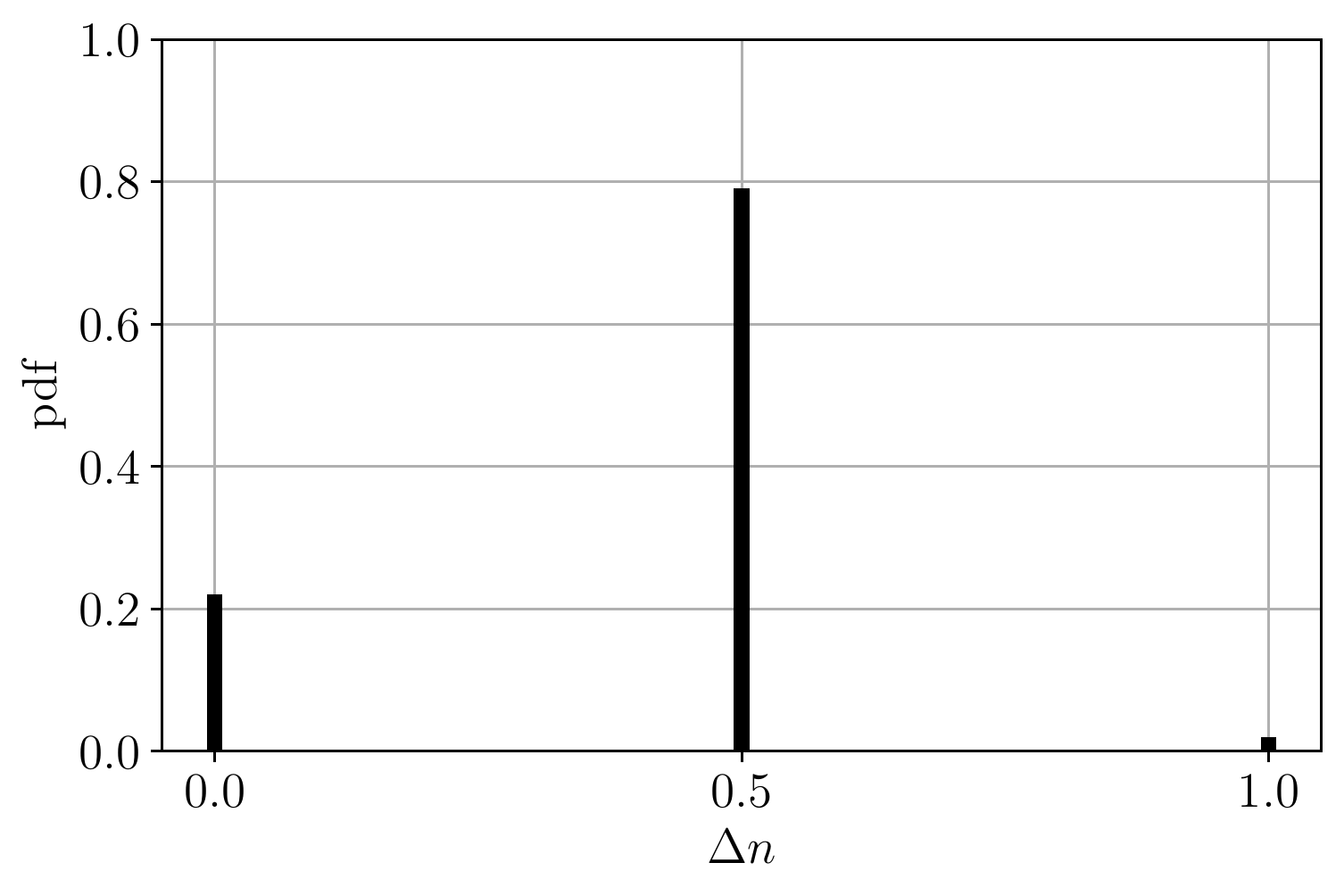}
\caption{The difference in Morse factor for the GW191230--LGW200104 event pair according to the \textsc{GOLUM} pipeline. The preferred value is 0.5 but there is also some support for 0.}
\label{fig:posts_super_sub}
\end{figure}

\begin{figure}
\centering
\includegraphics[keepaspectratio, width=\linewidth]{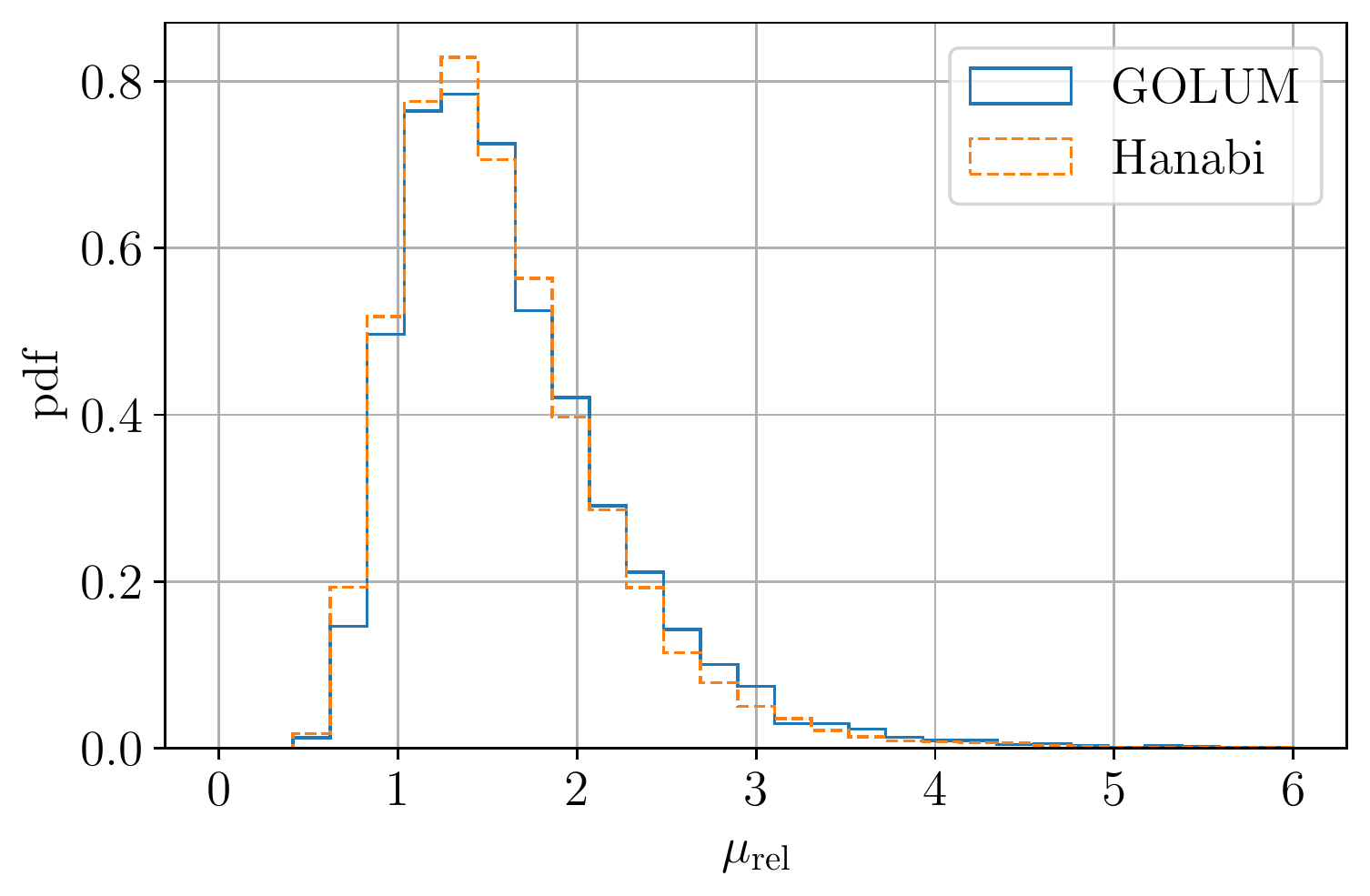}
\caption{Posterior distribution for the relative magnification for the GW191230--LGW200104 event pair measured with the \textsc{GOLUM} (solid blue) and \textsc{Hanabi} (dashed orange) pipelines. One sees that the measured values are consistent between the two pipelines.}
\label{fig:mu_rels_golum_hanabi}
\end{figure}

The \textsc{GOLUM} analysis also offers the possibility to gain insight into the lensing parameters. In particular, it gives information about the difference in Morse factor and relative magnification\footnote{It also gives the possibility to constrain the time delay, but since the arrival times are very well measured already in GW data analysis, this does not provide much additional information.}. Their posterior distributions are given in Fig.~\ref{fig:posts_super_sub} and~\ref{fig:mu_rels_golum_hanabi}, respectively, in which it can be seen that the relative magnification peak is at $\sim$1.5, meaning that its value is close to the highest-probability region expected for an SIE lens model (see for example Fig.~\ref{fig:CompaMgal}). On the other hand, the difference in Morse factor is less well constrained, with the main support being for $\Delta n=0.5$, but also some support for the $\Delta n=0$ case. We note that, generally, for well-detectable lensing events, the difference in Morse factor is well recovered~\citep{Janquart:2021qov, Janquart:2023osz}. This observation may indicate that the event is unlensed but also simply that the lower SNR of the signal makes the identification harder. These lensing parameters and the time delay, however, are consistent with expected values for a galaxy-scale lens.

Based on the \textsc{GOLUM} results, we may also investigate how the coherence ratio and the FAP evolve with the inclusion of expected parameter values from a lens model, as was done in Sec.~\ref{Sec:GW191103_GW191105}. Using the same background, and the same models as within that section, we compute the population-reweighted coherence ratios. These values are reported in Table~\ref{tab:clus_pop_super_sub}. Notably, the coherence ratio found for the SIE model including both the relative magnification and the time delay is now higher than that for the GW191103--GW191105 event pair, meaning that the observed characteristics are more in agreement with the expected value for a galaxy-lens model and the currently observed population than for that pair. This is a demonstration of the fact that the candidate pair---even though the sub-threshold event is not confirmed to be of astrophysical origin---is interesting. 

\begin{table}
    \centering
    \begin{tabular}{l c r}
    \hline
    \hline
    Statistic & $\log_{10}$ value & $\textrm{FAP}_{\textrm{PP}}$ \\
    \hline 
    \hline
    $\mathcal{C}_U^L$ & 1.105 &  $1.401\times10^{-2}$\\
    $\mathcal{C}_{\mathcal{M}_{\mu, t}}$ & 3.427 & $1.167\times10^{-3}$ \\ 
    $\mathcal{C}_{\mathcal{M}_{t}}$ & 1.915 & $2.017\times10^{-3}$\\
    \hline
    \end{tabular}

    \caption{Values of the detection statistic for the GW191230--LGW200104 lensed candidate pair without lens model ($\mathcal{C}_U^L$), and with an SIE lens, with ($\mathcal{C}_{\mathcal{M}_{\mu, t}}$) and without ($\mathcal{C}_{\mathcal{M}_{t}}$) relative magnification accounted for. The $\rm{FAP}_{\rm{PP}}$ is decreased when using an SIE model, showing that the observed characteristics are in line with the expected behaviour for the given model and population.}
    \label{tab:clus_pop_super_sub} 
\end{table}

\begin{figure}
\centering
\includegraphics[width=0.75\columnwidth]{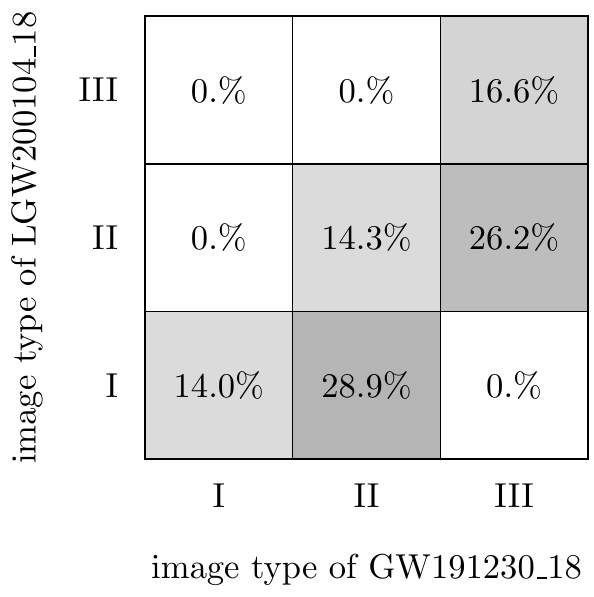}
\caption{\label{fig:super_sub_hanabi}
Posterior probability mass function for the image type of GW191230 and the image type of LGW200104 from \textsc{hanabi}. It is consistent with the \textsc{GOLUM} result that it is more likely for the difference in Morse factor to be $\Delta n = 0.5$ (i.e. the II-I and III-II configuration) than to be $\Delta n = 0$ (i.e. the I-I, II-II and III-III configuration).}	
\end{figure}

The GW191230-LGW200104 pair was also analyzed by the full JPE package \textsc{hanabi} \citep{Lo:2021nae} where the joint parameter space of the two events was simultaneous explored by the stochastic sampler \textsc{dynesty} \citep{10.1093/mnras/staa278} with settings identical to those used in \cite{LIGOScientific:2023bwz}. The parameters found are consistent with the ones found using the \textsc{GOLUM} framework---see Fig.~\ref{fig:mu_rels_golum_hanabi} for a comparison of the relative magnifications. In particular Fig.~\ref{fig:super_sub_hanabi} shows the posterior probability mass function for the possible image types of the GW191230-LGW200104 pair. We see that the image type configurations for the two events that have non-zero support have the difference in the Morse phase factor $\Delta n$ either $0$ (i.e. the I-I, II-II and III-III configuration) or $0.5$ (i.e. the II-I and III-II configuration). Again, this is consistent with the values shown in Fig.~\ref{fig:posts_super_sub} obtained using \textsc{GOLUM}.

We also performed the Bayes factor calculation comparing the probability ratio of the lensed versus the unlensed hypothesis for this pair in the same fashion that we did for the GW191103--GW191105 pair as in Sec.~\ref{Sec:GW191103_GW191105}. Again, we use the same set of source population models as in \cite{LIGOScientific:2023bwz}, e.g. the \textsc{powerlaw + peak} model for the source masses from the GWTC-3 observations \citep{LIGOScientific:2021psn} and three models for the merger rate density: Madau-Dickinson~\citep{Madau:2014mmd}, $\mathcal{R}_{\rm min}(z)$, and $\mathcal{R}_{\rm max}(z)$. Table~\ref{tab:super_sub_hanabi_astro} shows the $\log_{10}$ Bayes factors computed using the three merger rate density models with the simple SIS lens model \citep{LIGOScientific:2023bwz} and the SIE + external shear model \citep{hanabi_astro}. We see that the values calculated using the SIE + external shear model are positive but only mildly ($<1$), and they are also consistently higher than the values computed using the SIS model (which are all negative), indicating that the pair is more consistent with a more realistic strong lensing model. It should be noted that the calculations assumed that both GW events are astrophysical of origin and the second is treated as a supra-threshold event. 

\begin{table*}
\centering
\begin{tabular}{lccc}
\hline \hline
\backslashbox{Lens model}{Merger rate density} & Madau-Dickinson & $\mathcal{R}_{\rm min}(z)$ & $\mathcal{R}_{\rm max}(z)$ \\
\hline \hline
  SIS	& $-0.76$ & $-0.35$ & $-0.57$ \\
  SIE + external shear & $\;\;0.14$ & $\;\;0.57$ & $\;\;0.30$ \\
\hline
\end{tabular}
\caption{\label{tab:super_sub_hanabi_astro}$\log_{10} \mathcal{B}^{\rm L}_{\rm U}$ for the GW191230 and LGW200104 pair from \textsc{hanabi} assuming three different (source) merger rate density models and two different lens models. We see that the values with the SIE + external shear model are all positive (but only mildly) and consistently higher than that with the SIS model which are all negative, indicating a higher compatibility of the pair with a more realistic strong lensing model. Note that the calculations assumed that both GW events are astrophysical of origin. These values are not sufficient to claim the event pair to be lensed as we would require a positive $\log_{10}$ posterior odds, and the observed Bayes factors are not high enough to balance the low prior odds for strong lensing.}
\end{table*}

Despite some of the evidence for this event aligning relatively well with the expectations for a lensed event, there remain several key arguments against a claim of lensing for this pair. The first is that whilst it is the case that the event has the highest currently observed Bayes factor, it is insufficient to yield a positive log posterior odds considering that the $\log_{10}$ prior odds is between $-2$ and $-4$~\citep{Ng:2017yiu, Oguri:2018muv, 2018MNRAS.476.2220L, Buscicchio:2020cij, Mukherjee:2020tvr, Wierda:2021upe}. The second argument, is the nature of the trigger itself. The sub-threshold event is not convincing---consider for instance the extremely low $p_{\mathrm{astro}}$ and FAR---and there is no clear evidence to claim that the event is a genuine GW detection. 

In the end, although the event pair is unlikely to be lensed, the analyses performed on this event pair serve as a powerful demonstration of the necessity for searching for such sub-threshold counterparts and the kinds of information that they may yield.

\subsection{Electromagnetic Follow-Up}

\begin{figure}
\centering
\includegraphics[keepaspectratio, width=\linewidth]{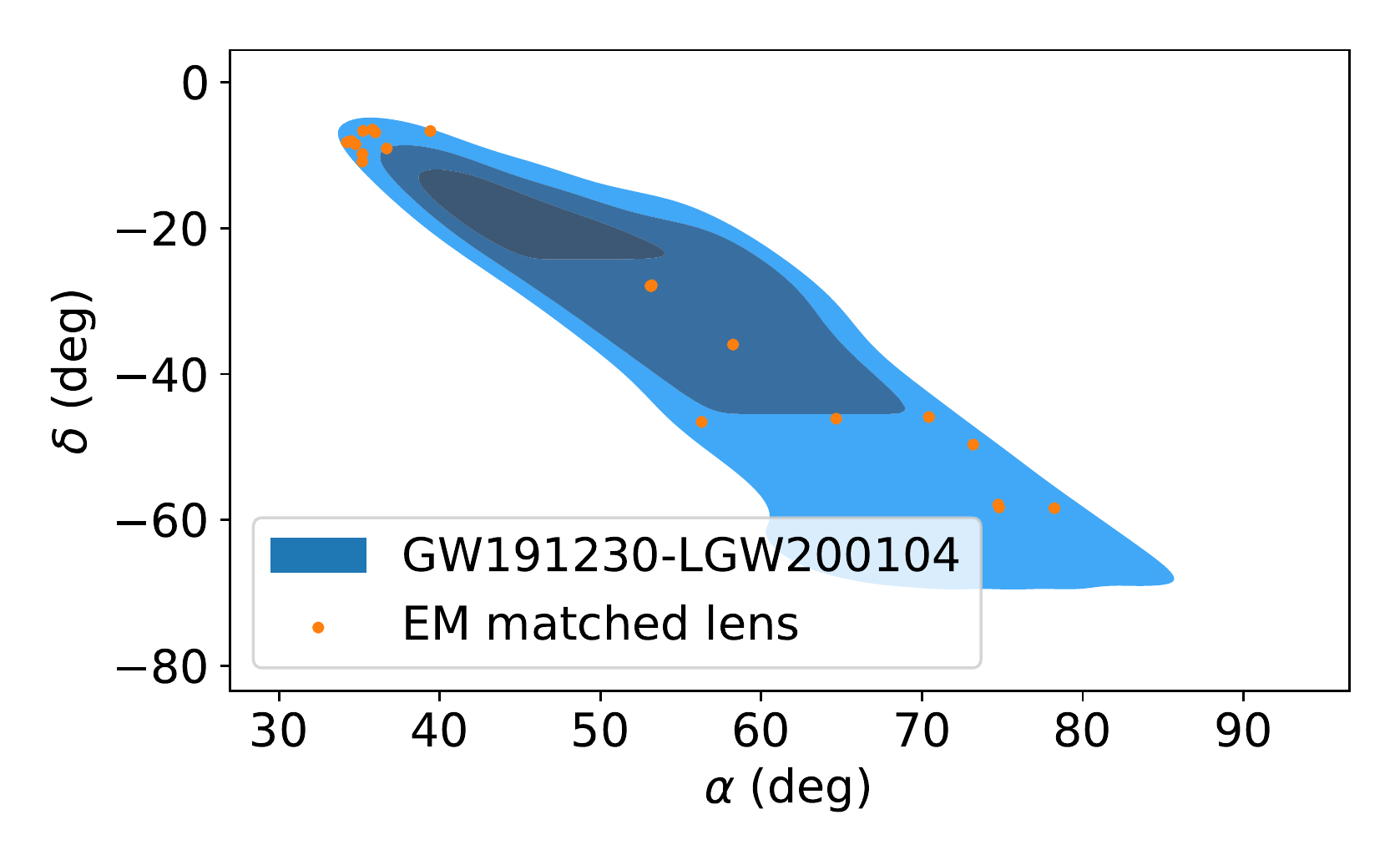}
\caption{Sky localisation 10\%, 50\%, and 90\% confidence region (from dark to light) for the GW191230-LGW200104 pair. Overlaid are the cross-matched 21 candidates from the Master Lens Database.}\label{fig:sup_sub_skyloc}
\end{figure}

Even though the lensing hypothesis is disfavored, we investigate if there are any EM lens systems with consistent lensing properties from the literature for this event pair. As in Sec.~\ref{subsec:GW191103_GW191105_EM_follow}, we cross-matched with the MLD.\@ The grade A and grade B lenses selected from the catalog at galaxy scales showed 21 matches (see Fig.~\ref{fig:sup_sub_skyloc}). There are two major lens samples that fall within the 90\% confidence region of the sky localisation in addition to a handful of systems from heterogeneous studies~\citep[e.g.,][]{Treu2006,Shu2016}. The Strong Lensing Legacy Survey (SL2S) lens systems are those with RA$<40$~deg$^2$~\citep{Gavazzi2012, More2012, Sonnenfeld2013} and South Pole Telescope (SPT)--ALMA lens systems are those with RA$>50$~deg$^2$~\citep[e.g.,][]{Weiss2013,Spilker2016}. Some of these lens systems do not have sufficient information (for instance, are lacking source redshift) and the others do not have best-fit mass model parameters or sufficiently high-resolution imaging to identify the multiple lensed image positions. In the absence of these constraints, the time delay of a few days (i.e.  $\sim 5$ days for the GW191230--LGW200104 pair) can easily be consistent with many of the lens systems. As a result, a rudimentary analysis to determine the likelihood of these lens systems and to select the most likely EM counterparts is not possible, and a more detailed mass modelling for all of the 21 systems would be necessary to find the most promising EM counterpart. Whilst the observation of a third or a fourth GW image would also help further in constraining the characteristics, the lack of high resolution imaging to clearly and accurately measure the image positions are still anticipated to be the limiting factor. Lastly, we again caution the reader that any incompleteness of the survey (both telescope imaging and subsequent lens searches) may mean that additional potential EM counterparts may have been missed from our initial list of 21 candidates.  We found more candidates as compared to the 
EM follow up of GW191103--GW191105 (Sec.~\ref{subsec:GW191103_GW191105_EM_follow}) merely because there are more optical surveys that have looked towards the sky region of interest here with respect to the sky overlap region of GW191103--GW191105 which is nearer to the poles. Hence in order to have a robust association one needs to incorporate the selection effects for both the EM and GW observations (see~\cite{Wempe:2022zlk} for possible avenues).

%% file: GW200208.tex
GW200208\_130117, denoted GW200208 from here on, was selected for follow-up in this paper for two reasons. The first was because it was the event with the highest $\log_{10} \mathcal{B}^{L}_{U}$ in the O3 microlensing analysis~\citep{LIGOScientific:2023bwz}, with a value of 0.8 which, whilst positive, is still inconclusive. The secondary reason was that the event had a relatively narrow posterior on the redshifted lens mass which is atypical of unlensed events. In the O3 lensing paper it was considered that the cause of the apparent favouring of microlensing for the event could be due to short duration noise fluctuations causing an apparent dip in the signal, mimicking the beating pattern of microlensing~\citep{LIGOScientific:2023bwz}. 

\subsection{Microlensing Model Investigation}

As has been done with a selection of the other events within this paper, GW200208 was re-examined using the \textsc{Gravelamps} pipeline~\citep{Wright:2021cbn} to investigate the potentiality of model selection in the case of a microlensing candidate. Whilst the isolated point mass model used by \textsc{Gravelamps} is similar to that used by the O3 microlensing search pipeline, there are sufficient implementation differences to warrant re-examination with \textsc{Gravelamps} for this model rather than simply comparing the results of the SIS investigation with those of the O3 microlensing analysis pipeline.

For all of the models examined, \textsc{Gravelamps} had increased favouring for the microlensing hypothesis with this event as compared to the O3 microlensing anlysis pipeline. In the point lens case, the $\log_{10} \mathcal{B}^{L}_{U}$ increases to $1.20$. This confirms the result from the O3 analysis and shows the event warrants additional investigations. In the SIS case, the preference increased further with the $\log_{10} \mathcal{B}^{L}_{U}= 1.77$. This again is sufficiently high as to warrant additional scrutiny, but not high enough to make a statement by itself.

One stage of preparatory work that would shed additional light on the potential significance of the $\log_{10} \mathcal{B}^{L}_{U}$ figures would be a detailed background study to determine the range over which unlensed events may appear as microlensing candidates. Such a study had been done for the microlensing search in~\cite{LIGOScientific:2021izm, LIGOScientific:2023bwz} and allowed the statement above that for the case of that pipeline, the $\mathcal{B}^{L}_{U}$ recorded for GW200208 was within the expected range for an unlensed event. Due to computational constraints, this has not yet been performed for the \textsc{Gravelamps} models, although it is one of the steps that should be taken during O4 so that the significance of candidates may be evaluated quickly. What we note is that the increase between the two pipelines would not necessarily have rendered GW200208 outside of expectations for the O3 microlensing search pipeline, and the general trend of events analysed would appear to indicate that in general SIS is preferred over the point mass model. This is likely due to a lens with similar parameters producing lower peak amplifications in the SIS model as compared with the point mass model which would yield smaller deformations from the unlensed template. 

\begin{figure}
    \includegraphics[width=\linewidth]{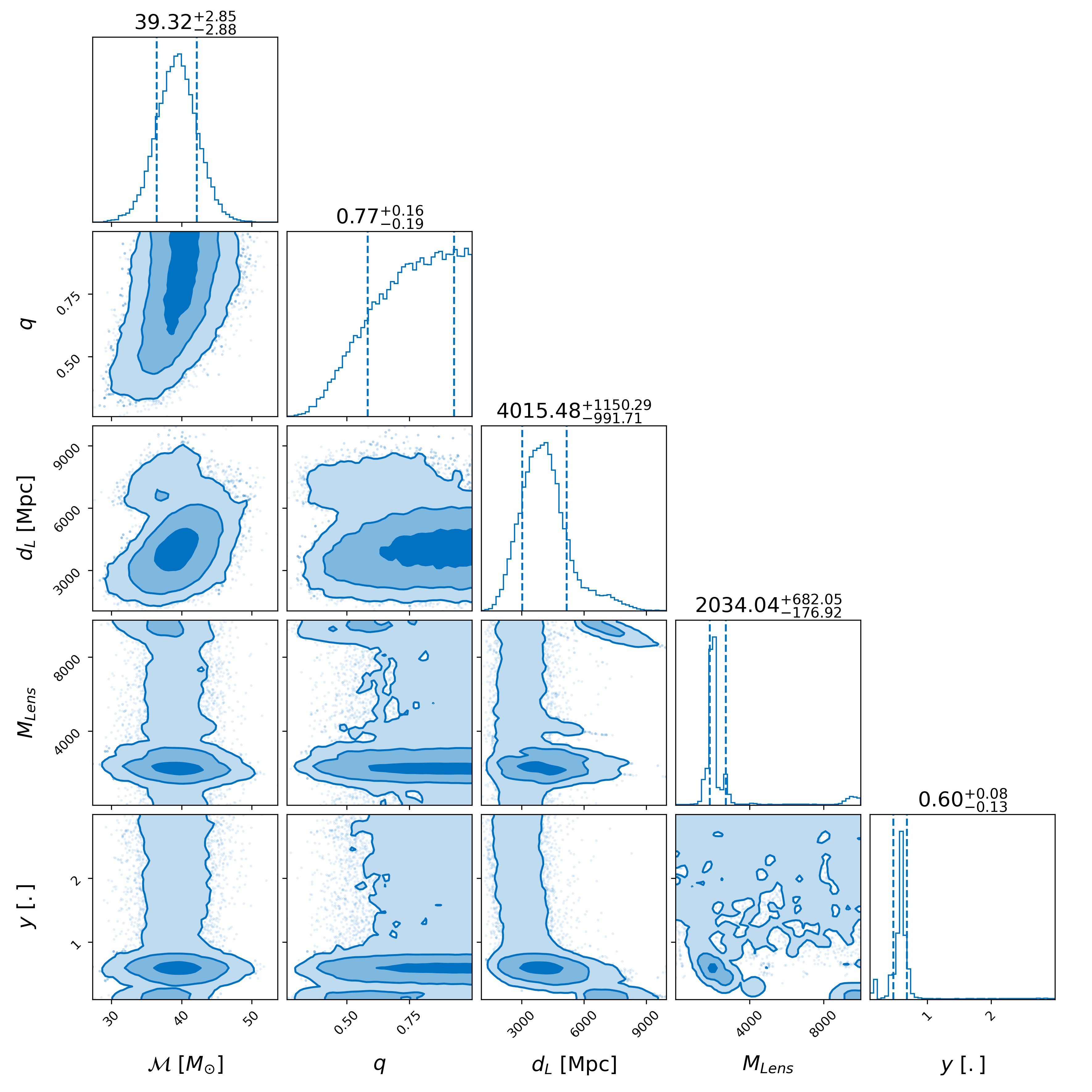}
    \caption{Posteriors of a subset of source parameters as well as lensing parameters for GW200208 in the SIS microlensing case. Unlike the other events that have been examined within this work, the lensing parameters for the model are well constrained, even if the 3-$\sigma$ confidence interval is a bit noisy. This means that this event, unlike the others, cannot be immediately ruled out as a lensing candidate from this test.}\label{fig:GW200208-microlensing-posterior}
\end{figure}

\begin{figure}
    \includegraphics[width=\linewidth]{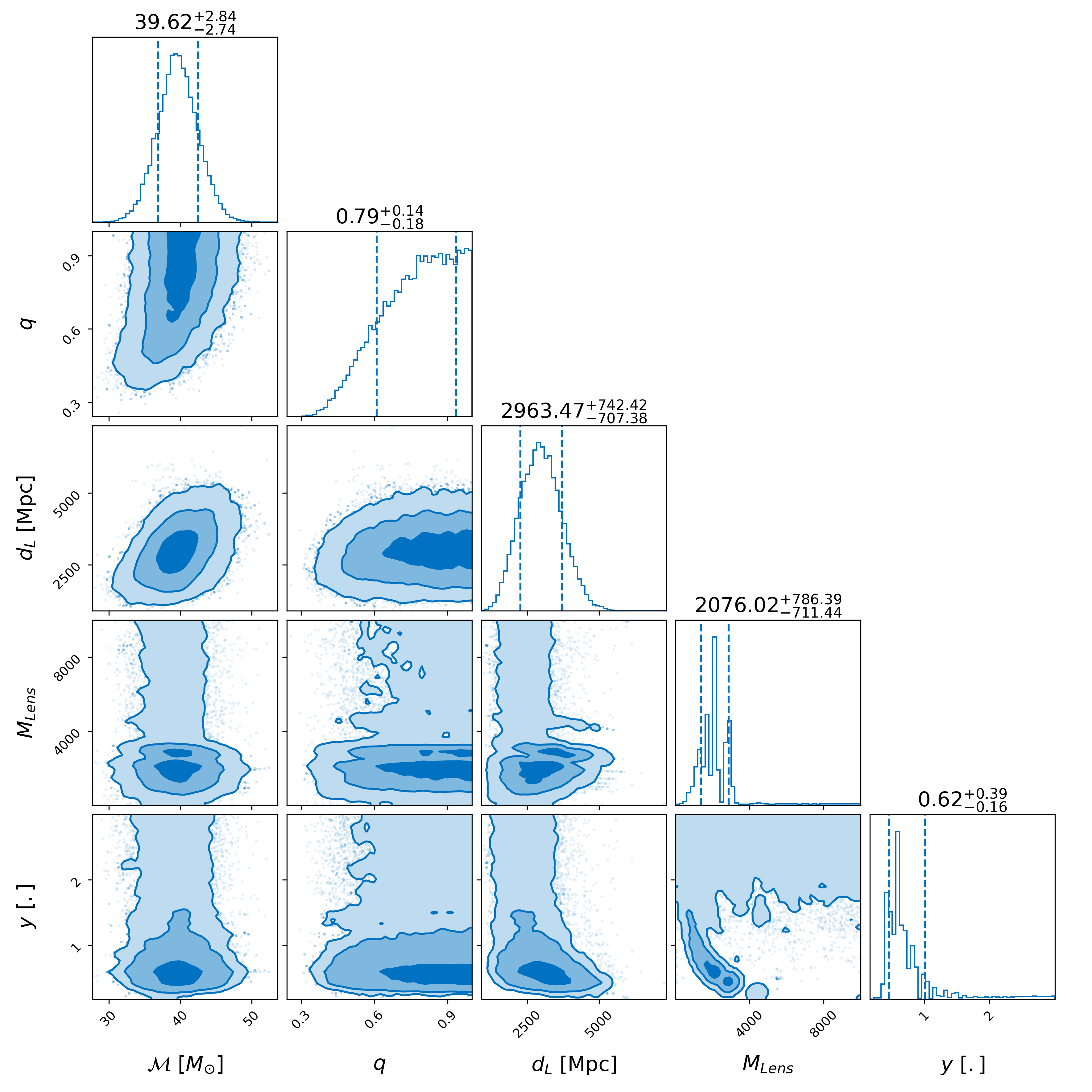}
    \caption{Posteriors of a subset of source parameters as well as lensing parameters for GW200208 in the point mass microlensing case. For this case, the posteriors are similarly constrained, notably arriving at similar lensing parameters as the SIS case even if the 3-$\sigma$ confidence interval is a bit noisy.}\label{fig:GW200208-microlensing-pessimistic-posterior}
\end{figure}

For the other events that have been examined, the posteriors for lensing parameters have been a factor in determining that the microlensing hypothesis is unlikely. However, in the case of GW200208---the posteriors of which in the SIS case are shown in Fig.~\ref{fig:GW200208-microlensing-posterior}---this same test yields results more consistent with the microlensing hypothesis. As can be seen in the figure, the lensing parameter posteriors are relatively narrowly constrained around a $2000M_{\odot}$ lensing object with a source position value of $0.60$. Fig.~\ref{fig:GW200208-microlensing-pessimistic-posterior} shows that in the more pessimistic point lens case, the lensing parameters are constrained to similar values which further cements the need for additional scrutiny of this event. For the two models, we see that the 3-$\sigma$ confidence intervals for the lensing parameters are a bit noisy. However, the peaks in the density distributions remain clearly visible. 

\begin{table*}
\centering
\begin{center}
\begin{tabular}{c c c c c c c}
 \hline
 \hline
 Waveform & $f_\mathrm{low}$ & $f_\mathrm{high}$ & duration & $p(M^{z}_{\textrm{Lens}})$ & $p(y)$ &$\log_\mathrm{10} \mathcal{B}^\mathrm{L}_\mathrm{U}$ \\
 \hline
 \hline
\texttt{IMRPhenomXPHM} & $20$ & $448$ & $4$ & L.U (min=$1$, max=$10^5$) &  P.L ($\alpha=1$, min=$0.1$, max=$3.0$) & $0.89$  \\
\texttt{IMRPhenomXPHM} & $20$ & $1024$ & $4$ &  L.U (min=$10$, max=$10^5$) & P.L ($\alpha=1$, min=$0.01$, max=$5.00$) & $0.63$  \\ 
\texttt{IMRPhenomXPHM} & $20$ & $896$ & $8$ & L.U (min=$10$, max=$10^5$) & P.L ($\alpha=1$, min=$0.01$, max=$5.00$) & $0.46$ \\ 
\texttt{IMRPhenomXPHM} & $15$ & $448$ & $4$ & L.U (min=$10$, max=$10^5$) &  P.L ($\alpha=1$, min=$0.1$, max=$3.0$) & $1.02$  \\ 
\texttt{IMRPhenomXPHM} & $15$ & $448$ & $4$ & L.U (min=$10$, max=$10^5$) &  P.L ($\alpha=1$, min=$0.01$, max=$5.00$) & $0.53$  \\ 
\texttt{IMRPhenomXPHM} & $15$ & $448$ & $4$ &  L.U (min=$10$, max=$10^5$) &  Uniform (min=$0.1$, max=$3.0$) & $1.04$  \\ 
\texttt{IMRPhenomXPHM} & $15$ & $448$ & $4$ & L.L.U (min=$10$, max=$10^5$) &  P.L ($\alpha=1$, min=$0.1$, max=$3.0$) & $0.70$  \\ 
\texttt{IMRPhenomXPHM} & $15$ & $448$ & $4$ & L.L.U (min=$10$, max=$10^5$) &  Uniform (min=$0.1$, max=$3.0$)& $0.95$  \\ 
\texttt{IMRPhenomXPHM} & $15$ & $448$ & $4$ & Uniform (min=$10$, max=$10^5$) &  Uniform (min=$0.1$, max=$3.0$) & $0.50$  \\ 
\texttt{NRSur7dq4} & $20$ & $448$ & $4$ & L.U (min=$1$, max=$10^5$) &  P.L ($\alpha=1$, min=$0.1$, max=$3.0$) & $0.96$  \\ 
\texttt{NRSur7dq4} & $18$ & $448$ & $4$ & L.U (min=$1$, max=$10^5$) &  P.L ($\alpha=1$, min=$0.1$, max=$3.0$) & $0.90$  \\
\hline
\end{tabular}
\caption{This table presents the results of a Bayesian model comparison study between the unlensed and the microlensed hypotheses for $\mathrm{GW}200208$, with microlensing model corresponding to an isolated point-lens mass. The study was conducted for different configurations and sampler settings, as indicated by the different columns, to verify for possible noise artifacts and check the influence of the sampler settings on the results. The table includes the waveform approximant used, the lower and higher frequency cutoffs used for likelihood evaluation $(f_{\rm low},~f_{\rm high})$, duration of the data segment used, and the priors on the redshifted microlens mass $(M^{z}_{\textrm{Lens}})$ and the impact parameter $(y)$, represented by $p(M_{\rm Lz})$ and $p(y)$ respectively. The Bayes factor for the support of microlensing over the unlensed waveform model is given by $\log_{10}\mathcal{B}^\mathrm{L}_\mathrm{U}$. The range of the priors is also indicated. The terms `L.U' and `L.L.U' refer to log-normal and log-log-normal distributions respectively, while `P.L' refers to a power law profile with the index given by $\alpha$. }
\label{table:GW200208_GWMAT}
\end{center}
\end{table*}

\begin{figure*}
    \includegraphics[width=0.9\textwidth]{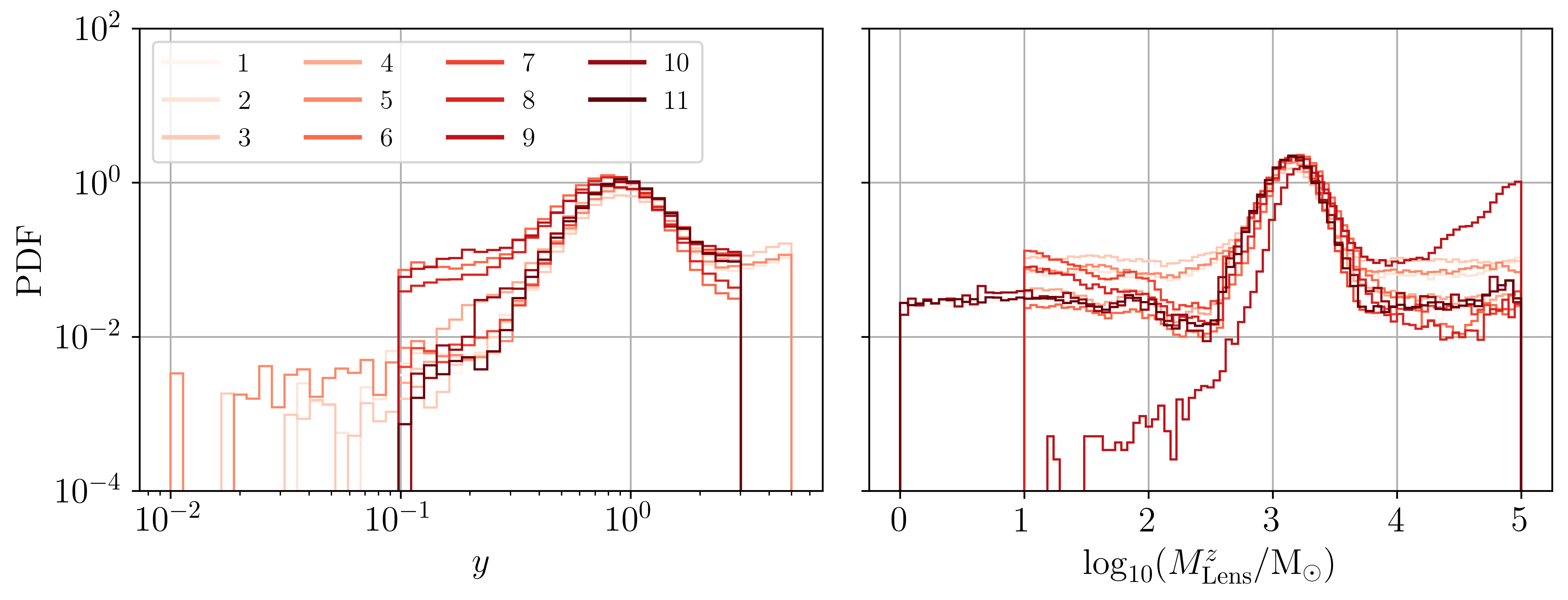}
    \caption{The posterior densities of the recovered microlensing parameters for different parameter estimation runs, as presented in Table~\ref{table:GW200208_GWMAT}. The results are visualized with varying colors from light to dark (numbered from 1 to 11, indicating different runs as we move down the table.}
    \label{fig:GW200208_ML_prms_posteriors_GWMAT}
\end{figure*}

We further investigate GW200208 using various sampler and prior settings, as well as testing different waveform models, as listed in Table~\ref{table:GW200208_GWMAT}. These tests are designed to verify whether noise artifacts could be at the root of the observed support for microlensing, and whether the results are robust for different sampler settings. We assume the microlensing model of an isolated point-lens and do parameter estimation using the \textsc{GWMAT} framework~\citep{GWMAT2023}. 

This pipeline utilizes a \textsc{Cython} implementation of the amplification factor calculation for the isolated point mass model serves as an independent cross-check for the study. Additionally, the pipeline incorporates a dynamic cutoff based on the source position $y$ to transition to a geometric optics approximation. The resulting probability density functions representing the recovered microlensing parameters are illustrated in Fig.~\ref{fig:GW200208_ML_prms_posteriors_GWMAT}.

Firstly, we observe that the posteriors for both parameters, $y$ and $\log_\mathrm{10}M^{z}_{\textrm{Lens}}$, are consistent across different runs, with median values and 1-sigma errors of $1.07^{+0.61}_{-0.32}$ and $3.15^{+0.18}_{-0.21}$, respectively.
However, the posteriors for $y$ show signs of railings at the upper end, particularly in runs with an upper limit of 5 in the prior.  Since the SNR is around $11$, we limit the prior on $y$ to a maximum of 5, which would already require an SNR $\gtrsim 40$ to make the microlensing signatures detectable\footnote{The minimum SNR required to distinguish two waveforms with a match value of $m$ is roughly $\sqrt{2 / (1 - m^2)}$, where we used an estimate of the Bayes factor and set a threshold of unity for distinguishability  \citep{cornish2011gravitational, vallisneri2012testing, del2014testing}. For $M^{z}_{\textrm{Lens}} = \mathcal{O}(10^3~\mathrm{M}_\odot)$, the match value between the unlensed and lensed waveform with $y=5$ comes out to be $\sim 0.9993$, which implies a minimum SNR of $\sim 40$.}. It is worth noting that the posteriors for $\log_\mathrm{10}M_\mathrm{Lz}$ are relatively well-converged, with a sharp peak, except in the case where we used a uniform prior in $M^{z}_{\textrm{Lens}}$, which shows a tendency towards bimodality with another peak at $\log_\mathrm{10}M^{z}_{\textrm{Lens}}\gtrsim 5$. 

As shown in Table \ref{table:GW200208_GWMAT}, we primarily use $f_\mathrm{high}=448$ Hz\footnote{We choose $f_\mathrm{high} \leq 0.875*(f_\mathrm{s}/2)$, where $f_\mathrm{s}$ is the sampling frequency (see Appendix E of \cite{LIGOScientific:2021djp}).} and a duration of $4$ seconds due to the heavy mass nature of the $\mathrm{GW}200208$ event, which has a total binary mass of approximately $90$ M$_\odot$ and negligible spin content.
Comparing the first entry in the table with the second-to-last entry, we find that \texttt{NRSur7dq4}~\citep{varma2019surrogate} yields a slightly higher Bayes factor value than \texttt{IMRPhenomXPHM} for similar settings, except that the \texttt{NRSur7dq4} case imposes a total mass constraint of greater than $66$ M$_\odot$, considering its region of validity for $f_\mathrm{low}=20$ Hz. However, since the event has a total mass $\geq 74$ M$_\odot$ with $3\sigma$ certainty, we also analyzed the event with $f_\mathrm{low}=18$ Hz (last row), resulting in a slight decrease in $\log_\mathrm{10} \mathcal{B}^\mathrm{L}_\mathrm{U}$. On the other hand, when we lowered the value of $f_\mathrm{low}$ to $15$ Hz for the \texttt{IMRPhenomXPHM} case, the $\log_\mathrm{10} \mathcal{B}^\mathrm{L}_\mathrm{U}$ increased compared to a similar run with $f_\mathrm{low}=20$ Hz (see 1st and 4th entry).

When we choose $f_\mathrm{high}=896$ Hz and a duration of $8~$s (3rd row), both $\mathcal{B}^\mathrm{U}_\mathrm{noise}$ and $\mathcal{B}^\mathrm{L}_\mathrm{noise}$ decrease, as does $\log_\mathrm{10} \mathcal{B}^\mathrm{L}_\mathrm{U}$, resulting in the lowest value among all the different settings used in the table. 
Additionally, $\log_\mathrm{10} \mathcal{B}^\mathrm{L}_\mathrm{U}$ also decreases when we broaden the prior on $y$ or $\log_\mathrm{10}M_\mathrm{Lz}$ (compare, for example, the 4th and 5th row), which could be additionally lowered due to railing and bimodalities at the higher values of $y$ and $\log_\mathrm{10}M^{z}_{\textrm{Lens}}$, as shown in Fig.~\ref{fig:GW200208_ML_prms_posteriors_GWMAT}. 

The apparent railing and the bimodality can be attributed to the fact that if the likelihood fails to exhibit strong unimodality, the posterior densities may vary depending upon the prior beliefs. A higher upper limit in the prior of $y$ with a power law profile $p(y) \propto y$ will assign more weight to higher values of $y$. Similarly a uniform prior in $M^{z}_{\textrm{Lens}}$ places a higher weight on heavier microlenses than a log-uniform or log-log-uniform prior, thereby increasing the posterior density in that region. However, if the SNR is high, or if the event is truly microlensed, the likelihood values are better constrained and the posterior densities would not be expected to change much with the priors. 

We also note that the Bayes factors presented in Table~\ref{table:GW200208_GWMAT} show more variability. These results indicate that we cannot make a firm conclusion on whether the event is microlensed or not based on the Bayes factor, and the event can only be deemed interesting probabilistically depending on the prior beliefs we choose.

\subsection{Maximum Likelihood Injection}

One avenue of investigation to determine whether an event with the parameters that are suggested by the lensing models within \textsc{Gravelamps} would be detected, and if it was detected, how significant a detection would we expect is to examine a simulated waveform with the maximum likelihood parameters injected into simulated detector noise.

Whilst as stated above, a full-scale injection campaign was not undertaken for the \textsc{Gravelamps} analysis due to temporal and computational constraints, we can investigate if the $\mathcal{B}^{L}_{U}$ figures would be plausible for a genuine microlensing event of the suggested parameters by injecting a signal with the parameters of the maximum likelihood sample of the \textsc{Gravelamps} analysis into a realisation of Gaussian noise assuming a representative PSD for the noise around the time of detection and analysing this injection with the \textsc{Gravelamps} models in the same fashion as the real event. 

Performing this analysis yields value for the $\log_{10} \mathcal{B}^{L}_{U}$ of $0.37$ and $0.79$ for the isolated point mass and SIS profiles respectively which are lower than those given for the event. This suggests that it would be difficult to confidently confirm an event with these parameters, and therefore this test does not rule out either the possibility of a genuine microlensing event or a noise fluctuation in the data. This again highlights the need for additional investigations such as the aforemnetioned full scale injection campaign to given greater context to the significance of the calculated Bayes factors.

\subsection{Residual Power Examination}

\begin{figure*}
    \centering
    \includegraphics[width=0.7\linewidth]{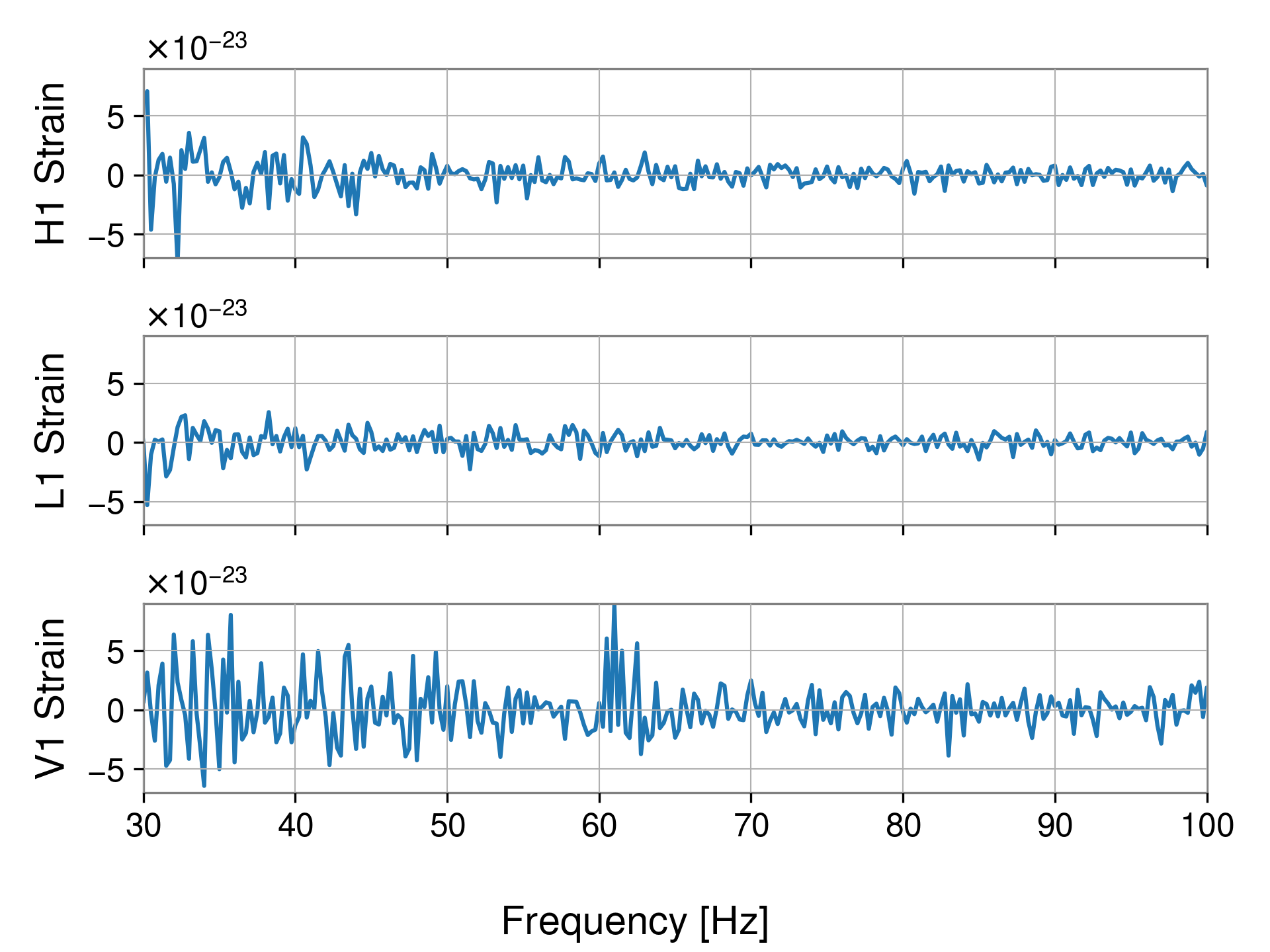}
    \caption{The residual power remaining, from top-to-bottom, in LIGO Hanford, LIGO Livingston, and Virgo when subtracting the best fit non-lensed waveform template for GW200208 as determined by the unlensed PE from the detector strain, over a subset of the total frequency range. As can be seen, there is no obviously coherent oscillatory behaviour in the residual power which would be expected in the case of the microlensing hypothesis. This absence would suggest that what remains is noise related rather than signal related.}\label{fig:GW200208-residual-power}
\end{figure*}

An additional means of scrutinising a microlensing candidate event is to examine the residual power that is left within the data when the maximum-likelihood waveform fit from the non-lensed PE carried out on the event is removed from the data. In the case of a genuine microlensing event, one would expect to see remaining oscillating amounts of power in each of the detectors due to the unaccounted for oscillating behaviour of the amplification factor. In the case of a non-lensed event, absent any other systematic errors, one would expect to see the fluctuations associated with the noise of the detector. This type of analysis is also performed when looking for deviations from general relativity, where one looks at the residual power in the data after the maximum-likelihood general-relativity template---equivalent to the unlensed one---has been subtracted from it~\citep{LIGOScientific:2021sio}. The residual power investigation carried out in~\citet{LIGOScientific:2021sio} for this event yielded a $p$-value, corresponding to the probability of obtaining a background event with a residual SNR higher than the event, of 0.97. This suggests the remaining power is within expectations for residual noise.

The residuals from performing this subtraction are shown for each of the detectors, for a subset of the total frequency range, in Fig.~\ref{fig:GW200208-residual-power}. As can be seen, none of the detectors display an obvious coherent oscillation in the residual power that would be expected in the microlensing hypothesis. These residuals are more typical of the noise which may indicate that the event is unlikely to be a microlensing event. Hence, despite the increased favouring of the microlensing hypothesis under the SIS case in terms of raw PE analysis, this work draws the same conclusion as that of the lensing searched conducted by the LVK:\@ GW200208, whilst interesting, is not a genuine microlensing event---though it does highlight the need for more systematic studies on the imapct of the noise on microlensing searches in the future. 

\subsection{Millilensing Analysis}
\label{ssub:millilensing_analysis}

The range of masses favoured by the microlensing analysis both in~\citet{LIGOScientific:2023bwz} and within this work would also be within the millilensing regime as described in Sec.~\ref{Sec:TheoryIntro}. In the analysis performed here, four millilensing waveform models were used---three with fixed numbers of millilensing signals (two, three, and four signals respectively), and the fourth being a variable multi-signal waveform allowing any number of signals from 1 to 6. 

With each of the millilensing waveform models, we performed parameter estimation of the source and lensing parameters using the Bayesian inference library \textsc{Bilby}~\citep{2019ApJS..241...27A, 2020MNRAS.499.3295R} with the \textsc{dynesty}~\citep{10.1093/mnras/staa278} sampler and the \texttt{IMRPhenomXPHM} waveform approximant~\citep{Pratten:2020ceb}, following the method developed in~\citet{Liu:2023ikc}.

The plots resulting from these PE runs are presented in this section and Appendix~\ref{app:milli_lensing}. Before commenting on each of the results individually, we note the terminology used commonly for each of the figures. The millilensing parameters are described by a series of effective luminosity distances, $d^{\textrm{eff}}_{j}$, time delays with respect to the first image $t_{j+1}$, and Morse phase $n_{j}$ for the $j^{th}$ image. The convention for this work is that the images are referred to in time ordering.

\begin{figure}
    \centering
    \includegraphics[width=\linewidth]{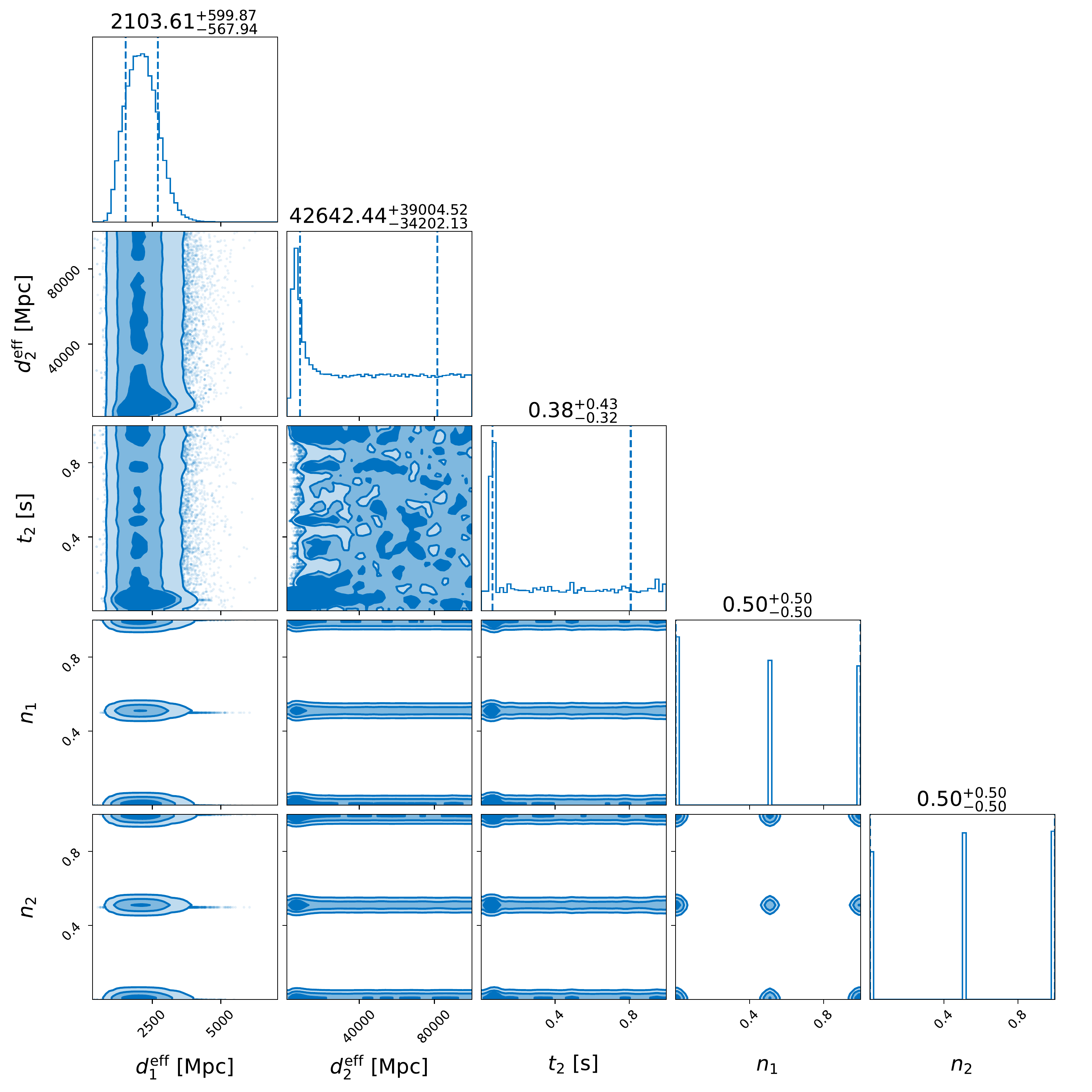}
    \caption{Corner plot of the millilensing parameters obtained from a two-signal analysis of GW200208. Notable is that there is a peak within the effective luminosity distance and time delay parameters for a potential second millilensing signal. However, this peak could be explained by the presence of detector noise.}\label{fig:milli_two}
\end{figure}

Turning to the specific results, we begin with Fig.~\ref{fig:milli_two} detailing the two-signal case. The posterior distribution of the effective luminosity distance of the first signal, $d^{\mathrm{eff}}_1$, displays a clear peak as would be expected of a real signal. The posteriors for the second signal parameters---i.e.\ the effective luminosity distance and the time delay---both display peak-like features but also have an extended underlying posterior. Without additional evidence this would be insufficient to claim millilensing of the signal and could be explained by the presence of noise, as discussed in the microlensing analysis of the event. 

The three-signal analysis results (detailed in Appendix~\ref{app:milli_lensing}) are in agreement with the two-signal case, where the effective luminosity distance, $d^\mathrm{eff}_2$, and time delay of the second signal, $t_2$, display a peak-like feature in the posterior distribution. The corresponding parameters of the third signal, however, do not show significant peaks in their distributions which disfavours the presence of a third millilensing component signal. The four-signal analysis (detailed in Appendix~\ref{app:milli_lensing}) also lacks any peaking features in the parameters of either the third or fourth signal---returning their uniform priors and giving additional evidence for the disfavouring of any more than two signals. 

\begin{table}
    \centering
    \begin{tabular}{l l}
    \hline
         Model & $\log_{10}(\mathcal{B}^{\mathrm{Milli}}_{\mathrm{U}})$ \\
         \hline
         \hline 
         Two signals & 0.86  \\
         Three signals & 0.92  \\
         Four signals & 0.96  \\
         Multi-signal & 1.10 \\
         \hline  
    \end{tabular}
    \caption{Comparison of Bayes factors for the evidence against the unlensed hypothesis from the millilensing runs for GW200208.}
    \label{tab:milli_bayes_factor}
\end{table}

\begin{figure}
    \centering
    \includegraphics[width=\linewidth]{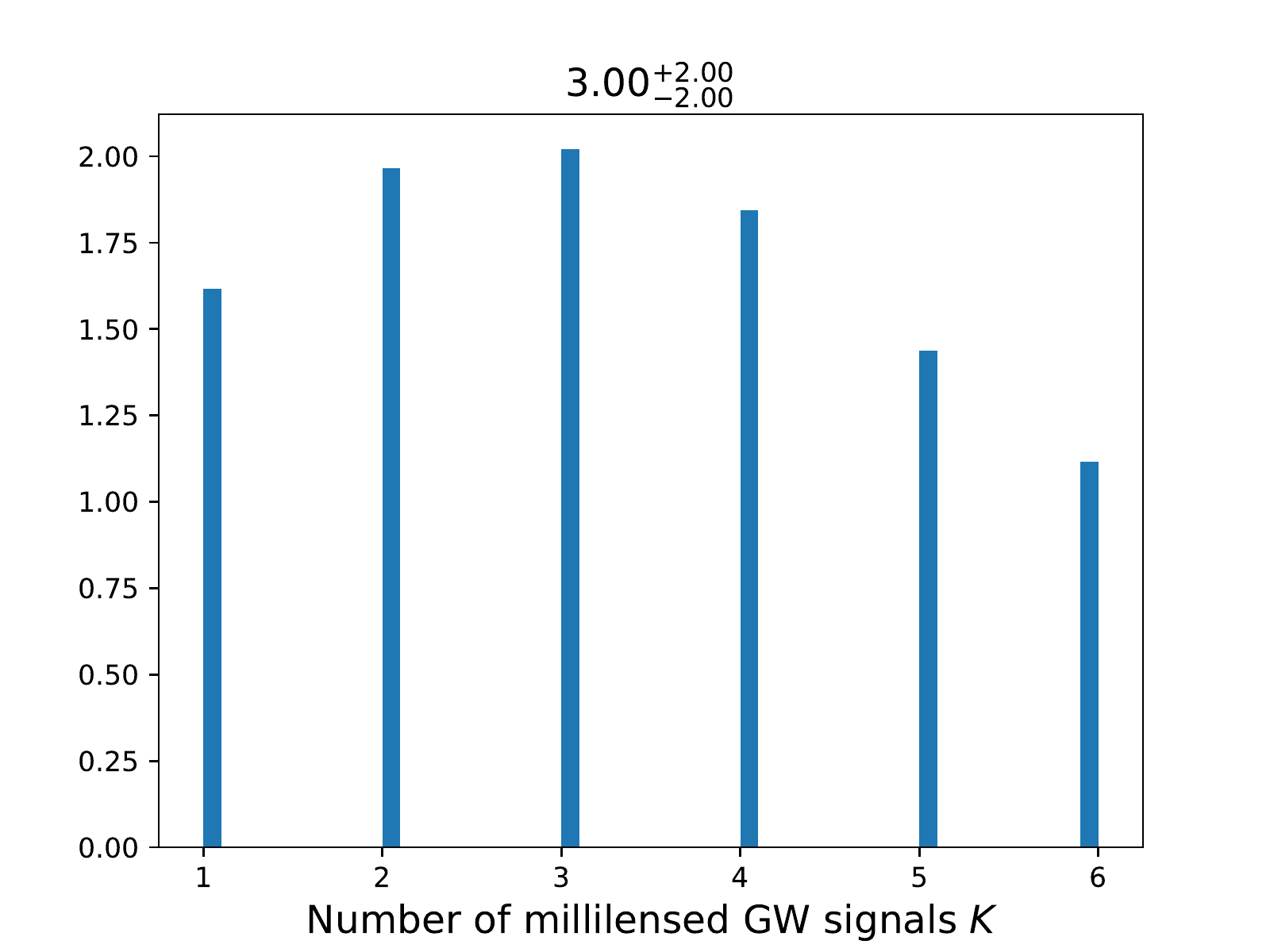}
    \caption{Discrete posterior of the number of signals in the multi-signal analysis for GW200208. This posterior is insufficient to confidently assert a number of signals present.}\label{fig:milli_K}
\end{figure}

Lastly, a multi-signal analysis, making the number of millilensing components signals itself a free parameter, was performed. In this analysis, the number of signals was allowed to range from 1 to 6. The posterior distributions for the millilensing parameters are shown in Appendix~\ref{app:milli_lensing}. These posteriors are again consistent with the assertion that there is no favouring for any number of signals above a possible second one. The additional results of attempting to infer the number of signals are shown in Fig.~\ref{fig:milli_K}. The discrete posterior here is notably ambiguous disallowing confident constraints on the number of signals here---despite only the posteriors of the second image having any notable features. This serves to underline the fact that the features within the second image posteriors are insufficient to claim a millilensing detection. 

In addition to the posterior plots, we also compute the Bayes factors for the millilensing hypothesis against the unlensed one. The values are given in Table~\ref{tab:milli_bayes_factor}. They slightly favor the millilensing case, not significantly enough to truly favor this hypothesis when accounting for the astrophysical information, the prior odds, and the observed posteriors.

It is thus the conclusion of the millilensing analysis that there is insufficient evidence to support millilensing within GW2000208 despite a favouring of the millilensing hypothesis when comparing the Bayes factors of signal versus noise for each of the models.

%% file: ConclusionsAndProspects.tex
In this work, we have analyzed candidates found to be interesting by the LIGO-Virgo-KAGRA lensing searches in the full O3 data~\citep{LIGOScientific:2023bwz} as though they were genuinely lensed. We considered three main types of lensing: strong lensing, millilensing, and microlensing with the types being defined by the effects they have on observed gravitational waves. Though the events investigated do not display strong evidence of being lensed, the analyses done here demonstrate possible follow-up strategies for future observing runs in order to assess the significance of any lensing candidate event. 

Firstly for the GW190412 event---which displayed the greatest support for being a type II image---, we analysed the data with two other waveform models, showing that these do not show as strong a feature. Therefore, the observed support is most likely due to combined noise and waveform systematics. Additionally, we study the event with a microlensing pipeline to see if such lensing could lead to the apparent deviations in the overall phasing, finding no evidence for this hypothesis. 

Next considered was the GW191103--GW191105 pair, which was flagged as interesting because of its relatively high coherence ratio and the consistency of the relative amplitudes and time separation with the expectations for the relative magnification and time delay of galaxy lenses. Testing the effect of waveform systematics on the posterior overlap analysis showed that the lensing hypothesis is favoured regardless of waveform choice. We then went on to demonstrate that whilst the event is compatible with galaxy-lens models, inclusion of models in the coherence ratio ultimately does not yield a significant increase in support, as seen by the low coherence ratio. The disfavouring of the lensed hypothesis is further shown by including a more realistic SIE model in our analysis pipeline, still finding a negative log Bayes factor. Furthermore, we demonstrated how an electromagnetic counterpart to the host galaxy could be searched for and showed that no confident counterpart could be found. We also demonstrated that neither of the individual events has any indications of microlensing effects. Finally, we looked for a subthreshold lensed counterpart but found no promising candidate.  

A new ranking scheme for the sub-threshold counterparts of detected supra-threshold events found a new interesting candidate pair: the GW191230\_180458 supra-threshold and the LGW200104\_180425 sub-threshold events. We performed investigations with additional dedicated sub-threshold searches which confirmed interest in the event pair. As was done for the other event pairs, we then analysed it using the standard and follow-up tools. First with the posterior overlap analysis, we saw the event pair is an interesting candidate. Again, a waveform systematic study yielded consistent results for the various waveforms considered in this work. In this case, analysis with the joint parameter estimation tools showed that upon the inclusion of a galaxy-lens model, the coherence ratio was higher than for the previous pair though only to the extent that 40 unlensed events can produce a pairing with similar results by coincidence. Additionally, performing the computation of the proper Bayes factor with an SIS model leads to negative log Bayes factor, disfavouring the lensing hypothesis. On the other hand, the inclusion of an SIE lens model leads to a marginally positive log Bayes factor. However, it is not high enough to compensate for the prior odds, and therefore the posterior odds is disfavoring the lensing hypothesis. Besides, the low $p_{astro}$ and FAR cast doubt on the astrophysical origin of the sub-threshold event. Finally, as with the previous pair, a search for possible electromagnetic counterparts yielded no confident matches which is in line with the expectation for the events not being lensed. Let us also re-iterate the absence of clear evidence for the sub-threshold event to be genuine in the first place.

The last event analysed was GW200208\_130117 which was flagged as the event closest to the expectations for a microlensing event. First, we re-examined the event using different lens models. We found the Bayes factors to be slightly higher than those computed in~\citet{LIGOScientific:2023bwz}, but still compatible with values found for background unlensed events. We also studied the variation of the Bayes factor for a point-mass lens model depending on the priors used. We found values ranging from slightly disfavoring lensing to favoring this hypothesis, in line with other analyses performed on this event. To deepen the investigation, we perform a maximum-likelihood injection, recovering the injection with a Bayes factor comparable to the one found for the event, showing the difficulty to confidently identify microlensing at this sensitivity. We then looked at the residual power remaining when subtracting the best-fit unlensed  waveform. This test did not yield any particular evidence for the remaining power being consistent between the detectors, which is in line with the event being most likely not microlensed. Finally, since millilensing may also lead to beating patterns in the waveform, we searched for millilensing features in this event. These searches demonstrated that there was no additional evidence for any more than two lensed waveforms comprising the event, and the combination of posteriors and Bayes factors were not sufficient to conclusively favour the millilensing hypothesis in general. 

In the end, the conclusions from the additional tests are in line with those given by the LIGO-Virgo-KAGRA Scientific Collaboration~\citep{LIGOScientific:2023bwz}, showing that none of the events or event pairs is likely to be genuinely lensed, regardless of their initially intriguing characteristics. By doing these additional studies, we have shown some important points for future lensing searches, such as the possibility of having waveform systematics, the impact of the lens model in the analysis, the difficulties one may have to distinguish between events resembling each other by chance and genuinely lensed ones, the interplay between microlensing and millilensing, and other additional avenues to further investigate lensing candidates in the future. These follow-up methods should be valuable in the future when more intriguing lensed candidates are found.

%% file: Acknowledgments.tex
The authors thank Aditya Vijaykumar for the useful discussion, providing some scripts to generate plots, and carefully re-reading the manuscript. The authors thank Christopher Berry for useful discussions on $p_{\mathrm{astro}}$.  

The authors are grateful for computational resources provided by the LIGO laboratory and Cardiff University and supported by the  National Science Foundation Grants PHY-0757058 and PHY-0823459, and the STFC grant ST/I006285/1, respectively. The authors are also grateful to the Inter-University Center for Astronomy \& Astrophysics (IUCAA), Pune, India for additional computational resources.

This material is based upon work supported by NSF’s LIGO Laboratory which is a major facility fully funded by the National Science Foundation. The authors also gratefully acknowledge the support of the Science and Technology Facilities Council (STFC) of the United Kingdom, the Max-Planck-Society (MPS), and the State of Niedersachsen/Germany for support of the construction of Advanced LIGO and construction and operation of the GEO 600 detector. Additional support for Advanced LIGO was provided by the Australian Research Council. The authors gratefully acknowledge the Italian Istituto Nazionale di Fisica Nucleare (INFN), the French Centre National de la Recherche Scientifique (CNRS) and the Netherlands Organization for Scientific Research (NWO), for the construction and operation of the Virgo detector and the creation and support of the EGO consortium. The authors also gratefully acknowledge research support from these agencies as well as by the Council of Scientific and Industrial Research of India, the Department of Science and Technology, India, the Science \& Engineering Research Board (SERB), India, the Ministry of Human Resource Development, India, the Spanish Agencia Estatal de Investigaci\'on (AEI), the Spanish Ministerio de Ciencia e Innovaci\'on and Ministerio de Universidades, the Conselleria de Fons Europeus, Universitat i Cultura and the Direcci\'o General de Pol\'itica Universitaria i Recerca del Govern de les Illes Balears, the Conselleria d’Innovaci\'o, Universitats, Ci\'encia i Societat Digital de la Generalitat Valenciana and the CERCA Programme Generalitat de Catalunya, Spain, the National Science Centre of Poland and the European Union – European Regional Development Fund; Foundation for Polish Science (FNP), the Swiss National Science Foundation (SNSF), the Russian Foundation for Basic Research, the Russian Science Foundation, the European Commission, the European Social Funds (ESF), the European Regional Development Funds (ERDF), the Royal Society, the Scottish Funding Council, the Scottish Universities Physics Alliance, the Hungarian Scientific Research Fund (OTKA), the French Lyon Institute of Origins (LIO), the Belgian Fonds de la Recherche Scientifique (FRS-FNRS), Actions de Recherche Concer\'ees (ARC) and Fonds Wetenschappelijk Onderzoek – Vlaanderen (FWO), Belgium, the Paris Ile-de-France Region, the National Research, Development and Innovation Office Hungary (NKFIH), the National Research Foundation of Korea, the Natural Science and Engineering Research Council Canada, Canadian Foundation for Innovation (CFI), the Brazilian Ministry of Science, Technology, and Innovations, the International Center for Theoretical Physics South American Institute for Fundamental Research (ICTP-SAIFR), the Research Grants Council of Hong Kong, the National Natural Science Foundation of China (NSFC), the Leverhulme Trust, the Research Corporation, the National Science and Technology Council (NSTC), Taiwan, the United States Department of Energy, and the Kavli Foundation. The authors gratefully acknowledge the support of the NSF, STFC, INFN and CNRS for provision of computational resources. This work was supported by MEXT, JSPS Leading-edge Research Infrastructure Program, JSPS Grant-in-Aid for Specially Promoted Research 26000005, JSPS Grant-inAid for Scientific Research on Innovative Areas 2905: JP17H06358, JP17H06361 and JP17H06364, JSPS Core-to-Core Program A. Advanced Research Networks, JSPS Grantin-Aid for Scientific Research (S) 17H06133 and 20H05639 , JSPS Grant-in-Aid for Transformative Research Areas (A) 20A203: JP20H05854, the joint research program of the Institute for Cosmic Ray Research, University of Tokyo, National Research Foundation (NRF), Computing Infrastructure Project of Global Science experimental Data hub Center (GSDC) at KISTI, Korea Astronomy and Space Science Institute (KASI), and Ministry of Science and ICT (MSIT) in Korea, Academia Sinica (AS), AS Grid Center (ASGC) and the National Science and Technology Council (NSTC) in Taiwan under grants including the Rising Star Program and Science Vanguard Research Program, Advanced Technology Center (ATC) of NAOJ, and Mechanical Engineering Center of KEK.

J. Janquart and C. Van Den Broeck are supported by the research programme of the Netherlands Organisation for Scientific Research (NWO). S.Goyal is supported by the Department of Atomic Energy, Government of India. J.M. Ezquiaga is supported by the European Union’s Horizon 2020 research and innovation program under the Marie Sklodowska-Curie grant agreement No. 847523 INTERACTIONS, and by VILLUM FONDEN (grant no. 53101 and 37766).{\'A}.\,Garr{\'on}, D.\,Keitel, P.\,Cremonese and S.\,Husa are supported by the Universitat de les Illes Balears (UIB); the Spanish Ministry of Science and Innovation (MCIN) and the Spanish Agencia Estatal de Investigaci{\'o}n (AEI) grants PID2019-106416GB-I00/MCIN/AEI/10.13039/501100011033, RED2022-134204-E, RED2022-134411-T; the MCIN with funding from the European Union NextGenerationEU (PRTR-C17.I1); the FEDER Operational Program 2021--2027 of the Balearic Islands; the Comunitat Aut{\`o}noma de les Illes Balears through the Direcci{\'o} General de Pol{\'i}tica Universitaria i Recerca
with funds from the Tourist Stay Tax Law ITS 2017-006 (PRD2018/24, PDR2020/11);
the Conselleria de Fons Europeus, Universitat i Cultura del Govern de les Illes Balears; and EU COST Action CA18108. {\'A}.\,Garr{\'on} is supported through SOIB, the Conselleria de Fons Europeus, Universitat i Cultura and the Conselleria de Model Econ{\`o}mic, Turisme i Treball with funds from the Mecanisme de Recuperaci{\'o} i Resili{\`e}ncia (PRTR, NextGenerationEU).
D.\,Keitel is supported by the Spanish Ministerio de Ciencia, Innovaci{\'o}n y Universidades (ref. BEAGAL 18/00148) and cofinanced by UIB. The authors thank
the Supercomputing and Bioinnovation Center (SCBI) of the University of Malaga
for their provision of computational resources and technical support (www.scbi.uma.es/site) and thankfully acknowledge the computer resources at Picasso and the technical support provided by Barcelona Supercomputing Center (BSC) through grants No. AECT-2022-1-0024, AECT-2022-2-0028, AECT-2022-3-0024, and AECT-2023-1-0023 from the Red Espa{\~n}ola de Supercomputaci{\'o}n (RES). J. Garcia-Bellido acknowledges support from the Spanish Research Project PID2021-123012NB-C43 [MICINN-FEDER], and the Centro de Excelencia Severo Ochoa Program CEX2020-001007-S at IFT. Prasia P. would like to thank Prof. Sukanta Bose for his support and IUCAA, Pune for providing computational facilities. A. K.Y. Li and R. K. L. Lo are supported by the National Science Foundation Grants PHY-1912594  and PHY-2207758. A. Mishra would like to thank the University Grants Commission (UGC), India, for financial support as a research fellow.

%% file: appendix_lensID.tex
In~\citet{LIGOScientific:2023bwz} the pairs which had false positive probability (FPP) less than 0.01 either with PO or \textsc{LensID} were passed on for the follow-up analysis.  According to PO the GW191103--GW191105 pair is found to be one of the most significant candidates. However, the \textsc{LensID} FPP is found to be 0.16. We cannot determine for certain why \textsc{LensID} did not find the pair significant for follow-up analysis, however, we can identify some possible contributing factors. Before detailing these, we briefly summarise how \textsc{LensID} works. \textsc{LensID} is made up of two ML models, one which takes Q-transforms input, and another which takes skymaps as input. On the basis of the Q-transforms, the network outputs the probability for the event pair to be lensed for each detector. Additionally, there is one output lensing probability based on the sky map. The entire probability for lensing is then computed by taking the four individual probabilities mentioned above. For more details we refer the reader to~\citet{Goyal:2021hxv}.

\begin{figure*}
\centering
\includegraphics[width= 0.49\linewidth]{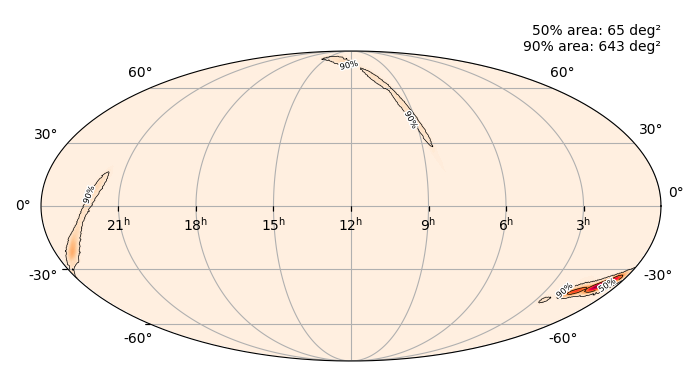}
\includegraphics[width=  0.49\linewidth]{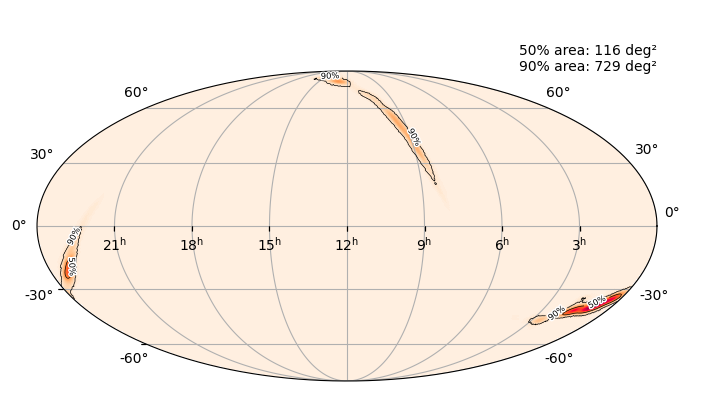}
\caption{Sky map for GW191105 from GraceDB created using \textsc{LALInference} (left) and from GWOSC created using \textsc{Bilby} after de-glitching the Virgo data (right). The \textsc{LALInference} sky maps are narrower as compared to \textsc{Bilby} ones, probably because of the glitch present in the data. \label{fig:sky}}
\end{figure*}

GW191103 was observed only in two detectors, LIGO Hanford (H1) and LIGO Livingston (L1), whereas,  GW191105 was observed in all three detectors but was contaminated by a glitch in the Virgo detector.  As seen in Fig.~\ref{fig:sky}, the final PE skymap of the event (right panel in Fig.~\ref{fig:sky}), which is made after deglitching the data~\citep{LIGOScientific:2021djp}, is different from the initial skymap (left panel in Fig.~\ref{fig:sky}), reducing the sky map FPP from 0.08 to 0.02 after using the PE sky map, still not crossing the threshold.  

\begin{figure*}
\centering
\includegraphics[width= 0.7\linewidth]{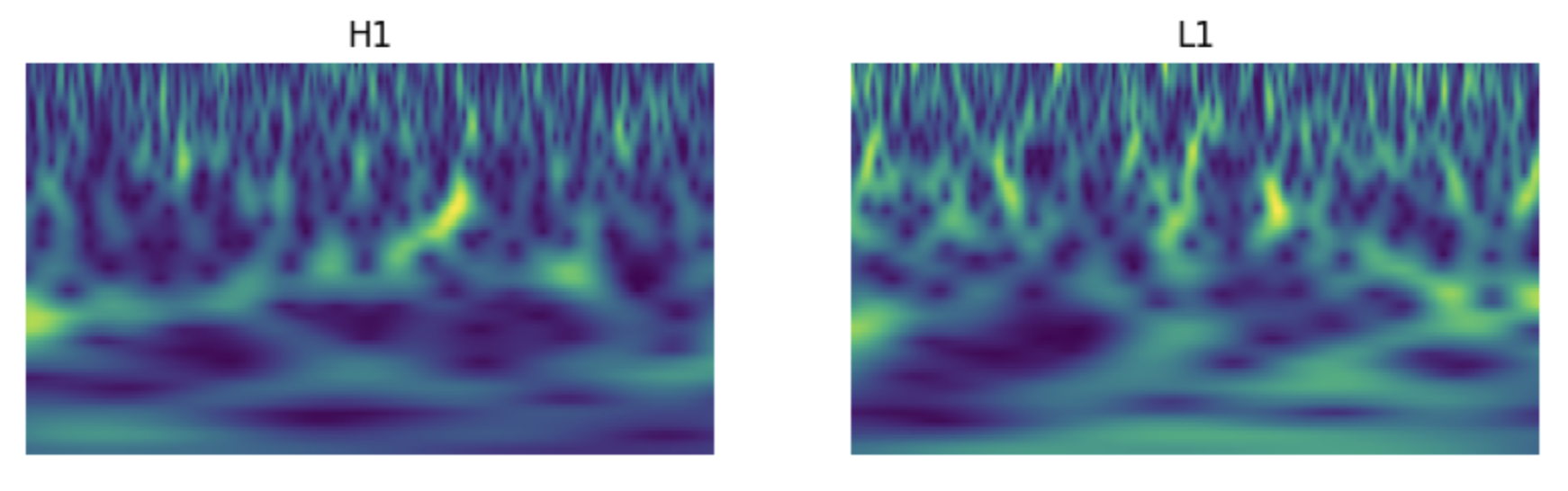}
\includegraphics[width= 0.7\linewidth]{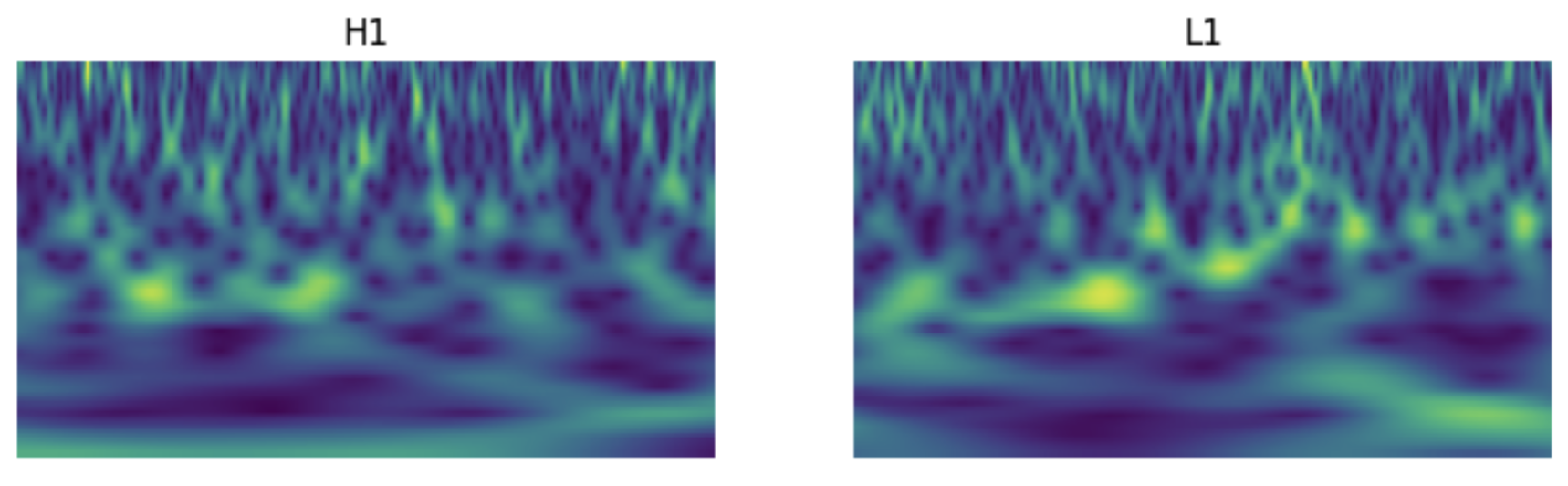}
\caption{Q-transforms images input to the \textsc{LensID} pipeline for GW191103 (top panel) and GW191105 (bottom panel). The chirping feature for GW191105 is broken in both the LIGO detectors, whereas for GW191103 the chirp signal is fairly visible in Hanford, and not so visible in Livingston. This could be one of the reasons why \textsc{LensID} did not identify this pair as significant\label{fig:qts}}
\end{figure*}

For the Q-transforms,  only the H1 and L1 detectors data are used by the framework.  We notice that the Q-transforms for the events, especially for GW191105, are visually poor.  They seem to be broken in the middle, as shown in the Fig.~\ref{fig:qts}.  Notice that the Q-transform of GW191105 in the L1 detector has a gap in the middle of the signal with peaks of power on both sides of the gap. This is not expected from a GW chirp signal. We checked that even though the SNRs are similar for both the events in the H1 and L1 detectors, the estimated probability for lensing varies a lot between the two detectors, 0.86 for H1 and 0.12 for L1.  This indicates that the ML algorithm is not robust to real noise fluctuations, which is expected as it is trained using simulated Gaussian noise signals. Additionally, from an injection study we found that \textsc{LensID} is more prone to misclassifying lensed signals with low chirp masses ($<20 M_{\odot}$) which is the case here.  In the future, to mitigate these problems, the ML models will be trained and tested on data containing real noise and lower chirp masses. 

%% file: app_pycbcsub.tex
\begin{figure*}
    \centering
    \includegraphics[width=0.7\linewidth]{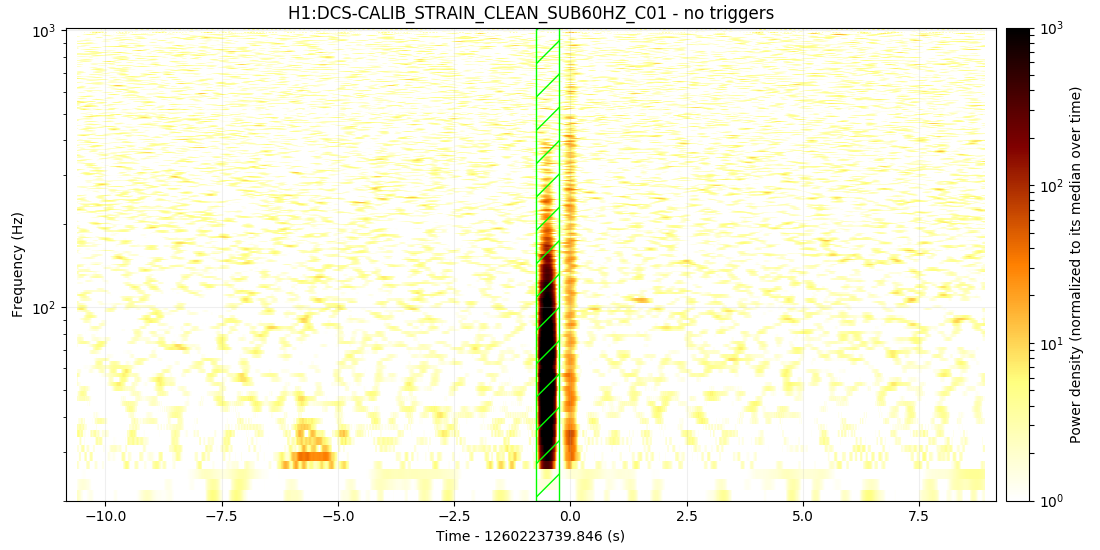}
    \caption{Time-frequency map of the third-ranked PyCBC candidate.
     This shows the glitch present near to GPS time 1260223739. 
     This represents the kind of quieter glitches that gets skipped in the normal autogating procedures.\label{fig:event_3_glitch}}
\end{figure*}
\begin{figure*}
\centering
    \includegraphics[width=0.7\linewidth]{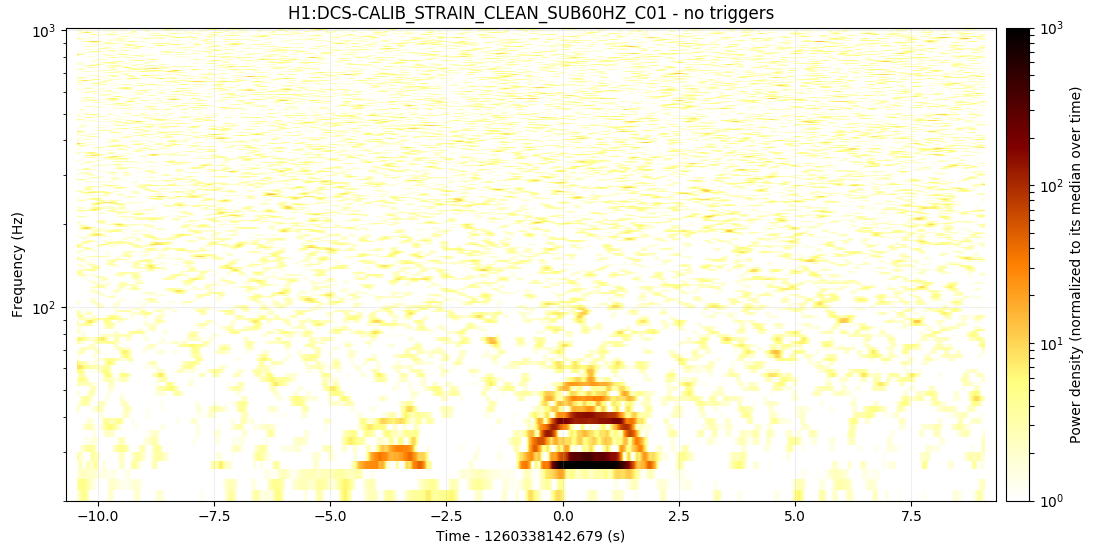}
    \caption{Time-frequency map of the fourth-ranked PyCBC candidate,
     consistent with a scattered-light glitch~\citep{Soni:2021lde, Soni:2021def, tolley2023archenemy}.\label{fig:event_4_glitch}}
\end{figure*}

In this section, we show the time-frequency maps for the two discarded (third and fourth ranked) candidate triggers found as possible sub-threshold counterparts for the GW191230 event by the PyCBC-based pipeline. The two can clearly be identified as glitches, with the third-ranked clearly having a power excess across a broad frequency band at the same time without presenting a time-frequency evolution similar to the one expected for a genuine GW signal (see Fig.~\ref{fig:event_3_glitch}), and the fourth in ranking (see Fig.~\ref{fig:event_4_glitch}) clearly matching a scattered-light glitch~\citep{Soni:2021lde, Soni:2021def, tolley2023archenemy}.

%% file: appendix_millilensing.tex
This Appendix presents additional details of the millilensing investigation for GW200208. 

\begin{figure*}
	\centering
	\includegraphics[width=\linewidth]{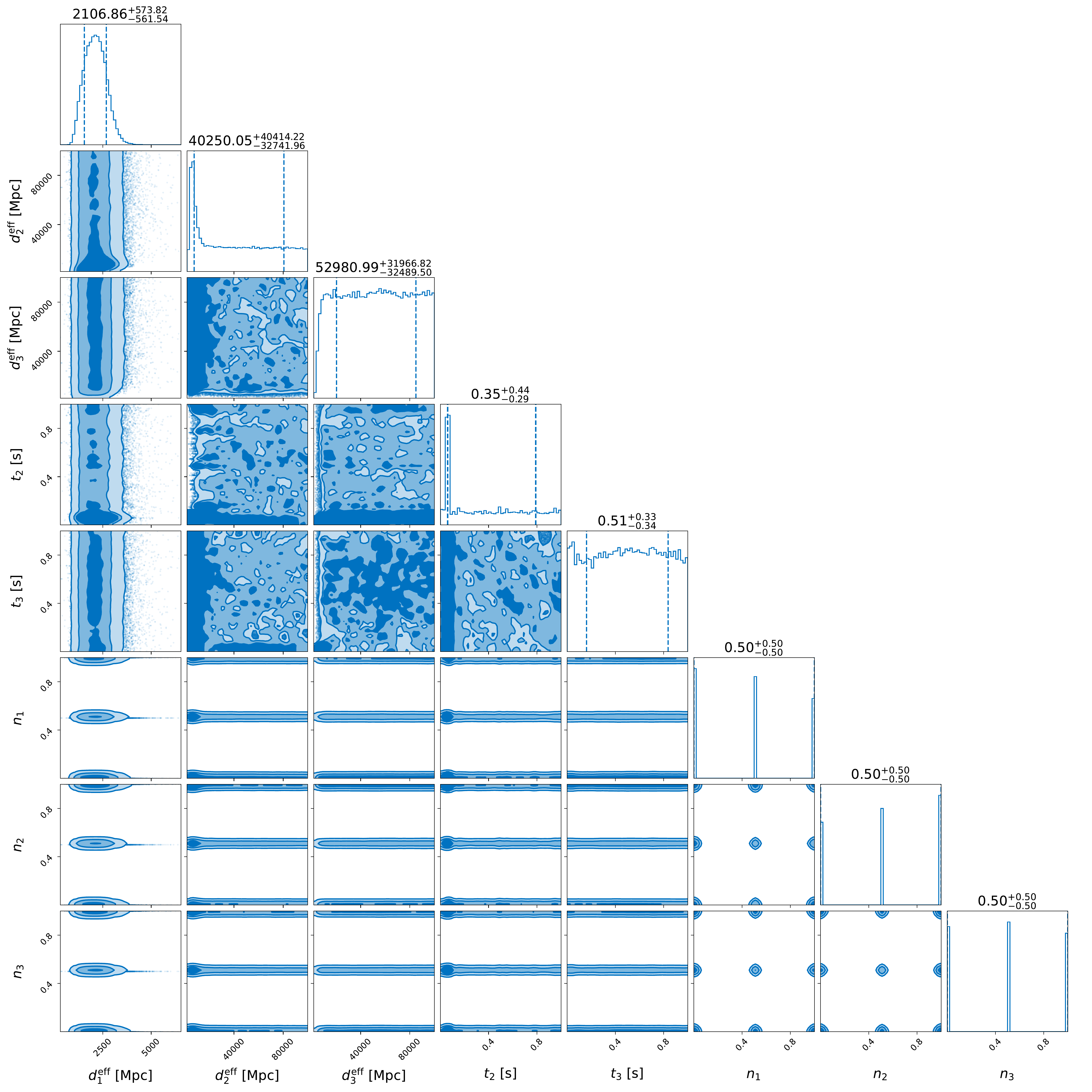}
	\caption{Corner plot of the millilensing parameters obtained from a three-signal analysis. A similar peak appears in the second image parameters as was present within the two-signal analysis shown in Fig.~\ref{fig:milli_two}. However, no such features are present within the posteriors of the third millilensing image.}\label{fig:milli_three}
\end{figure*}

\begin{figure*}
    \centering
    \includegraphics[width=0.49\linewidth]{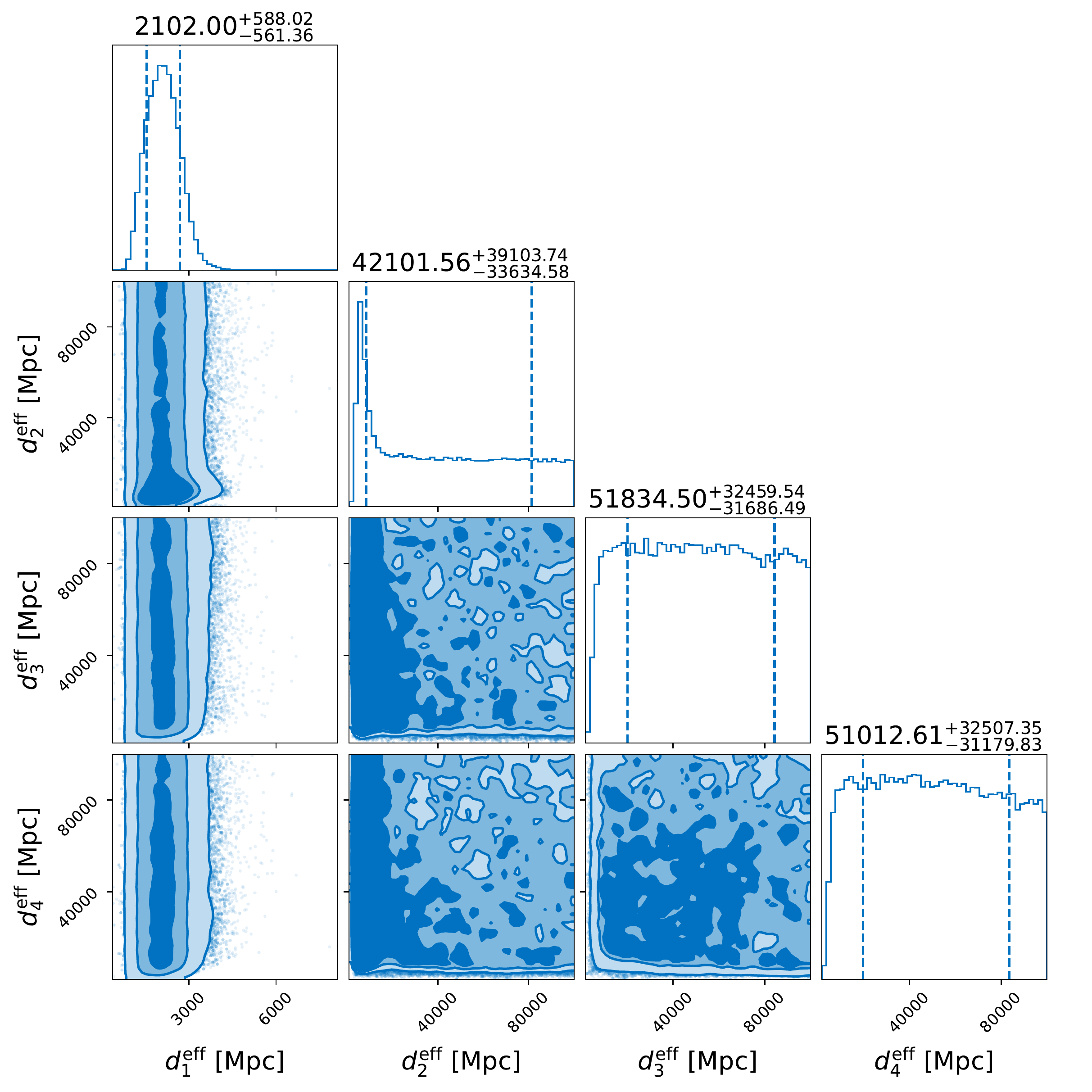}
    \includegraphics[width=0.49\linewidth]{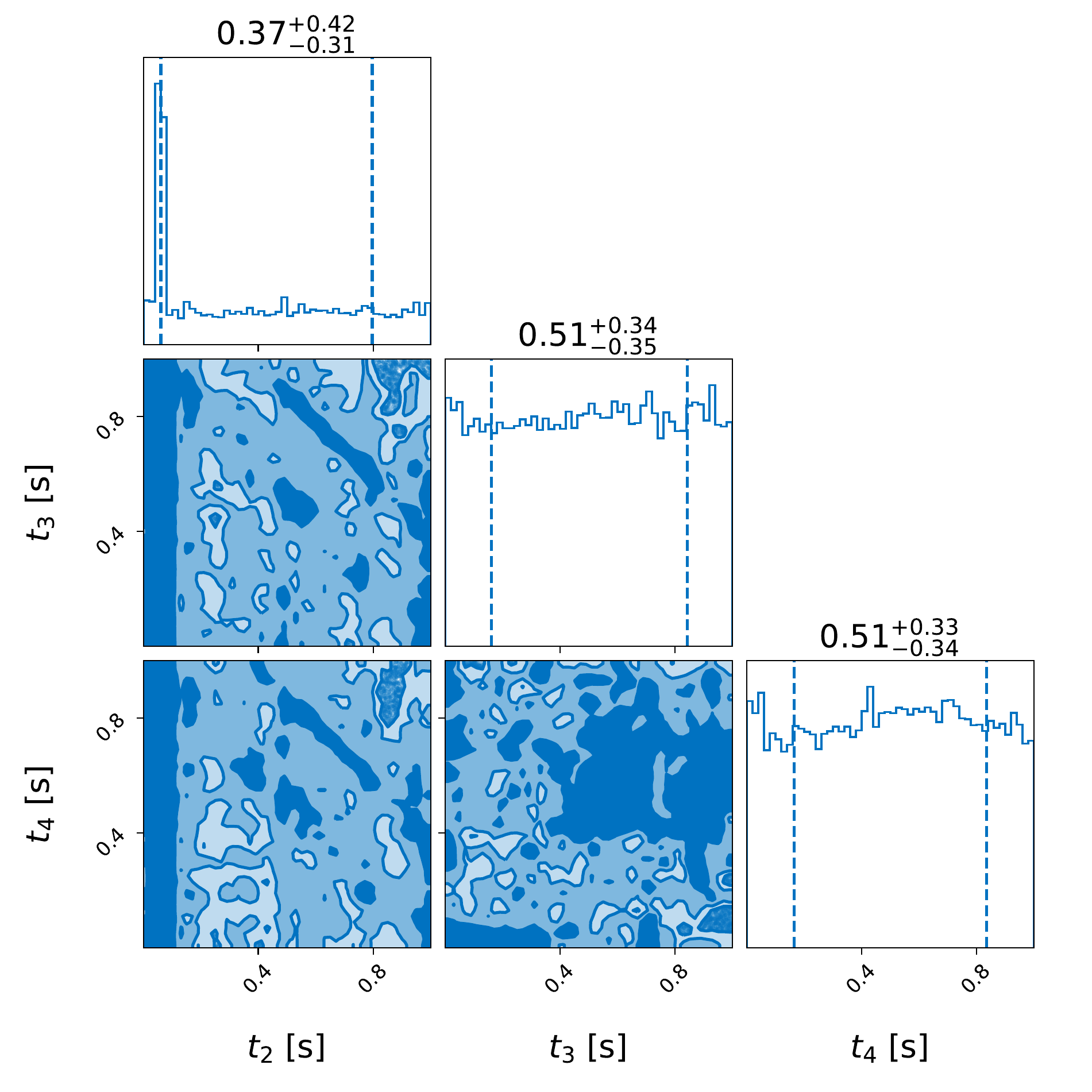}
    \caption{Corner plots of the effective luminosity distance (left) and time delay (right) parameters obtained from the four-signal analysis. Consistent with the previous analyses, there are no peaking features in the third or fourth signal posteriors.}\label{fig:milli_four}
\end{figure*}

Fig.~\ref{fig:milli_three} represents the result for a millilensing run done with 3 possible superposed images. It shows that in comparison with Fig.~\ref{fig:milli_two} the addition of a third image is not leading to the recovery of an extra possible image since the posterior for its lensing parameters are flat and uninformative. Similarly, the posteriors for the four-image analysis (Fig.~\ref{fig:milli_four}) show flat posteriors for the lensing parameters of the third and fourth possible images, leading to the conclusion that no more than two images can be identified in the data.

\begin{figure*}
    \centering
    \includegraphics[width=0.5\linewidth]{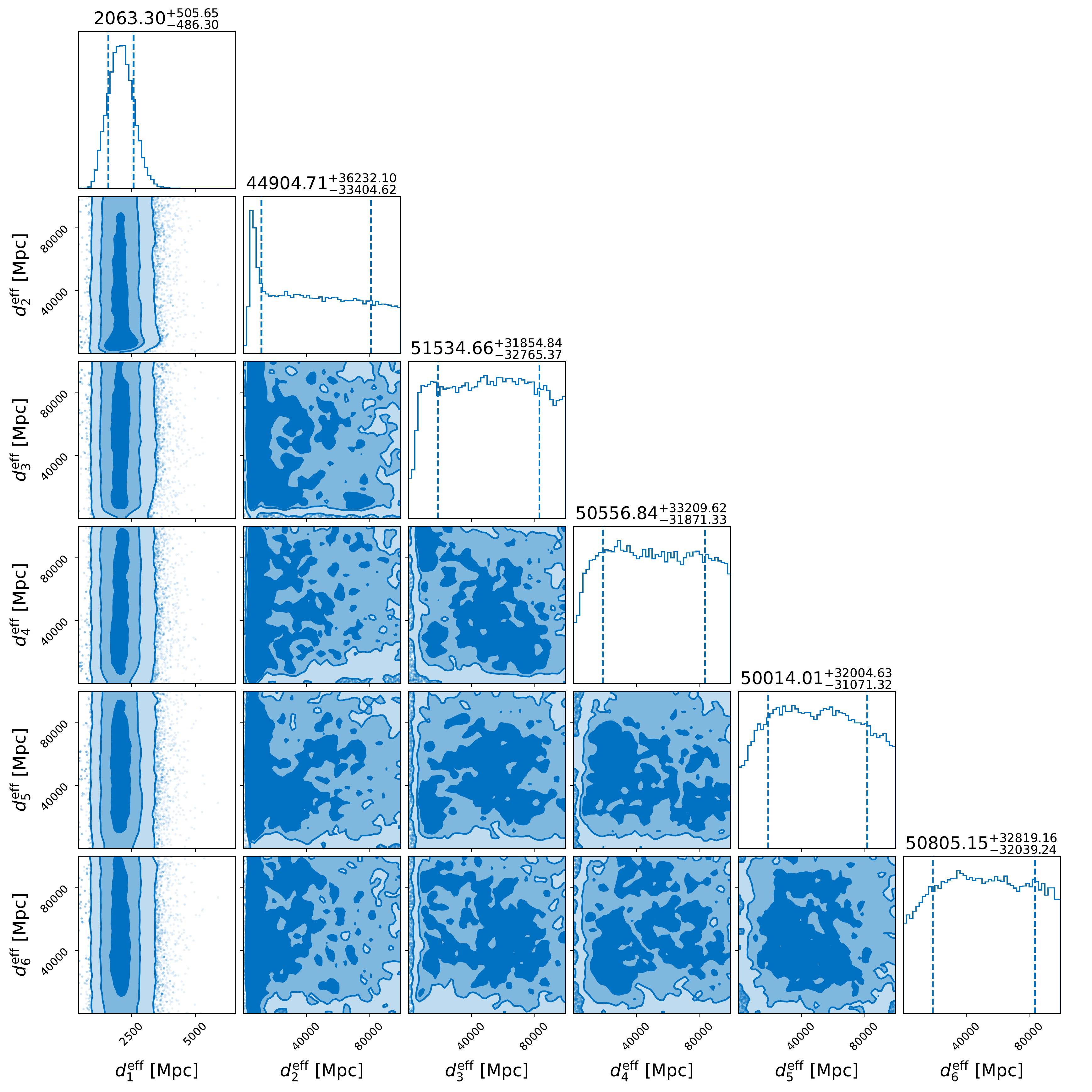}
    \includegraphics[width=0.49\linewidth]{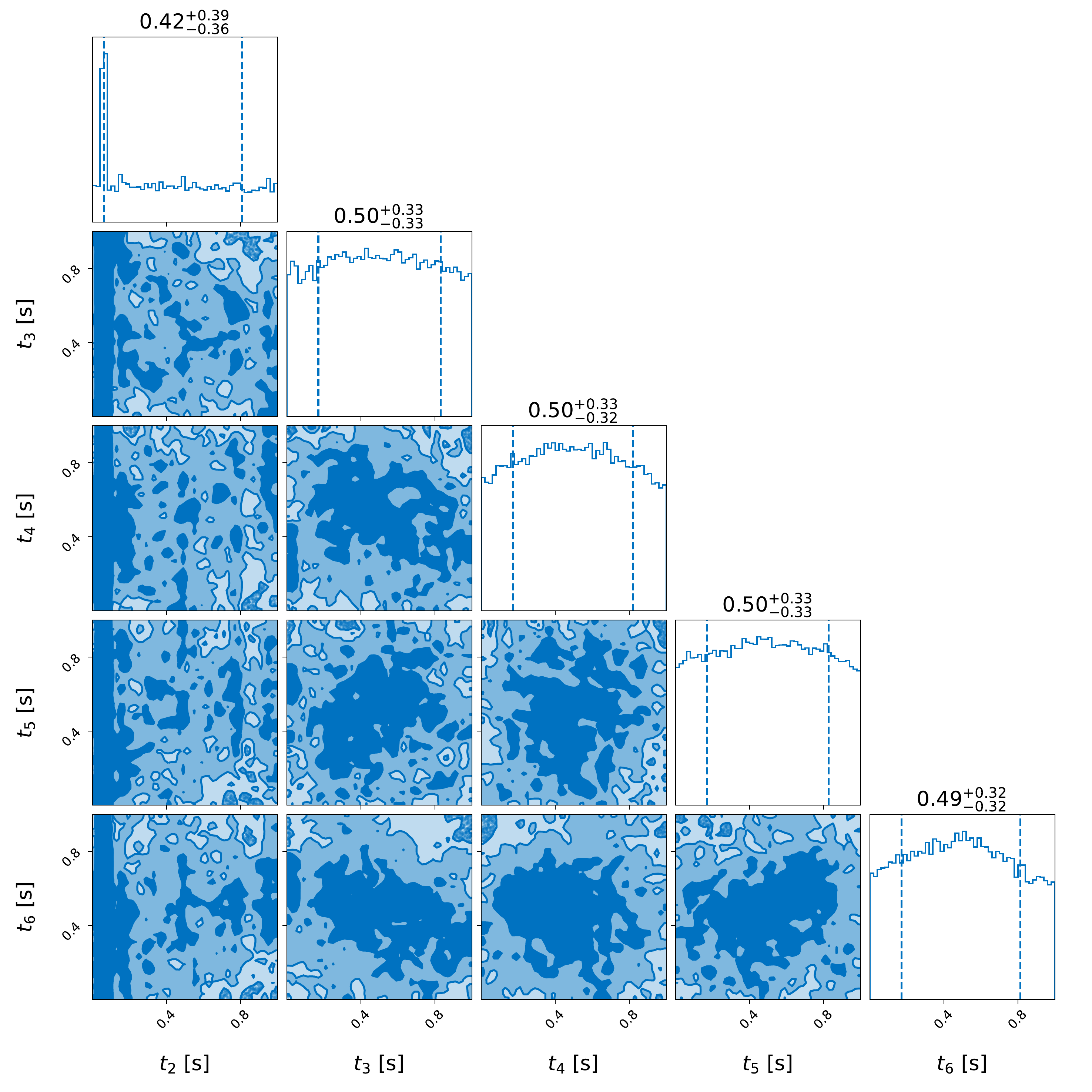}
    \caption{Corner plots of the effective luminosity distance (left) and relative time delays (right) obtained from a multi-signal analysis. Again, these are consistent with what has been seen in the previous analyses with no favouring for any number of signals above two.}\label{fig:milli_multi}
\end{figure*}

In addition to a posterior on the possible number of images, the run where the number of images is left free also returns posteriors for the lensing parameters of the different images. These are shown in Fig.~\ref{fig:milli_multi}. Only the posteriors for a possible second image are not completely uninformative. The others are flat, meaning that the analysis does not favour anything with more than two signals. 

